\documentclass{iopart}
\newcommand{\eq}{\begin{eqnarray}} 
\newcommand{\en}{\end{eqnarray}}
\def\bra#1{\mathinner{\langle{#1}|}}
\def\ket#1{\mathinner{|{#1}\rangle}}
\newcommand{\braket}[2]{\langle #1|#2\rangle}
\usepackage{bbm,graphicx,amssymb,amsfonts,units,bm,amsthm,mathrsfs,hyperref}

\begin{document}
\topical[Essential entanglement]{Essential entanglement for atomic and molecular physics}
\author{Malte C. Tichy$^{1}$, Florian Mintert$^{1,2}$, Andreas Buchleitner$^{1}$}
\address{(1) Physikalisches Institut - Albert-Ludwigs-Universit\"at Freiburg - Hermann-Herder-Strasse 3, D-79104 Freiburg im Breisgau, Germany}
\address{(2) Freiburg Institute for Advanced Studies (FRIAS), Albert-Ludwigs-Universit\"at Freiburg, Albertstrasse 19, D-79104 Freiburg, Germany}

\ead{a.buchleitner@physik.uni-freiburg.de}
\date{\today}
\begin{abstract}
Entanglement is nowadays considered as a key quantity for the understanding of correlations, transport properties, and phase transitions in composite quantum systems, and thus receives interest beyond the engineered applications in the focus of quantum information science. We review recent experimental and theoretical progress in the study of quantum correlations under that wider perspective, with an emphasis on rigorous definitions of the entanglement of identical particles, and on entanglement studies in atoms and molecules.
\end{abstract}
\date{\today}

\submitto{\JPB}
\maketitle

\tableofcontents

\section{Introduction}
While first considered as an indicator of the incompleteness of quantum physics \cite{PhysRev.47.777}, entanglement \cite{Schrodinger:1935ly} is today understood as one of the quantum world's most important and glaring properties. It contradicts the intuitive assumption that any physical object has distinctive individual properties that completely define it as an independent entity and that the result of measurement outcomes on one system are independent of any operations performed on another space-like separate system, an attitude also known as \emph{local realism}.

Thereby it poses important epistemological questions \cite{Schrodinger:1935fk}. 
Since an experimental test of the scenario suggested in \cite{PhysRev.47.777} to prove that incompleteness was long considered unfeasible, the interest in entanglement was long rather restricted to the philosophical domain. Not less than 30 years after the formulation of the Einstein Podolsky Rosen (EPR) paradox \cite{PhysRev.47.777,Bohm:1957vn}, a proposal for the direct, experimental violation of local realism, paraphrased in terms of a simple inequality, re-anchored the discussion on physical grounds. The \emph{Bell inequality} \cite{Bell:1964pt,Clauser:1969qa}, which sets strict thresholds on classical correlations of measurement results, was then proven to be violated in experiments which employed entangled states of photons. Stronger correlations than permitted by local realism were thus testified \cite{Freedman:1972fk,Aspect:1981zr,Salart:2008uq,Groblacher:2007kx,Tittel:1998ve}. 

Possible technological applications have triggered enormous interest in quantum correlations. With the discovery of Shor's factoring algorithm \cite{shor:303}, which relies on entanglement, quantum correlations became a topic of the information sciences, since they hold the potential to very considerably speed up quantum computers with respect to classical supercomputing facilities \cite{Jozsa:2003fk} and may thus jeopardize classical data encryption. Also other quantum technologies such as quantum imaging \cite{al:2002uq}, certain key-distribution schemes in quantum cryptography \cite{Ekert:1991kx} or quantum teleportation \cite{Bennett:1993hc} rely on  entanglement. Extensive research activity in these diverse areas by now led to considerable progress in our understanding of quantum correlations when associated with engineered systems with well-defined substructures. 

More recently, the theory of entanglement has penetrated into other fields of physics, to gain a fresh perspective on naturally occurring, often, rather complex systems, and to understand the role of quantum correlations for their spectral and dynamical properties, or for their functionality, even in biological structures \cite{Scholak:2011fk,scholak3560,Scholak:2010fk,Cai:2010qf,Briegel:2008ff}. It was shown, {\it e.g.},  that entanglement yields a versatile characterization of quantum phase transitions in many-body systems \cite{Amico:2008ph,Wu:2004ly,Osterloh:2002ve,MintertRey:2009uq,Gu:2004bs}, and that simple concepts like the \emph{area law} are indicative of an efficient numerical treatment of certain types of many-particle systems \cite{Eisert:2010fk}.

In atomic and molecular physics experimental progress  nowadays permits the detailed analysis and control of various coherent phenomena in few-body dynamics where entanglement is once again a potentially very useful tool. For example, it was recently proposed to solve the long-standing problem of a core vacancy (de)localization during a molecular ionization process by analyzing the entanglement between the photo- and Auger electrons born in such process \cite{al.:2007df}. 

The physical objects encountered in these latter fields are, however, much more difficult to control and to describe than the designed and engineered systems familiar from quantum information science. Atoms and molecules do not naturally exhibit definite and clear entanglement properties, nor well-separated entities such as photons in different optical modes \cite{Wieczorek:2009ff,PhysRevLett.101.010503}  or strongly repelling ions in radio frequency traps \cite{Blatt:2008cl} do: The identification of subsystems which can carry entanglement therefore becomes a non-trivial question, possibly complicated by the indistinguishability of particles, and typical Hilbert space dimensions tend to be rather large. Additionally, the interaction between system constituents is typically of long-range type, thus rendering entanglement a dynamical quantity, difficult to grasp, and without any unambiguous straightforward definition.  Moreover, it remains to be clearly defined to which extent phenomena like macroscopic quantum superpositions imply the existence of entanglement \cite{PhysRevA.81.010101,PhysRevLett.65.1838}, and vice-versa. This defines a challenging task for conceptual research in quantum information theory, which ought to respond to novel experimental challenges.

The scope of the present Topical Review is threefold: First, we will present an overview over the different facets entanglement can have, together with a conceptual framework which permits to compare the entanglement properties of distinct physical systems. Second, we will illustrate these theoretical concepts  with specific examples from experiments. Finally, we will discuss recent experimental and theoretical developments, in areas where entanglement  receives attention only very recently. In order to keep the presentation most intuitive, we will focus on specific physical realizations of entanglement rather than on abstract mathematical properties, thus always stressing the \emph{physical} impact of entanglement on measurement results.

To avoid redundancies with earlier reviews, we will not cover studies of entanglement in established fields, but refer the interested reader to \cite{RevModPhys.81.865,Bengtsson:2006fu}, where the mathematical and quantum-information aspects of entanglement are reviewed extensively. More specialized reviews focus on the specific properties of multipartite entanglement \cite{Eisert:2006il}, on the entanglement in continuous-variable systems \cite{Adesso:2007ul,Braunstein:2005mb}, and on the interconnection between entanglement and  violations of local realism \cite{Genovese:2005cr,Werner:2001dz}. An introduction to the quantification of entanglement, {\it i.e.}~on entanglement measures, can be found in \cite{Plenio:2006uq}, an overview on approximations to such measures of entanglement, especially on efficient lower bounds, together with applications to general scenarios of open system entanglement dynamics, is given in \cite{Mintert:2005rc}. Applications of entanglement to the simulation of many-body-systems are reviewed in \cite{Amico:2008ph,Latorre:2009fv}, focus on area laws can be found  in \cite{Eisert:2010fk}. The relevance of entanglement for decoherence theory is touched upon in  \cite{Buchleitner:2009fk,abucoherent}, entanglement in trapped ion systems is discussed in \cite{Haffner:2008wo}, and entanglement between electronic spins in solid-state devices in \cite{Burkard:2007ly}.

The text is organized as follows: In the next Section, we introduce fundamental notions of entanglement and related concepts, such as to define a common language for the subsequent Sections. In Section \ref{secentities}, we discuss the different possibilities to subdivide physical systems into subunits, and the influence of the specific choice of the partition on the resulting entanglement. Contact is made to state-of-the-art quantum optics experiments, for illustration. Finally, in Section \ref{atommol}, we move on to theoretical and experimental studies of quantum correlations in atomic and molecular physics which bear virtually all of the aforementioned complications. We conclude with an outlook on possible future directions for studies of atomic, molecular and biological systems from a quantum-information perspective. 
\section{Subsystem structures and entanglement quantifiers}
Let us now shortly recollect the required basic notions of entanglement. Since several reviews and introductory articles are available \cite{RevModPhys.81.865,Plenio:2006uq,Mintert:2005rc}, we introduce only those concepts necessary for the understanding of the subsequent Sections, to make this review self-containted. In particular, we define the different entanglement measures employed in atomic and molecular physics. We also explicitly discuss the requirements on the subsystem-structure, which are usually implicitly assumed. This is necessary to establish a general formalism that will allow us to classify the many diverse approaches to entanglement in different systems. 

\subsection{Quantum and classical correlations}
The central property that attracts broad attention are the ``nonclassical'' correlations of measurement results on different subsystems of an entangled state \cite{PhysRev.47.777}. The situation is easiest illustrated by a two-body system with two spatially well-separated subsystems. The Hilbert space $\mathscr{H}$ of the two-body system then naturally decomposes into a tensor product $\mathscr{H}_1 \otimes \mathscr{H}_2$
of the two Hilbert-spaces $\mathscr{H}_1$ and $\mathscr{H}_2$ of the two subsystems, which we both  assume to be of dimension $d$.

Now, consider the following exemplary bipartite state:
\eq
\ket{\Psi}=\frac{1}{\sqrt{d}}\sum_{j=1}^{d}\ket{j}\otimes\ket{j} \, .
\label{eq:psimm}
\en
A projective measurement in the (orthonormal) basis $\{\ket{j}\}$, performed on the first subsystem, has completely unpredictable results:
each of the possible outcomes will occur with the same probability $1/d$.
The same is true for the analogous measurement on the second subsystem. Yet, the measurement results on \emph{both} subsystems are perfectly correlated:
once a measurement result -- say $j=j_0$ -- is obtained on one subsystem, the two-body state is projected on the state $\ket{j_0}\otimes\ket{j_0}$,
so that a subsequent measurement on the other subsystem will yield the result $j_0$ with certainty.

Such correlations of measurement results might seem surprising. They could, however, be explained rather simply: Think, for example, of an experiment in which both subsystems are always prepared in the same (random) state $\ket{j}$.
In subsequent runs of the experiment (which are required to obtain reliable measurement statistics) the choice of $j$ is completely random. 
The experimentalist thus creates a mixed state
\eq \rho=\frac 1 d \sum_{j=1}^d \ket{\psi_j} \bra{\psi_j}\ , \ \mathrm{ with }\ket{\psi_j}=\ket{j}\otimes\ket{j}\ . \label{sepmixed} \en
This mixed state gives rise to exactly the same correlations of measurement results as found for our initial example (\ref{eq:psimm})  above. In fact, one does not even need a quantum mechanical system to observe such correlations --
also colored socks \cite{Bell:1987uq} or marbles will do.

The situation changes completely, however, if a measurement in a second, \emph{distinct} orthonormal basis set $\{\ket{\alpha_j}\}$ is performed: To be specific, let us assume that this measurement is performed on  the first subsystem.
Given the measurement outcome $\alpha_p$, the two-body state (\ref{eq:psimm}) is projected on the state
\eq
\ket{\alpha_p}\otimes \sum_j  \braket{\alpha_p}{j}\ket{j}=\ket{\alpha_p}\otimes\ket{\tilde\alpha_p}\ ,
\en
that is to say, the second subsystem is left in the state $\ket{\tilde\alpha_p}=\sum_i\langle\alpha_p|j\rangle|j\rangle$.
Quite strikingly, the states $\ket{\tilde\alpha_p}$ ($p=1,...,d$) are mutually orthogonal:
\eq
\langle\tilde\alpha_q|\tilde\alpha_p\rangle=
 \sum_{k,j} \langle j|\alpha_q\rangle \langle j|k\rangle \langle\alpha_p|k\rangle=\langle\alpha_p|\alpha_q\rangle=\delta_{pq}\ .
\en
The outcomes of a subsequent measurement on the second subsystem in the basis $\ket{\tilde\alpha_p}$ can therefore be predicted with certainty, although they were completely undetermined before the prior measurement on the first subsystem.
If the same is attempted with the state (\ref{sepmixed}), the first measurement projects the state on
\eq
\ket{\alpha_p}\bra{\alpha_p}\otimes \left(\sum_{j=1}^d\left|\langle\alpha_p\right|j\rangle|^2\ket{j}\bra{j}\right)\ .
\en
In other words, the second subsystem is left in a mixed state, with incoherently added components $\ket{j}\bra{j}$. Unless the basis $\{\ket{\alpha_p}\}$ coincides with the basis $\{\ket{j}\}$, results of a projective measurement on this second system component remain uncertain.

In conclusion, a mixed state such as (\ref{sepmixed}) can only explain correlations that are observed in \emph{one specific}  single-particle basis, whereas the entangled state (\ref{eq:psimm}) exhibits correlations for \emph{all} possible choices of orthogonal local basis settings. 
We therefore see that quantum physics hosts a type of correlations which cannot be classically described. These correlations are often referred to as \emph{quantum correlations} and understood as the ultimate physical manifestation of ``entanglement''.

\subsection{Separable and entangled states} \label{sepaent}
In order to assign quantum states the label ``entanglement'' in a systematic fashion, we first need to define this new quality on a formal level, to subsequently introduce measures for the amount of entanglement that is carried by a quantum state.

\subsubsection{Pure states}
A bipartite quantum state \eq \ket{\Phi} \in \mathscr{H}_1 \otimes \mathscr{H}_2 \en is \emph{separable} if it can be written as a product state, {\it i.e.}~if one can find single particle states $\ket{\phi_i} \in \mathscr{H}_i$ such that \eq \ket{\Phi}=\ket{\phi_1} \otimes \ket{\phi_2}. \label{separable} \en 

Separable states are completely determined by the single-particle states
%A \emph{separable state} $\ket{\Phi}$ is completely determined in terms of the single-particle states 
$\ket{\phi_1}$ and $\ket{\phi_2}$, which  contain all information on possible measurement outcomes.
Unlike the situation described by (\ref{eq:psimm}), a measurement performed on one subsystem has no effect on the other subsystem, {\it i.e.}~the subsystems are uncorrelated. Consequently, the \emph{reduced density matrices},
\eq
\varrho_1=\mathrm{Tr}_2\left( \ket{\Phi}\bra{\Phi} \right)\hspace{1cm}\mathrm{and}\hspace{1cm}\varrho_2=\mathrm{Tr}_1 \left( \ket{\Phi}\bra{\Phi} \right) \ ,\label{eq:reddmat}
\en 
where 
\eq \mathrm{Tr}_{1(2)}(\rho) = \sum_{j=1}^d \bra{\chi_j}_{1(2)} \rho \ket{\chi_j}_{1(2)} ,  \label{partialtrace} \en denotes the partial trace over the first (second) subsystem, describe pure states, and 
\eq \varrho_1=\ket{\phi_1} \bra{\phi_1} , \hspace{1cm} \varrho_2=\ket{\phi_2} \bra{\phi_2} , \en
such that the compound state can be written as a tensor product, $  \ket{\Phi}\bra{\Phi}=\varrho_1 \otimes \varrho_2  . $ This is equivalent to the statement that the first particle is prepared in $\ket{\phi_1}$, while the second particle is prepared in $\ket{\phi_2}$, which corresponds to the assignment of a \emph{physical reality}, as will be discussed in more detail in Section \ref{properties} below. 

A state that is not separable is called \emph{entangled}. 
The information carried by an \emph{entangled} state $\ket{\Psi}$, which cannot be written as tensor product as in (\ref{separable}),  is not completely specified in terms of the states of the subsystems: For the reduced density matrices, we have $\mathrm{Tr}{\varrho_{1/2}^2}\neq 1$, {\it i.e.}~the subsystems' states are mixed. Furthermore,  one finds $\ket{\Psi}\bra{\Psi}\neq\varrho_1\otimes\varrho_2$, {\it i.e.}~the two-particle state $\ket{\Psi}$ contains more information on measurement outcomes than is contained in the two single-particle states $\varrho_1$ and $\varrho_2$ together, in contrast to the above separable state (\ref{separable}). 
Moreover, distinct entangled states can give rise to the same reduced density matrices: The states $\ket{\Phi^{+}}=\left( \ket{1,1}+\ket{0,0} \right)/\sqrt{2}$ and $\ket{\Psi^{+}}=\left( \ket{0,1}+\ket{1,0} \right)/\sqrt{2}$ lead to the same, maximally mixed, reduced density matrices, $\varrho_{1}=\varrho_{2}=\mathbbm{1}_2/2$, where $\mathbbm{1}_l$ denotes the identity in $l$ dimensions. The state $\ket{\Phi^{+}}$, however, describes a \emph{correlated} pair, whereas $\ket{\Psi^+}$ is \emph{anti-correlated.}  

Formally, the attribute of separability (and, correspondingly, entanglement) boils down to the question whether the coefficient matrix $c_{j,k}$ in the state representation 
\eq \ket{\Psi}=\sum_{j,k=1}^d c_{j,k} \ket{j,k} , \label{generalrepresent} \en
admits a product representation, {\it i.e.}~$c_{j,k}=c^{(1)}_j \cdot c^{(2)}_k$ 
-- in this case, the state is separable. 

\subsubsection{Mixed states}
Separable and entangled states can also be defined for mixed states of a bipartite quantum system, which have to be described in terms of density matrices. In this case, the mixedness of the reduced density matrices is not equivalent to entanglement. 

A state that can be expressed as a tensor product of single-body density matrices,
\eq
\rho_p=\varrho_1\otimes\varrho_2\ , \label{productstate}
\en
bears no correlations between local measurement results at all, and is called a \emph{product} state.

\emph{Separable} states are defined by sets of single particle states $\varrho_1^{(i)}$ and $\varrho_2^{(i)}$ of the first and second subsystem, respectively, and by associated probabilities $p_i$ ({\it i.e.}~$p_i\ge 0$, and $\sum_ip_i=1$), such that \cite{Werner:1989ve} 
\eq
\rho_s=\sum_i\ p_i\ \varrho_1^{(i)}\otimes\varrho_2^{(i)}\ .\label{separho}
\en
Such separable states imply correlations between measurement results on the different subsystems, but these correlations can be explained in terms of the probabilities $p_i$, and, therefore, do not qualify as quantum correlations.
The tag \emph{entanglement} is, thus, reserved for those states that \emph{cannot} be described in terms of product states of single-particle states and of classical probabilities as in (\ref{separho}). They thus need to be described as
\eq
\rho_{\mathrm{ent}} = \sum_j p_j \ket{\Psi_j}\bra{\Psi_j} \label{nonsepdd}
 ,\en 
where, in \emph{any} pure-state decomposition of $\varrho_{\mathrm{ent}}$, at least one state $\ket{\Psi_j}$ is entangled.

\subsection{Bell inequality violation and nonlocality} \label{Bellin}
So far, we identified the difference between classical and quantum correlations with the help of the exemplary states given by Eqs.~(\ref{eq:psimm}) and (\ref{sepmixed}), and we have given formal definitions for entangled and separable states. On the other hand, we still need to establish how to unambiguously identify and quantify the exceptional, ``non-classical'', correlations inscribed in (\ref{eq:psimm}) in an experimental setting. This is required to provide a connection to probability theory and to provide a benchmark for the verification of entanglement in an experiment. 

For this purpose, we first need to specify what is understood as ``classical'' in our present context. We call a theory ``classical'' if it is local and realistic. The principle of \emph{locality} states that any object is influenced directly only by its immediate surroundings, and that there can be no signals between space-like separated events. \emph{Realism} denotes the assumption that any physical system possesses intrinsic properties, {\it i.e.}~an experimentalist who measures the value of an observable merely reads off a predefined value \cite{PhysRev.47.777} (as, {\it e.g.}, for the mixed state (\ref{sepmixed}) that is created by a random, however, realistic mechanism). That is to say, although one may ignore the value of a certain observable, each observable still possesses a definite value at any moment. In practice, ``local realism'' implies that measurement outcomes at one subsystem are independent of the measurements performed on the other subsystem, provided both subunits are spatially separated. Theories that obey local realism can be described by classical random variables. 

A rigorous way to establish a correspondence between ``non-classical'' correlations and entanglement is provided by \emph{Bell inequalities} \cite{Bell:1964pt}. These are defined in terms of correlations between measurement results of different single-partice observables, and they give threshold values for the correlations to be describable in terms of classical probability theory. An experimental violation, {\it i.e.}~any excess beyond the threshold value, indicates that the corresponding observables cannot be described as classical random variables. The community jargon also speaks of the unavailability of a ``local realistic description''. 
A widely used Bell inequality is the one presented by Clauser, Horne, Shimony and Holt (CHSH) \cite{Clauser:1969qa}. It is formulated for \emph{dichotomic observables} such as polarization, {\it i.e.}~observables which only take the values $+1$ and $-1$, and reads 
\eq | \langle A_1 \cdot B_1 \rangle + \langle A_1 \cdot B_2 \rangle + \langle A_2 \cdot B_1 \rangle - \langle A_2 \cdot B_2 \rangle | \leq 2 \label{CHSH}, \en
where $A_j$ and $B_k$ are different observables acting on the two subsystems. This inequality is derived under the assumption of local realism, which means that measurement outcomes at one subsystem are independent of the measurements performed on the other subsystem, provided both subunits are spatially separated. In order to violate the inequality, the local observables need to be non-commuting, $[A_1,A_2] \neq 0 \neq [B_1,B_2]$.  It is hence necessary to implement a rotation of the measurement basis to assess such non-commuting observables. This is also illustrated in the polarization rotation effectuated on the photons in Figure \ref{Bellillu}:
\begin{figure}[h] \center \includegraphics[width=12.5cm,angle=0]{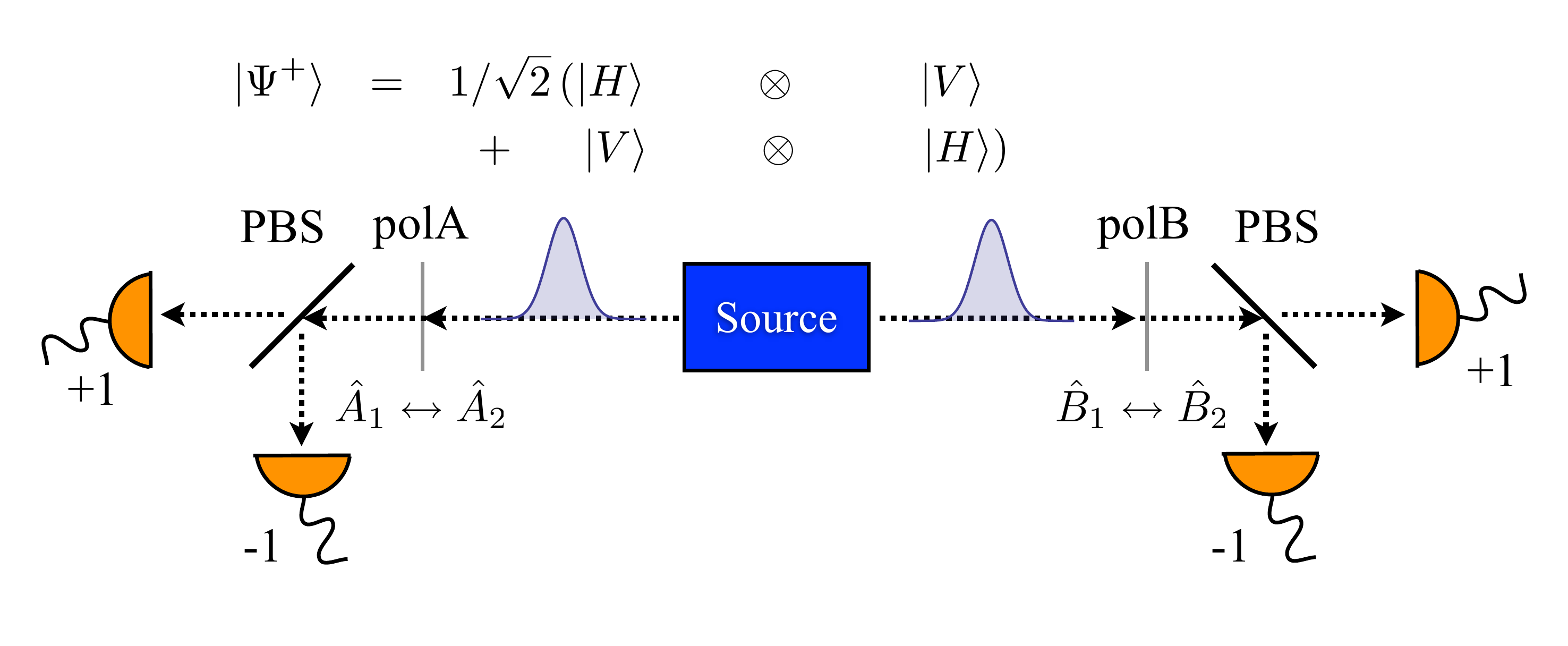} \caption{Bell-type experiment. A source creates photons in an entangled state, here the maximally entangled $\ket{\Psi^+}$ Bell state (\ref{Psimp}). The polarization of the photons is rotated at polA and polB, where at least two different settings, $\hat A_{1}/\hat A_2$ and $\hat B_{1}/\hat B_2$, are available. The photons then fall onto polarizing beam splitters (PBS) where horizontally polarized photons are reflected and vertically polarized ones are transmitted. The subsequent detection of the photons at either one of the two detectors yields the distinct values $\pm 1$.}  \label{Bellillu}  \end{figure}
In a polarization-entanglement experiment, it is not sufficient to measure the correlations in a given basis, say the $z$-basis, but the orientation of the quantization axis needs to be chosen locally. Similarly, correlations in the position or in the momentum alone \emph{do not} rigorously prove that a two-particle state is entangled in the external degrees of freedom (see also the discussion in Sections \ref{atomphotonent},\ref{twoelectronsmol}). The difficulty of the implementation of mutually distinct measurement bases is an impediment for the direct assessment of the entanglement properties of many naturally occurring, multicomponent quantum systems such as atoms \cite{Grobe:1994fk,Fedorov:2004kx}, or biological structures \cite{Arndt:2009uq}. It is then necessary -- and still largely open an issue -- to conceive alternative indicators that distinguish entanglement from classical correlations \cite{PhysRevLett.106.210501}.

For a \emph{maximally entangled bipartite qubit} (or \emph{Bell-}) \emph{state} as defined below in (\ref{Psimp}), (\ref{Phimp}) below, the expectation value of the left hand side of (\ref{CHSH}) can reach values up to $2 \sqrt 2\approx 2.82$, for a suitable choice of the measurement settings $A_j$, $B_k$. When such experimental conditions are met, the notion of non-locality as defined by Bell inequalities is qualitatively in agreement with the definition of separability and entanglement in Section \ref{sepaent} above: A violation of a Bell inequality proves that the state under consideration is entangled. The reverse is, however, not true: There are states which are entangled according to Section \ref{sepaent}, but do not violate any Bell inequality \cite{PhysRevLett.72.797}, since a description in terms of classical probabilities is available \emph{despite their non-separbility according to} (\ref{nonsepdd}) \cite{Werner:1989ve}. An example is given by the Werner-state $\rho_{\mathrm{W}}(p)$ \cite{Werner:1989ve}. For bipartite qubit systems, it reads 
\eq 
\rho_{\mathrm{W}}(p)=(1-p) \ket{\Psi^-}\bra{\Psi^-} + \frac{p}{4} \mathbbm{1}_4 , \label{WernerState}
\en
where $p\in [0,1]$, {\it i.e.}~the state is a mixture of the maximally entangled antisymmetric state $\ket{\Psi^-}$ given below in (\ref{Psimp}) and the maximally mixed, fully uncorrelated state $\mathbbm{1}_4/4= \mathbbm{1}_2/2 \otimes  \mathbbm{1}_2/2$ (see (\ref{productstate})). The entanglement and the non-local properties of the state depend on the parameter $p$: $\rho_{\mathrm{W}}(0)$ is a pure, maximally entangled state, which also violates (\ref{CHSH}) maximally. This violation persists for $0<p<1-1/\sqrt{2}$. For $p\ge 1-1/\sqrt{2}$, a local realistic description is available, but the state is still entangled as long as $p<2/3$. In other words, for $1-1/\sqrt{2}<p<2/3$, the state is entangled according to (\ref{nonsepdd}), but it does not violate local realism. This enforced qualitative distinction between non-locality and entanglement needs to be uphold for mixed states, whereas for pure bipartite qubit states, any non-product state violates a Bell inequality \cite{Gisin1991201}.

When we conclude from the violation of a Bell inequality that no local realistic description of a given experiment exists, we rely, among others, on the strong assumptions that the measurements which are performed on the subsystems are space-like separated \cite{PhysRevLett.81.5039}, and that possible detector inefficiencies are unbiased with respect to the measurement outcomes \cite{PhysRevD.2.1418,PhysRevD.35.3831}. These requirements represent serious challenges for experiments, such that tests of Bell inequalities are often plagued by \emph{loopholes}: The failure to fulfill the aforementioned assumptions may allow a description of the experiment outcomes by classical theories, and additional experimental effort is required to ``close the loopholes'' \cite{PhysRevLett.93.130409,Rowe:2001vn,Merali18032011}.

\subsection{Entanglement witnesses} \label{witnessesslabl}
Since the above discussion implies that non-locality is a stronger criterion than entanglement, Bell inequalities are not a universal means to experimentally detect quantum correlations. A general method to verify entanglement is given by 
\emph{entanglement witnesses} \cite{Horodecki:1996vn,Terhal:2002fk}, {\it i.e.}~operators which detect, or ``witness'', entanglement \cite{Barbieri:2003kx,Bourennane:2004bh}. A hermitian operator $\hat W$ is an entanglement witness if it fulfills 
\eq
\mathrm{Tr} (\rho_s \hat W ) \ge 0 \label{eq:witness_pure},\en
for all separable states $\rho_s$, and 
\eq
\mathrm{Tr} (\rho_e \hat W ) < 0 \label{eq:witness_purev},\en
for at least one entangled state $\rho_e$. 

\begin{figure}[h] \center \includegraphics[width=8.5cm,angle=0]{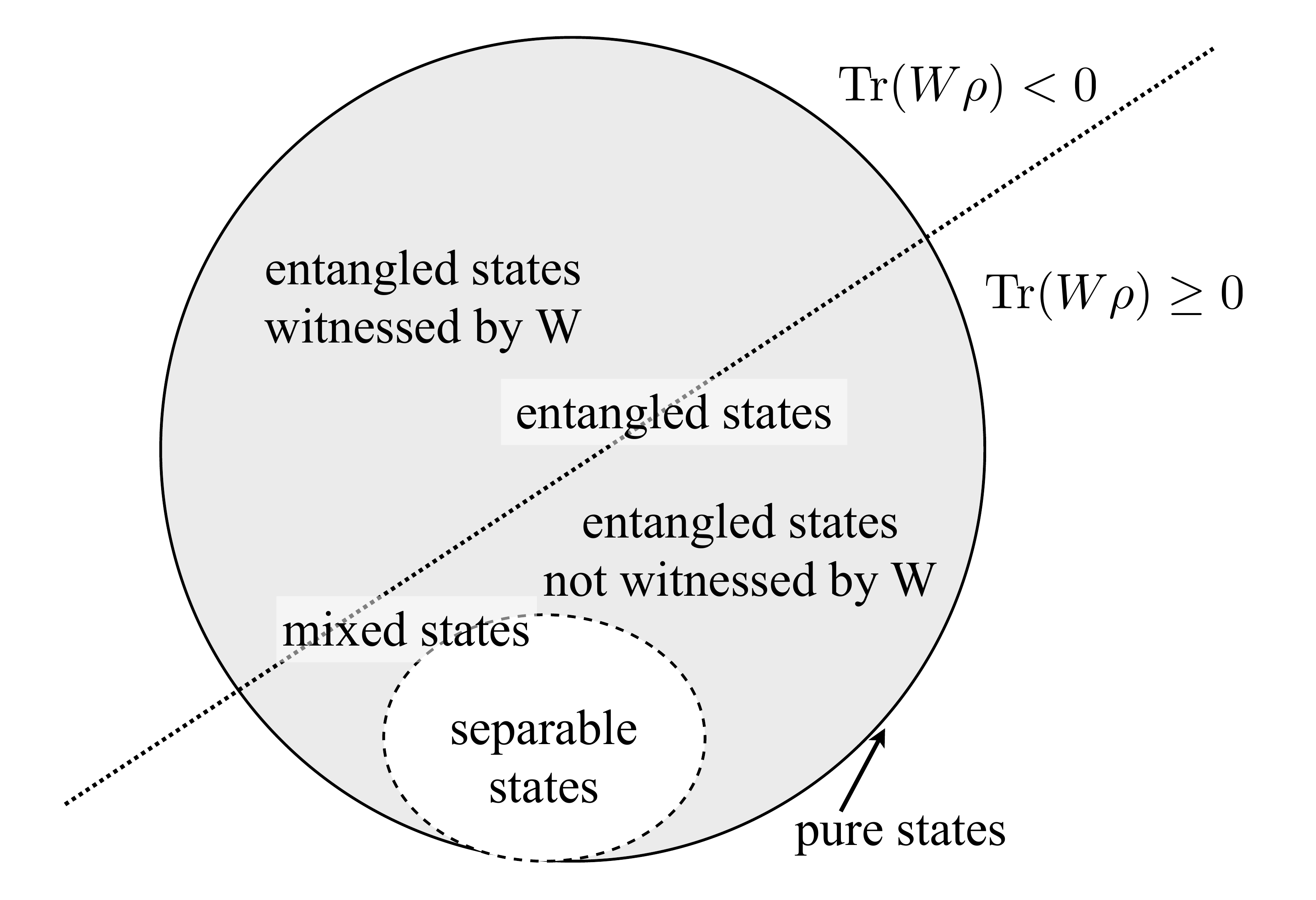} \caption{Illustration of the structure of the set of density matrices and of the concept of entanglement witnesses. Pure states lie on the surface of the convex set of density matrices; all states that are not on the surface are mixed. Separable states are a subset which is itself convex. The set of entangled states is shaded in gray, and not convex. In this geometrical picture, an entanglement witness defines a hyperplane (dotted line): States above that plane are detected to be entangled, while states below that plane cannot be unambiguously classified by this specific choice of witness/hyperplane. Adapted from \cite{RevModPhys.81.865}.}  \label{witnessfig}  \end{figure}

Consequently, any quantum state $\rho$ for which Tr$(\rho \hat W)<0$, {\it i.e.}~which yields a negative expectation value of the witness in the experiment, is thereby verified to be entangled. Witnesses are universal in the sense that one can find a suitable witness for any entangled state. Bell inequalities can be seen as a specific class of entanglement witnesses \cite{Hyllus:2005dq}, which only detect states that violate local realism. We summarize the concept of separable and entangled, pure and mixed states in Figure \ref{witnessfig}.

\subsection{Local operations and classical communication} \label{measures}
The distinction between separable and entangled states has attracted substantial interest both from the theoretical \cite{Peres:1996ys,Horodecki:1997zr,Horodecki:1996vn,Lewenstein:1998ve,Alcaraz:2002ij}  and from the experimental side \cite{Kampermann:2010ly,Barbieri:2004qf} (see, {\it e.g.}, \cite{Ghne:2009ys}, for a recent review), but this distinction alone is largely insufficient to draw a complete picture of the physics of entanglement.
The next step towards a deeper understanding is the ability to compare the entanglement content of different states. Given the rather abstract nature of entanglement, it is, however, not obvious how such comparison should work. 
By now, some consensus has been reached in the literature that the concept of {\em local operations and classical communication} (LOCC) provides an appropriate framework \cite{Nielsen:1999uq}.

\emph{Local operations} include all manipulations that are allowed by the laws of quantum mechanics -- including measurements, coherent driving, interactions with auxiliary degrees of freedom -- under the condition that they are restricted to either one of the individual subsystems. Operations that require an interaction between the subsystems do not fall into this class, and quantum correlations thus cannot be generated by local operations. 
Only classical communication is here admissible to create correlations: The result of a measurement on one subsystem can be communicated to a receiver which is ready to execute a local operation on the other subsystem, and this subsequent operation can be conditioned on the prior measurement result.
One can thereby indeed induce correlations: For example, the exemplary state (\ref{sepmixed}) could have been prepared by randomly preparing one of the basis states $\ket{i}$ of the first subsystem, and subsequent communication of  the choice of $i$ to a receiver which controls the second subsystem.
If this person prepares its subsystem in the same state, then many repetitions of this procedure yield (\ref{sepmixed}). It is, however, impossible to create an entangled state through the application of LOCC, starting out from an initially separable state.
Therefore, it is justified to consider  a state $\rho_2$ equally or less entangled than a state $\rho_1$,
if $\rho_2$ can be obtained from $\rho_1$ through the application of LOCC.

\subsubsection{Maximally entangled states} \label{maxienttt}
This immediately implies the notion of a \emph{maximally entangled state}, from which any other state can be generated through LOCC: For bipartite qubit states, the four Bell states
\eq 
\ket{\Psi^\pm} = \frac{1}{\sqrt 2} \left( \ket{0,1}\pm \ket{1,0} \right) \label{Psimp} ,\\ 
\ket{\Phi^\pm} = \frac{1}{\sqrt 2} \left( \ket{0,0}\pm \ket{1,1} \right)  \label{Phimp},
\en
are maximally entangled, since any bi-qubit state can be realized through LOCC applied to either one of them \cite{Nielsen:2000fk,Kaye:2007uq}.
Also on higher dimensional subsystems such states can be constructed, and, indeed, are precisely of the form (\ref{eq:psimm}).
The application of suitable LOCC allows to convert (\ref{eq:psimm}) into arbitrary bipartite $d$-level states.

If, however, the number of system components is increased,
 the concept of a maximally entangled state cannot be generalized unambiguously.
The most illustrative example is that of the tripartite Greenberger-Horne-Zeilinger (GHZ) state 
\eq \ket{GHZ}=\frac{1}{\sqrt 2}\left(\ket{0,0,0} + \ket{1,1,1} \right)  \label{GHZstate} , \en
on the one hand, and of the W-state
\eq \ket{W}=\frac{1}{\sqrt 3}\left( \ket{0,0,1}+ \ket{0,1,0} + \ket{1,0,0} \right)\ ,  \label{Wstate} \en on the other.
Both states are strongly entangled, and are certainly candidates to qualify as maximally entangled.
Though, neither can a $\ket{W}$ state be obtained through the application of LOCC on a $\ket{GHZ}$ state, nor is the inverse possible \cite{Duerr:2000pj}. Consequently, since there is no state in a tri-qubit system from which all other states can be obtained, one has to get acquainted with the idea that there is no unique maximally entangled state in multi-partite systems, but that there are inequivalent classes, or families, of entangled states. The classification of these is subject to active research (see \cite{RevModPhys.81.865} for an overview, and \cite{Hiesmayr:2008dz,Huber:2010kx,PhysRevLett.104.020504,Salwey:2010fk} for recent results), and still far from being accomplished. 

\subsection{Entanglement quantification} \label{quantif}
In order to define a \emph{measure} of entanglement, we need to specify under which conditions two quantum states can be regarded as equivalent, {\it i.e.}~when they carry the same amount of entanglement. This can be certified if the states are related to each other via invertible LOCC, {\it i.e.}~via unitary operations that are applied locally and independently on the subsystems: 
\eq  \ket{\Phi_1} = {\cal U}_{\mathrm{local}} \ket{\Phi_2} =   {\cal U}_1 \otimes  {\cal U}_2 \otimes \dots \otimes {\cal U}_N \ket{\Phi_2} , \nonumber \\ 
 \ket{\Phi_2} = {\cal U}_{\mathrm{local}}^\dagger \ket{\Phi_1} =  {\cal U}_1^\dagger \otimes  {\cal U}_2^\dagger \otimes \dots \otimes {\cal U}_N^\dagger \ket{\Phi_1} ,  \en
where local unitaries $ {\cal U}_{\mathrm{local}}= {\cal U}_1 \otimes  {\cal U}_2 \otimes \dots {\cal U}_N$ are induced by single-particle Hamiltonians $ H_1, \dots  H_N$, such that (with the convention $\hbar=1$) 
\eq
{\cal U}_l={\cal U}_1\otimes{\cal U}_2 \otimes \dots \otimes {\cal U}_N =e^{-i\tau(H_1\otimes{\mathbbm 1}\otimes\dots \otimes {\mathbbm 1}+{\mathbbm 1}\otimes H_2\otimes\dots \otimes {\mathbbm 1} + \dots )}\ . \label{hamilnointeraction}
\en
This equivalence relation is motivated by the fact that the parties which are in possession of the subsystems can perform such local invertible operations without any mutual communication or other infrastructure.  Any entanglement quantifier, therefore, ought to be independent of such local unitaries. 
\subsubsection{Entanglement monotones for bipartite pure states} 
In the particular case of a pure state of a bipartite system, the invariants under local unitaries are precisely given by the state's \emph{Schmidt coefficients} $\lambda_j$, which are the squared weights of the state's \emph{Schmidt decomposition}  
\eq \ket{\Psi}=\sum_{i=1}^s \sqrt{\lambda_i} \ket{\tilde \phi_{1,i}} \otimes \ket{\tilde \phi_{2,i}}\ \label{SchmidtDecomp} , \en
with the characteristic trait that one summation index suffices, in contrast to a representation of $\ket{\Psi}$ in arbitrary basis sets on $\mathscr{H}_1$ and $\mathscr{H}_2$, as in (\ref{generalrepresent}). The Schmidt coefficients coincide with the eigenvalues of the reduced density matrices (\ref{eq:reddmat}), which can be deduced from the specific form of (\ref{SchmidtDecomp}): The partial trace (\ref{partialtrace}) on either subsystem directly yields    \eq\varrho_1=\sum_j\lambda_j \ket{\tilde \phi_{1,j}} \bra{\tilde \phi_{1,j}},  \hspace{.71cm}\mathrm{and}\hspace{.71cm}\varrho_2=\sum_j\lambda_j \ket{\tilde \phi_{2,j}} \bra{\tilde \phi_{2,j}} \ .  \label{reducedden} \en Since the states $\{\ket{\tilde \phi_{1,j}}\}$ and $\{\ket{\tilde \phi_{2,j}}\}$ form orthogonal bases, respectively, they are indeed the eigenstates of $\varrho_1$ and $\varrho_2$, and the $\lambda_j$ are the corresponding eigenvalues.

The $\lambda_j$ fully determine the entanglement of $\ket{\Psi}$, and functions $M(\ket{\Psi})$ of the $\lambda_j$ that are non-increasing under LOCC are called \emph{entanglement monotones} \cite{Vidal:2000kx}, which  quantify the state's entanglement. Under some additional requirements \cite{Vedral:1997uq,Plenio:2006uq} beyond the scope of our present discussion, entanglement monotones are also called \emph{entanglement measures}. The following entanglement monotones for bipartite systems will appear in the course of this review:
\begin{itemize}
\item The \emph{Schmidt rank} $s$ \cite{Terhal:2000uq}, which is the number of non-vanishing \emph{Schmidt coefficients} $\lambda_j$ in the expansion (\ref{SchmidtDecomp}). It ranges from unity (separable) to $d$. The matrix rank of the reduced density matrix of either subsystem equals $s$ (see (\ref{reducedden})). 
\item The \emph{Schmidt number} \cite{Grobe:1994fk}, \eq K = \frac{1}{\sum_{i=1}^s \lambda_i^2} =\frac{1}{\mathrm{Tr}(\varrho_1^2) }\ , \label{Schmidtnumber}\en
which estimates the number of states involved in the Schmidt decomposition. It can also be seen as the inverse participation ratio, and ranges from unity for separable states to $d$, with $d$ the dimension of the subsystems.
\item The \emph{concurrence} \cite{Wootters:1998fk,PhysRevA.64.042315}, defined by  
\eq C(\ket{\Psi})= \sqrt{\frac{d}{d-1}\left(1-\sum_{j=1}^d \lambda_j^2 \right) }\ ,  \label{concurrence} \en
is directly related to the \emph{linear entropy} of a probability distribution defined by the weights $\lambda_j$, 
\eq 
S_{\mathrm{lin}}(\{ \lambda_j \})=1-\sum_{j=1}^d \lambda_j^2 . \en 
\item[-] The \emph{entanglement of formation} or \emph{entanglement entropy} \cite{Bennett:1996fk}, is the von Neumann-entropy \cite{von-Neumann:1955ys} of one of the reduced density matrices $\varrho_j$, \eq E(\ket{\Psi})=S(\varrho_j)=-\mathrm{Tr}( \varrho_j ~ \mathrm{Log}_2(\varrho_j) ) ,  \label{eoff} \en 
which, given the spectral decomposition of $\varrho_{j}$, boils down to the Shannon entropy \cite{Shannon:1948fk} of a probability distribution defined by the Schmidt coefficients $\lambda_j$, 
\eq E(\ket{\Psi})= H(\{ \lambda_1 \dots \lambda_j \}) = - 
\sum_j \lambda_j \mathrm{Log}_2(\lambda_j) .\en 
\end{itemize}
Since all these quantities are given in terms of the Schmidt coefficients $\lambda_j$, they are readily evaluated for arbitrary bipartite pure states. 

In addition, the Schmidt coefficients allow us to decide whether a state $\ket{\Phi}$ can be prepared deterministically from an initially given state $\ket{\Psi}$ via LOCC: This is possible if and only if the Schmidt coefficients of $\ket{\Phi}$ are \emph{majorized} by the Schmidt coefficients of $\ket{\Psi}$ \cite{Nielsen:1999uq}. Majorization is defined by 
\eq  \forall k, 1\le k \le d: \sum_{j=1}^k \lambda^\Phi_j  \le \sum_{j=1}^k \lambda^{\Psi}_j ,\en
where the $\lambda^\alpha_j$ are the entries of the Schmidt vectors $\vec \lambda^\alpha$, $\alpha=\Phi,\Psi$, sorted in increasing order. 
Consistently with our definition of maximally entangled states (see Section \ref{maxienttt}), maximally entangled states as the one given by (\ref{eq:psimm}) majorize any other less entangled state.

\subsubsection{Entanglement monotones for mixed states}
It is more difficult to evaluate the entanglement content of a mixed state given by its pure state decomposition $\rho=\sum_jp_j\ket{\Psi_j}\bra{\Psi_j}$. Since this decomposition is not unique \cite{Hughston:1993ly}, a simple average over its pure states' entanglement $M(\ket{\Psi_j})$, with weights $p_j$, does not provide an unambiguous result (also see \cite{Nha:2004ve,Viviescas:2010qf,Carvalho:2007bh}). The problem is cured by taking the infimum over all pure-state decompositions \cite{Bennett:1996kx},
\eq M(\rho)= \mathrm{inf}_{\left\{ \Psi_j, p_j \right\} } \sum_j p_j M(\ket{\Psi_j}) , \label{purestatedecomp} \en
which, indeed, implies a variation over states $\ket{\Psi_j}$ \emph{and} weights $p_j$, since the cardinality of the sum is itself variable. The thus defined \emph{mixed state entanglement monotone} guaranties in particular that $M(\rho)$ vanishes on the separable states. 

A closed formula is available for the concurrence (\ref{concurrence}) of a mixed two-qubit system $\rho$ \cite{Hill:1997eu,Wootters:1998fk},
\eq 
C(\rho)=\mathrm{max}\left\{ 0, \lambda_1 - \lambda_2 - \lambda_3 - \lambda_4  \right\} , \label{concurrencemixedstate}
\en 
where the $\lambda_j$ are the eigenvalues of the matrix 
\eq R= \sqrt{\sqrt{\rho} (\sigma_y \otimes \sigma_y) \rho^\star (\sigma_y \otimes \sigma_y) \sqrt{\rho}}~ , \en
 in decreasing order.  In practice, the optimization problem implicit in (\ref{purestatedecomp}) renders its quantitative evaluation a challenging task for larger systems, beyond two qubits, and there is only limited insight and literature on approximations and rigorous bounds \cite{Mintert:2007ys,Borras:2009zr,Mintert:2005rc,Ma:2010vn,Ghne:2009ys,Guhne:2008zr,Gao:2006ly}.

\begin{figure}[h] \center \includegraphics[width=5.5cm,angle=0]{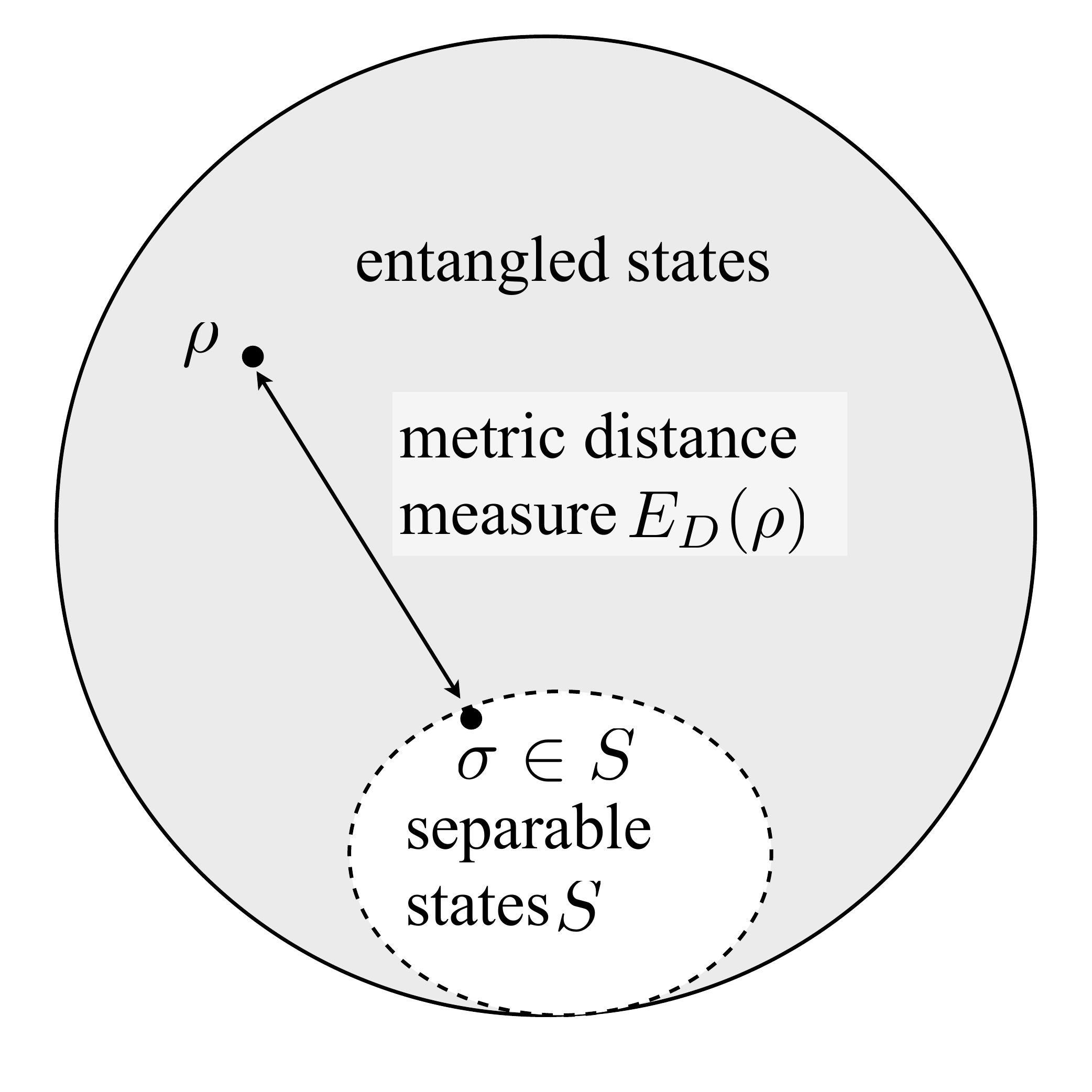} \caption{Illustration of the concept of the metric distance entanglement measure $E_D$, defined in (\ref{metricdistanceent}): The separable state $\sigma$ is chosen such that the metric distance to the entangled state of interest is minimal. }  \label{metricdistance}  \end{figure}

For systems with more than two subunits, no straightforward generalization of the Schmidt decomposition is available. Other concepts of entanglement measures have thus been designed, which can be generalized to such \emph{multipartite} states, {\it i.e.}~states with more than two subsystems. One example is the distance to the set of separable states \cite{Bengtsson:2006fu}, \eq E_D(\rho) = \mathrm{min}_{\sigma \in S}D(\rho,\sigma) ,  \label{metricdistanceent} \en where $D(\rho,\sigma)$ is a distance measure between two states  \cite{Reed:1980ly}, and the minimum is taken over all states $\sigma$ within the set of separable states $S$. This quantity possesses a straightforward geometrical interpretation, illustrated in Figure \ref{metricdistance}.
The evaluation of this and any other multipartite measure for mixed, multipartite states is, in general, computationally demanding, due to the reasons described above.

\subsection{Modeling physical systems}
\subsubsection{Subsystem structures} \label{substruct}
\begin{figure}[h] \center \includegraphics[width=12.5cm,angle=0]{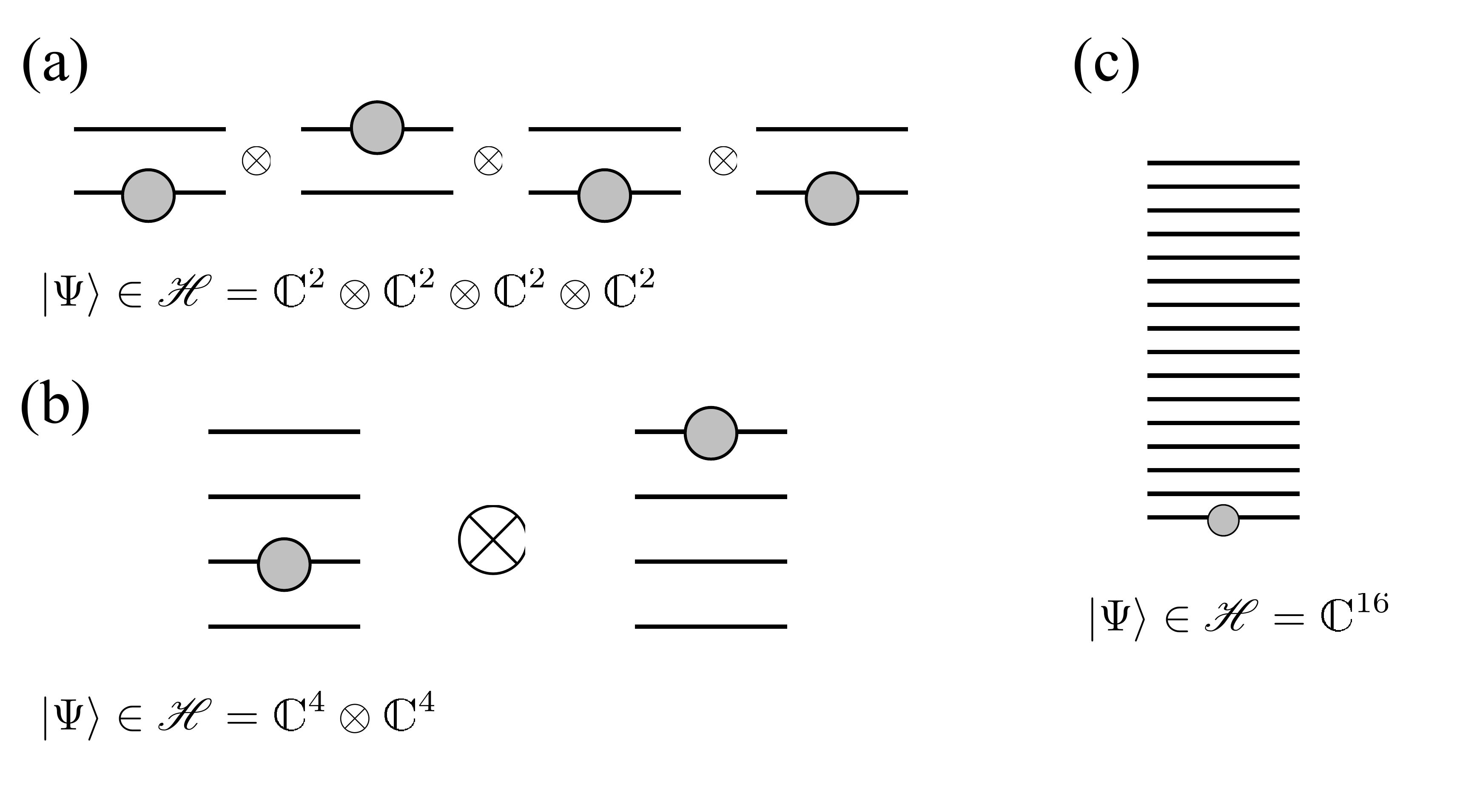} \caption{Illustration of possible subsystem-structures of $\mathscr{H}=\mathbbm{C}^{16}$. }  \label{partitions}  \end{figure}

In the modeling of entanglement inscribed into real physical systems, it is natural to ask how to partition the Hilbert space, {\it i.e.}~how to choose the subsystem structure on which entanglement is defined. This choice can be largely variable \cite{Zanardi:2004fk}: Consider, {\it e.g.}, a 16-dimensional space, $\mathscr{H}=\mathbbm{C}^{16}$. This can be seen as the tensor product of four Hilbert spaces that represent a two-level system each ($16=2^4, \mathbbm{C}^{16}=\mathbbm{C}^2\otimes \mathbbm{C}^2\otimes \mathbbm{C}^2\otimes \mathbbm{C}^2$, Figure \ref{partitions}a), or as
the tensor product of two Hilbert spaces  that each represent a particle with four discrete eigenstates ($16=4^2, \mathbbm{C}^{16}=\mathbbm{C}^4 \otimes \mathbbm{C}^4$, see Figure \ref{partitions}b), or as one, indivisible, Hilbert space of dimension 16 (Figure \ref{partitions}c). The specific physical situation is then clearly distinct, as evident from the illustrations.
In general, a natural partition is induced by the definition of the system degrees of freedom: A 16-dimensional Hilbert space is spanned by two four-level atoms, a four-ion quantum register where each ion bears two computational levels, or, {\it e.g.}, by a quantized single mode resonator field populated  by maximally 15 photons. 

The very definition of the system degrees of freedom thus also affects the expected entanglement between them. Consider, {\it e.g.}, the hydrogen atom, as an example for a continuous variable system with infinite-dimensional subsystem dimension: When partitioned into center-of-mass degree of freedom and relative coordinates, both degrees of freedom separate completely, and no entanglement is exhibited. In contrast, if we choose the partition into electron- and proton-coordinate, the subunits remain coupled, and exhibit non-vanishing entanglement \cite{Tommasini:1998qf}.  
Very distinct entanglement properties can thus be ascribed to the same physical system \cite{Barnum:2004wc,Barnum:2005xz,Zanardi:2004fk,Zanardi:2001kx} -- according to the choice of the system partition. As long as the partition is itself invariant under the system dynamics, it can be defined a priori once and forever, while the situation may turn more complicated if this condition is not fulfilled -- {\it e.g.} for identical particles that are scattered by a beam splitter (see Section \ref{entextr}), or in electrons bound by atoms (see Section \ref{elecelec})). 

\subsubsection{Superselection rules} \label{secSSR}
Even when given a fixed subsystem structure, constraints are possible on realizable operations and measurements on the system, which also restrict the verifiable or exploitable entanglement. The impossibility to measure coherent superpositions of eigenstates of certain operators, either due to fundamental reasons like, {\it e.g.}, charge conservation, or as a consequence of the lack of a shared reference frame that prevents the experimentalists to gauge their measurement instruments, is described by the formulation of \emph{superselection rules} (SSR) \cite{PhysRev.88.101,Greenberger:2009uq,Bartlett:2007bf}. Such rules strongly restrict the nature and outcome of possible measurements. In contrast to selection rules which give statements on the system evolution generated by some Hamilton with certain symmetries, SSRs postulate a much more strict behavior. Two states $\ket{\Psi_1}$ and $\ket{\Psi_2}$ are said to be \emph{separated by a SSR} if for \emph{any} physically realizable observable $ A$ (and not just for a specific Hamiltonian),
\eq \bra{\Psi_1}  A \ket{\Psi_2} =0 .\en
It is important to note that SSR are not equivalent to conservation laws and also do not restrict the accessible states of a system. Instead, a SSR is a postulated statement on the \emph{physical realizability of operators} \cite{Bartlett:1991ys}. For example, the direct implementation of an operator which projects on a coherent superposition of an electron and a proton is impossible. 

When a SSR applies, the executable operations and measurements are restricted, and the entanglement of a system possibly cannot be fully accessed. Entanglement constrained by SSR is indeed bounded from above by the entanglement evaluated according to the rules of the above sections \cite{Bartlett:1991ys,Schuch:2004kx,Schuch:2004xl}. A SSR that prevents a local basis rotation, {\it e.g.}, may enforce that the entanglement which is formally present in a state effectively reduces to a classical correlation. Suppose two parties share the state $\left(\ket{e,p}+\ket{p,e} \right)/\sqrt{2}$ where $\ket{e}$ and $\ket{p}$ indicate the wave-function of an electron and a proton, respectively. While this state is formally entangled, it is effectively impossible to measure a coherent superposition of an electron and a proton. Hence, measurements in other bases, {\it e.g.} in the basis $\left(\ket{e}\pm\ket{p}\right)/\sqrt{2}$, are unfeasible, and no \emph{quantum} correlations can be verified. Consequently, the very classification of pure entangled states becomes more complex due to the impossibility of performing certain measurements \cite{Bartlett:2006gb}.

\begin{figure}[h] \center \includegraphics[width=8.5cm,angle=0]{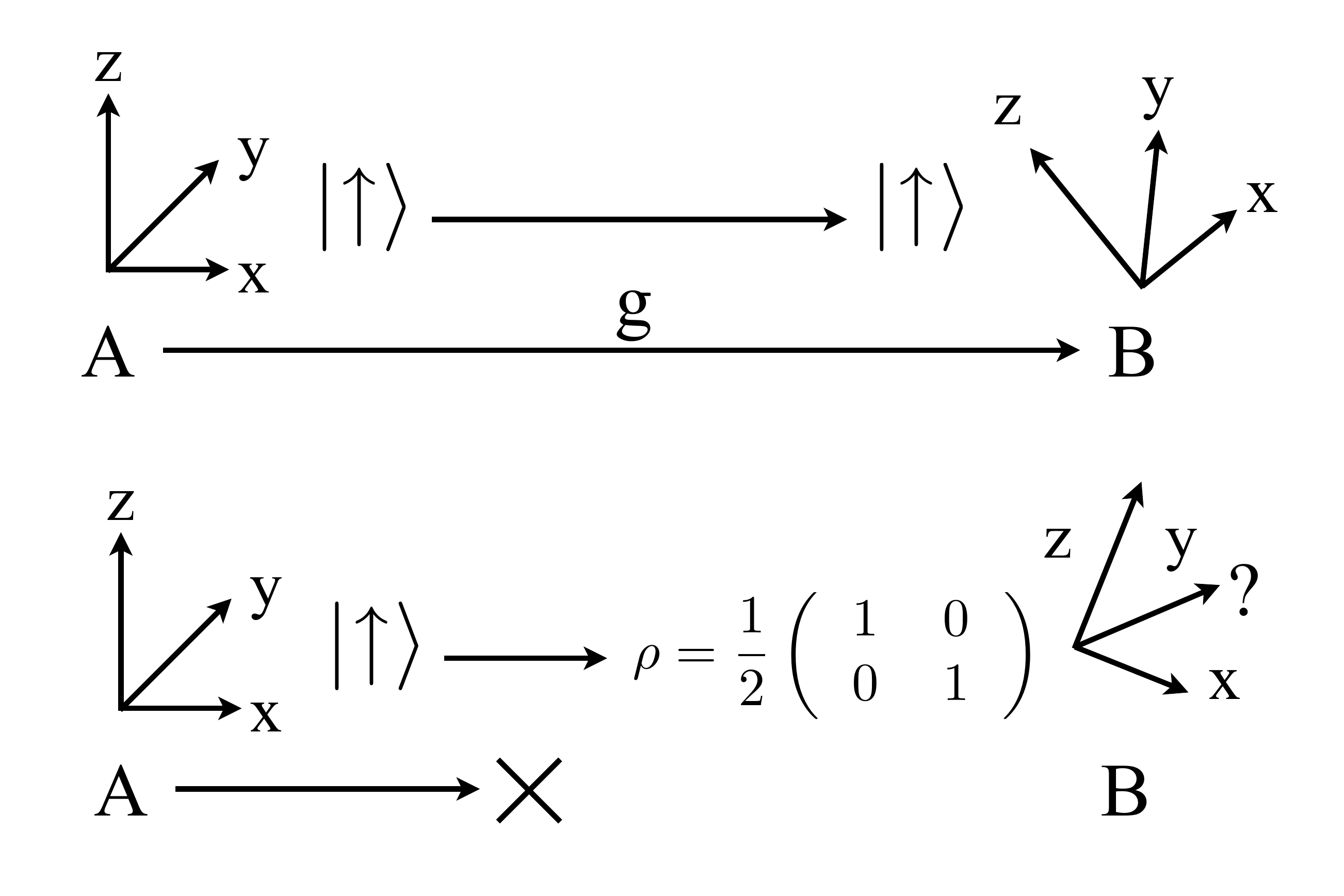} \caption{Transmission of the quantum state $\ket{\uparrow}$ between two parties which share a spatial reference frame (upper panel), or do not (lower panel). The party B has access only to an unbiased mixture when no such convention is available. } \label{referenceframe}  \end{figure}

Effective SSR do not only appear due to fundamental conservation rules, but also when two parties in possession of a compound quantum state do not share a perfect reference frame \cite{Bartlett:2007bf,Enk:2006hc}, {\it i.e.}~a convention on time, phase, and spatial directions. This connection between SSRs and reference frames can be seen in the following example: Suppose one party, A, prepares a quantum state $\rho$ and sends it to another party, B. The reference frames of A and B are related to each other by a transformation $T(g)$, where $g$ is an element of the transformation group, and $T(g)$ is its unitary realization. If, however, B does not know the relative orientation of the individual reference frames, the state B will effectively have access to is the one send by A, but averaged over all possible transformations. Such effectively accessible quantum state for B reads 
\eq \bar \rho = \int \mathrm{d} g ~T(g) \rho T^{-1}(g) \equiv G[\rho] ,\en
where the integral is taken over all elements of the group, and $G$ is baptized the \emph{twirling operation}  \cite{Bartlett:2007bf}. In general, $G$ is not injective,\footnote{In other words, $G$ does not preserve distinctness: There can be two distinct states $\rho_1\neq \rho_2$ for which $G[\rho_1]=G[\rho_2]$.} hence the accessible space becomes smaller under the twirling operation. Only quantum states that are fully invariant under arbitrary transformations, {\it i.e.}~states $ \rho$ with $[ \rho, T(g)]=0$ for any $g$, can be distinguishable building blocks, and only operations that commute with all transformations $g$ can be reliably performed on the subsystems \cite{Bartlett:1991ys}. The transformation group can, {\it e.g.}, represent spatial rotations and generate invariance under $O(3)$, as illustrated in Figure \ref{referenceframe}: If party A sends a particle in the spin-up state, $\ket{\uparrow}$, to party B, the latter has access to this pure state only when A and B share a convention regarding the spatial quantization axis, {\it i.e.}~the axis in space with respect to which the state has been prepared. 
Without such convention, B will observe a fully mixed state and cannot extract any information. In this case, B cannot distinguish any single-particle state by its spin, because all are equivalent to the completely unbiased mixture $G[\ket{\uparrow}]=\mathbbm{1}_2/2$.

On the other hand, given a fundamental SSR generated by a compact group,\footnote{In other words, a group whose elements are elements of a compact set, in contrast to, {\it e.g.} the Lorentz or the Poincar\'e group.} like the very $U(1)$ group which describes charge conservation \cite{Aharonov:1967ys}, the possibility to access a shared reference frame in which the underlying symmetry is broken can actually enable the preparation and measurement of superposition states \cite{Bartlett:2007bf}, and thereby to circumvent the SSR. The indirect observation of such coherences may be possible with the help of ancilla\footnote{\label{ancillad} Ancillae denote particles which help to perform certain tasks or operations, inspired by the latin translation of ``maiden''. Often, they fall in oblivion when the desired final state is obtained.} particles which interact with the system and thereby provide such symmetry breaking \cite{White:2009vn}. 

The problem of entanglement constrained by SSR will turn saliant in Section \ref{modenen}, where we will discuss in more detail the problem of assigning entanglement to mode-entangled states, {\it i.e.}~states in which the local particle number is not fixed. 

\section{Subunits and degrees of freedom} \label{secentities}
In the last chapter, we discussed entanglement as a property of a compound system that characterizes correlations between the system's subunits, which, from now on, will also be called \emph{entities}. In general, certain degrees of freedom of these subunits can be quantum-mechanically correlated, {\it i.e.}~entangled, to those of other entities, and the tensorial Hilbert space structure directly reflects this, possibly hierarchical, subsystem structure. 

For identical particles, however, the invariance of all physical observables with respect to the exchange of any two particles and the (anti)symmetry of the (fermionic) bosonic multiparticle wave-function lead to a new conceptual challenge, and to the failure of the above strategy. The strong constraints on allowed states and on possible measurements imply that direct access to the ``first'' or ``second'' particle is not possible when dealing with identical particles, since particle \emph{labels} are no physical degrees of freedom. Therefore, the identification of Hilbert spaces with subsystems, which we explicitly assumed in Section \ref{sepaent}, fails when dealing with identical particles, and a different formalism is required to characterize the subunits between which entanglement is considered.  As we will see below, upon closer inspection of  several, partially irreconcilable entanglement concepts \cite{Ghirardi:2004fk,Wiseman:2003mz,Dowling:2006kx,schliemann-cirac,Eckert:2002vn}, the debate on what is the proper formalism is still not completely settled in the literature. 

In order to find a suitable concept for the entanglement of identical particles, we start with the conceptually simpler case of non-identical or effectively non-identical particles. Once we will have established the physical ingredients of entanglement, and after relating these to the formal definitions introduced in Section 2, we will proceed towards more elaborate subsystem structures with identical particles involved. Finally, we give an overview of the diverse degrees of freedom that can exhibit entanglement in a real experimental setting.

\subsection{Non-identical particles}
We first shortly revisit the entanglement of distinguishable particles in order to establish the \emph{physical} meaning of entanglement and of separability in the spirit of the original EPR approach \cite{PhysRev.47.777}. 
We will thus establish the terminology which we also need for a rigorous definition of the entanglement of identical particles. We aim at an approach which allows for a continuous transition from the entanglement properties of indistinguishable to those of distinguishable particles (as quantified in Section \ref{measures}), {\it e.g.} under a dynamical evolution which transforms the system constituents from an initially indistinguishable to a finally distinguishable state (as possibly induced, {\it e.g.}, by an atomic or molecular fragmentation process). 

In the case of non-identical particles, the entities that carry entanglement are well defined: They correspond to the very particles, which exhibit some properties such as rest mass or charge, which allows to distinguish them without ambiguity. Other degrees of freedom of these particles that can be coherently superposed then eventually give rise to the particles' entanglement. 
In this situation, we can directly associate Hilbert space $\mathscr{H}_1$ with the Hilbert space of the first particle, and Hilbert space $\mathscr{H}_2$ with that of the second. Vectors in these Hilbert spaces directly describe the state of the dynamical degree of freedom, such as spin or momentum, for the respective particle. 

\subsubsection{Entanglement and subsystem properties according to the EPR approach} \label{properties}
A quantum state is not entangled \cite{PhysRev.47.777} if we can assign \emph{a complete set of properties} to each individual subsystem \cite{Ghirardi-statphys}, {\it i.e.}~if we can design a projective measurement on each subsystem with an outcome that can be predicted with certainty. In other words, there exists an observable such that its measurement on the given non-entangled state reveals the ``pre-existing'' system values, which we intuitively ascribe to it, as in classical mechanics. In the jargon, this is equivalent to finding a \emph{realistic description} of the system's constituents \cite{PhysRev.47.777}. 

Specifically, for a pure state $\ket{\Psi}$ which describes a composite two-particle system, the subsystem $S_1$ is non-entangled with the subsystem $S_2$ if there exists a one-dimensional projection operator $P$ (with eigenvalue 1, by definition) such that 
\eq \bra{\Psi} P \otimes \mathbbm{1} \ket{\Psi}= 1 , \en where $P$ acts on the Hilbert space of the first particle, and the identity $\mathbbm{1}$ on the second particle \cite{Ghirardi-statphys}. In this case, we can assert that the first subsystem possesses the physical properties defined by the operator $P$. The outcome of the measurement of $P$ is not subject to any uncertainty, the first subsystem was necessarily prepared in the unique eigenstate of $P$. 

This notion of ``possession of a complete set of properties'' is fully equivalent to the notion of separability that we fixed in (\ref{separable}): The properties of any separable state are naturally given by projections on the quantum states of its constituent particles; as soon as a state requires more than one product state for its representation (and thus is an entangled state), such well-defined constituent properties cannot be identified anymore.

Besides separability, we can also define \emph{partial} and \emph{total entanglement} in the above framework \cite{Ghirardi-statphys}. For example, the state
\eq \frac{1}{\sqrt 2} \left(\ket{\uparrow}_1 \ket{\downarrow}_2+  \ket{\downarrow}_1 \ket{\uparrow}_2 \right) \ket{L}_1 \ket{R}_2 ~,\label{partialent} \en
where $\ket{\uparrow}, \ket{\downarrow}$ denote internal degrees of freedom, while $\ket{L}$ and $\ket{R}$ denote orthogonal spatial wave-functions, is partially entangled: Whereas system 1 possesses properties associated with it being prepared in the  $\ket{L}$ state, we cannot specify all of its properties, since the particles' internal degrees of freedom remain fully correlated and locally unknown. This corresponds to the situation in most experiments: Particles do possess properties by which they can be distinguished -- {\it e.g.}, their position -- while their entanglement manifests in some other degree of freedom. In such a situation, the ``labeling'' degree of freedom (position $\ket{L}$ or $\ket{R}$ in (\ref{partialent})) can be simply dropped, and the entangled states, {\it e.g.}, of a photon's polarization and of the electronic degree of freedom of an atom, lives in the two-particle Hilbert-space 
\eq \ket{\Psi} \in \mathbbm{C}_1^2 \otimes  \mathbbm{C}_2^2 \label{qubithsp} .\en Each factor $\mathbbm{C}^2$ in (\ref{qubithsp}) describes the two-dimensional state-space of a polarized photon or of a two-level atom. 
This is the typical structure of a two-qubit state so frequently encountered in the quantum information context.

\subsection{Creation and dynamics of entanglement} \label{entint}
\subsubsection{Interaction and entanglement}
For distinguishable particles, the strict rule that \begin{quotation}``Only interaction between particles\footnote{The interaction can be direct or through an intermediate, ancilla degree of freedom (see footnote on p. \pageref{ancillad} above).} can lead to an entangled state.''\end{quotation} can be formulated, in contrast to indistinguishable particles, as we will see in Section \ref{measurementinduced}. Distinguishable particles, even if prepared in a pure separable state, will be entangled for almost all times \cite{Durt:2004tg}, if the unitary evolution which describes their common evolution contains parts which do not factorize as in (\ref{hamilnointeraction}), {\it i.e.}~if the Hamiltonian contains an interaction part. Given a many-particle Hamiltonian, one can deduce from its interaction part which degrees of freedom will be entangled. A Hamiltonian which only couples external degrees of freedom will, {\it e.g.}, not induce any entanglement in the particles' spin. 

In every many-particle bound state such as that of a simple hydrogen atom, the constituents are necessarily entangled in their external degrees of freedom \cite{Tommasini:1998qf}: The binding potential contains the operator $\vec r_1 - \vec r_2$, {\it i.e.}~it couples the position operators of the constituents (proton and electron, for the hydrogen atom). Inclusion of the spin-orbit interaction additionally induces entanglement between the external degrees of freedom and the spin, as we will discuss in more detail in Section \ref{ionizedelectrons}. The analogous reasoning applies for unbound systems: In the course of a scattering process, particles naturally entangle under the specified interaction \cite{Law:2004qf}. 

We retain the intuitive picture that, for distinguishable particles, entanglement is a direct consequence of interaction.

\subsubsection{Open system entanglement}
If a quantum system is closed, \emph{i.e.} if it is decoupled from uncontrolled degrees of freedom lumped together under the term ``environment'', all entanglement properties are encoded in the Hamiltonian and in its eigenstates, or in the dynamical evolution it generates.

The time evolution of entanglement under such strictly Hamiltonian dynamics can then be evaluated by application of pure state entanglement measures on the time dependent state vector $\ket{\Psi(t)}$. 
Purely Hamiltonian evolution is, however, untypical for experiments, where 
the environment cannot be screened away completely. Therefore, decoherence, {\it i.e.}~the gradual loss of the off-diagonal entries  of the density matrix \cite{Breuer:2006ud}, makes entanglement to fade away and limits its possible harvesting.

To account for such -- detrimental -- environmental influence, one can evaluate mixed state entanglement measures as defined in  (\ref{purestatedecomp}) on the system density matrix $\rho(t)$, for all $t$. However, this quickly turns into a tedious, if not unfeasible task with increasing system size, due to the optimization problem implied by  (\ref{purestatedecomp}). Efficiently evaluable lower bounds of entanglement alleviate the computational challenge \cite{Huber:2010kx,Mintert:2005rc,Mintert:2007ys,Borras:2009zr}, though cannot fully compensate the unfavorable scaling of optimization space with system size.

Alternative approaches try to circumvent this problem by incorporating the specific type of environment coupling into the analysis, to extract the entanglement evolution of representative ``benchmark'' states \cite{Konrad:2008vn,Tiersch:2008kx,Tiersch:2009uq,Li:2009ys,unravelling, vogelsberger}. The very knowledge of the source of decoherence effectively reduces the complexity of the problem (however, it remains to be quantified to which extent). For the simplest scenario of open system entanglement evolution, a qubit pair with only one qubit coupled to the environment, even an exact \emph{entanglement evolution equation} is available:  Given the pure two-qubit's initial state $\ket{\chi}$, and the  completely positive map\footnote{A positive map $\Lambda$ maps positive operators -- defined by a spectrum with strictly non-negative eigenvalues -- on positive operators, and, in particular, density matrices onto density matrices. For  $\Lambda_C$ to be completely positive, also all possible extensions of the map to larger systems of the form $\mathbbm{1}_N \otimes \Lambda_C$ need to be positive.} \$ \cite{Breuer:2006ud} that one qubit is exposed to, the system's final state reads
\eq \rho^\prime = (\mathbbm{1}\otimes \$ ) \ket{\chi}\bra{\chi} .\en
The entanglement of the evolved quantum state $\rho^\prime$, quantified by the concurrence (as defined in (\ref{concurrence})), is then given by the entanglement evolution equation  \cite{Konrad:2008vn} \eq C\left[ (\mathbbm{1} \otimes \$) \ket{\chi} \bra{\chi} \right] = C( \ket{\chi}) C\left[(\mathbbm{1}\otimes \$) \ket{\Phi} \bra{\Phi} \right] \ , \label{factorevoo} \en 
where $\ket{\Phi}$ is the (maximally entangled, see (\ref{Psimp},\ref{Phimp})) benchmark state, and $\ket{\chi}$ an \emph{arbitrary} initial state. The entanglement evolution of $\ket{\chi}$ thus factorizes into a contribution given by the benchmark state's entanglement upon the action of the map \$, and a second term given by the initial state's concurrence $C(\ket{\chi})$. This result can be generalized for bipartite systems with $d$ dimensional subspaces \cite{Tiersch:2008kx}, with $C$ in (\ref{factorevoo}) replaced by G-concurrence $G_d(\rho)$ \cite{Gour:2005zr}. The measure $G_d(\rho)$ quantifies entanglement of rank $d$ states, {\it i.e.}~of states that are obtained as a coherent superposition that exhausts all basis states. Whenever the Schmidt rank (see Section \ref{quantif}) of the evolved state drops below $d$, $G_d$ drops to 0, while lower-ranked entanglement may still be present in the system. For $k<d$, upper bounds for $G_k$ and thereby clear disentanglement criteria can be derived within the same formalism. Since the evolution of the quantum state is continuous, it is, however, guaranteed that any state with initially non-vanishing $G_d$ will remain $G_d$-entangled at least for short times.

\subsubsection{Entanglement statistics}
Given the difficulty to characterize the open system entanglement evolution for individual initial states, because of the exponential scaling of state space with system size, it is suggestive to employ statistical tools. It can then be shown, under rather general assumptions on the open system dynamics described by some time-dependent map $\Lambda_t$ and on the employed entanglement measure $E$,\footnote{The measure $E$ needs to be Lipschitz-continuous \cite{Ruskai:1994fk}, and the system and environment need to be initially uncorrelated or at most classically correlated such that $\Lambda_t$ does not depend on the initial state.} that deviations $\epsilon$, of the final entanglement of an arbitrary pure initial state $\ket{\Psi}$ under the action of $\Lambda_t$ from the mean entanglement $\langle E \rangle(t)$ of all pure states acted upon by $\Lambda_t$, are exponentially suppressed in $\epsilon$ and in the system dimension $d$ \cite{tiersch}:
\eq P\left( \left|E(\Lambda_t \ket{\Psi} \bra{\Psi} )-\langle E  \rangle (t) \right| > \epsilon \right) \le 4 \mathrm{exp}\left( - \frac{2d-1}{96 \pi^2 \eta_E^2 \eta_{\Lambda_t}^2} \epsilon^2 \right) ,  \label{boundedeps} \en
where $P$ quantifies the probability of the event specified by its argument, and $\eta_E$ and $\eta_{\Lambda_t}$ are parameters that characterize $E$ and $\Lambda_t$, with $\eta_{\Lambda_t}<1$.

Whereas the efficient evaluation of the entanglement evolution of individual initial states of large composite systems will at some point turn prohibitive, (\ref{boundedeps}) provides a statistical estimate which becomes ever tighter with increasing system size. In other words, the vast majority of initially pure multipartite states in a high-dimensional Hilbert space share the same entanglement properties \cite{tiersch}. Consequently, the entanglement evolution is similar for almost all initial states, and an asymptotic behavior emerges with typical traits that are independent of the exact quantum state, reminiscent of thermodynamic quantities.  

\subsection{Identical particles} \label{idparticlesee}
As anticipated above, a consistent treatment of the entanglement of identical particles is much more subtle than for distinguishable  particles. The Hilbert-space structure of two or more identical particles does not reflect any more a physical partition into subsystems, due to the (anti)symmetrization of the many-particle wave-function, as a result of the symmetrization postulate \cite{Ballentine:1998vn}.  The principle of indistinguishability \cite{Messiah:1964ys} that applies to all physical operators leads to an uncircumventable super-selection rule (see Section \ref{secSSR}). This problem has been at the origin of a long debate \cite{Ghirardi-statphys,Ghirardi:2004fk,Ghirardi:2003uq,Cavalcanti:2007vn,Dowling:2006kx,Shi:2003ys,Paskauskas:2001ly,Ghirardi:2004kx,al:2009cr,Zhou:2009fk,Li:2001uq} on how to define a useful measure for a given state of identical particles. In the following, we describe the current state of affairs, and elaborate on how to refine the above criterium of a complete set of properties (Section \ref{properties}) for the case of identical particles \cite{Ghirardi-statphys}.
\subsubsection{Entanglement of particles} \label{entidp}
In many cases, when two identical particles are well separated -- as in typical experiments with photons in different optical modes, or with strongly repelling trapped ions -- no ambiguity is possible, and the physical subsystem-structure is apparent from the preparation of the state and the accessible observables \cite{Herbut:1987hb}. This is already realized in quantum mechanics textbooks, {\it e.g.} \cite{Peres:1993jt} states that \begin{quotation}``No quantum prediction, referring to an atom located in our laboratory, is affected by the mere presence of similar atoms in remote parts of the universe.'' \end{quotation} 

Still, the formal notion of entanglement which we introduced for the case of distinguishable subsystems in Section \ref{measures} ought to be adjusted, since its naive application yields an unphysical form of entanglement, as can be seen by closer scrutiny of the following, exemplary state: 
\eq \ket{\Psi_{AB}}= \frac{1}{\sqrt{2}} \left(  \ket{A, 1} \otimes \ket{B, 0} \pm \ket{B,0} \otimes \ket{A,1} \right) \label{ABstate} , \en
where $\ket{A}$ and $\ket{B}$ describe orthogonal wave-functions. On a first glance, the wave-function appears to be entangled, since it cannot be written as product state, and it has Schmidt rank two (see (\ref{SchmidtDecomp})). However, a more careful analysis of the situation shows  that neither particle in the system is affected by any \emph{physical} uncertainty: The wave-function $\ket{\Psi_{AB}}$ describes two particles, with their positions in space described by $\ket{A}$ and $\ket{B}$, which are prepared in the internal states $\ket 1$ and $\ket 0$, respectively. Physically speaking, we can easily assign a physical reality and thereby \emph{properties} (in the sense of Section \ref{properties}) to the particles: A measurement of the particle located at $\ket{A}$ ($\ket B$) will always yield the internal state $\ket{1}$ ($\ket{0}$).  On the other hand, no physical operator can be conceived which refers unambiguously to the ``first'' or the ``second'' particle, since the particles are - by assumption - indistinguishable, which implies the permutation symmetry of all operators.  In other words, the merely \emph{formal} entanglement in the unphysical particle labels \emph{cannot} be exploited directly, and \emph{does not} correspond to a lack of information about the physical preparation of the system's constituents -- indeed we have just established that there are two particles in the system which both possess a physical reality \cite{Ghirardi:2004fk}. Hence, instead of the Hilbert spaces of the particles which do not allow any more to address the particles individually, some other, \emph{physical}, degrees of freedom need to be identified with the distinctive properties of the entities that carry entanglement. 

\subsubsection{Slater decomposition and rank} \label{slaterdecomp}
In order to differentiate between physical correlations and mere correlations in the particle labels, we use the \emph{Slater decomposition}  instead of the Schmidt decomposition (\ref{SchmidtDecomp}): Two fermions which occupy an $n$-dimensional Hilbert space can always be described by the following quantum state \cite{schliemann-cirac,Schliemann:2001uq}
\eq \ket{\Psi_{\mathrm{fermion}}}=\sum_{a,b} w_{a,b} f^\dagger_a f^\dagger_b \ket{0}, \en
with antisymmetric coefficients $w_{a,b}=-w_{b,a}$, and fermionic creation operators $f^{\dagger}_a$ ($f^{\dagger}_b$), which act on the vacuum state $\ket{0}$ and create a particle in the single-particle state $\ket{a}$ ($\ket{b}$). In analogy to any bipartite state of distinguishable particles that can be written in the Schmidt-decomposition (\ref{SchmidtDecomp}), the above state can be represented in the \emph{Slater decomposition} \cite{schliemann-cirac},
\eq \ket{\Psi_{\mathrm{fermion}}}= \sum_{i} z_i f^\dagger_{a^{(i)}} f^\dagger_{b^{(i)}} \ket{0} ,\en
where the single particle states $\ket{a^{(i)}}=f^\dagger_{a^{(i)}} \ket 0 $ and $\ket{b^{(i)}}=f^\dagger_{b^{(i)}} \ket 0 $ fulfill $\braket{a^{(i)}}{b^{(j)}}=\delta_{a,b} \delta_{i,j}$. The number $r$ of non-vanishing expansion coefficients $z_i$ defines the {\it Slater rank}, which is unity for non-entangled states, and larger than unity for entangled states. In other words, elementary Slater determinants that describe fermions are the analogues of product states in systems that consist of distinguishable particles. The state (\ref{ABstate}) represents -- in the antisymmetric case for fermions -- a single Slater determinant and is, therefore, correctly recognized as \emph{non-entangled}.  The entanglement measures introduced in Section \ref{quantif},  based on the distribution of Schmidt coefficients, can thus, in general,  be recovered for identical particles by consideration of the Slater coefficients instead. 

The convex roof construction (\ref{purestatedecomp}) for distinguishable particles can be imported directly, and allows the computation of entanglement measures for mixed state of identical particles. For example, the Schmidt rank of a mixed state of fermions can be obtained as follows: Given a mixed state of fermions decomposed into pure states, \eq \rho=\sum_{j} p_j \ket{\Psi_j^{(r_j)}}\bra{\Psi_j^{(r_j)}} ,\en where $r_j$ is the Slater rank of the respective pure state $\ket{\Psi_j^{(r_j)}}$, the Slater rank of $\rho$ is defined as $k=\mathrm{min}(r_{\mathrm{max}})$, where $r_{\mathrm{max}}$ is the maximal Slater rank within one decomposition, and the minimum is taken over all decompositions, in strict analogy to the case of distinguishable particles \cite{Terhal:2000uq}. Witnesses (see section \ref{witnessesslabl}) for the minimal number of Schmidt coefficients \cite{Sanpera:2001fk} for distinguishable particles can be imported to Slater witnesses \cite{schliemann-cirac} which witness states that require a certain minimal Slater rank. 

While the analogies between Slater rank and Schmidt rank, worked out in \cite{schliemann-cirac}, also suggest similarities for the properties and the interpretation of the reduced density matrix of one particle (see Section \ref{quantif}), it is important to note that the reduced density matrix of one particle, $\varrho_{1}$, still exhibits some intrinsic uncertainty due to the formal entanglement in the particle label. The relationship between Schmidt coefficients  and the eigenvalues of the reduced density matrix (see (\ref{SchmidtDecomp}),(\ref{reducedden}) in Section \ref{quantif}) therefore breaks down in the case of identical particles: The Slater coefficients are not directly related to the eigenvalues of the reduced density matrix. Entanglement measures based on the reduced density matrix therefore need to be interpreted carefully here, since they do not yield a result in full analogy to the case of distinguishable particles. Such interpretation follows below in Section \ref{propertiesidpa}.

\

Similarly as for fermions, a quantum state of two bosons \cite{Paskauskas:2001ly,Eckert:2002vn,Li:2001uq} can be written as
\eq  \ket{\Psi_{\mathrm{boson}}}=\sum_{i,j=1}^n v_{i,j} b_i^\dagger b_j^\dagger \ket {0} .\en The symmetric coefficient matrix $v_{i,j}=v_{j,i}$ can be diagonalized such that the state can be written as a combination of doubly-occupied quantum states, 
\eq \ket{\Psi_{\mathrm{boson}}}=\sum_{j=1}^n \tilde v_{j} \left( \tilde b_j^\dagger \right)^2 \ket {0} \en 
Again, we call the minimal number $r$ of non-vanishing $\tilde v_j$ the \emph{bosonic Slater rank}. As we will see below, the interpretation of the Slater rank is not directly analogous to the case of distinguishable particles, as it was for fermions. The reason lies in the fact that two bosons can populate the very same quantum state, which is impossible for fermions and cannot be modeled within the usual quantum-information framework laid out in Section 2.

\subsubsection{Subsystem properties and identical particles}  \label{subpropid}
In physical terms, the concept of properties of particles, discussed in Section \ref{properties}, can be adapted to the case of identical particles by taking into account that the particle label itself is \emph{not} a physical property \cite{Ghirardi:2004kx,Ghirardi:2004fk,Ghirardi:2003uq}. This fact needs to be included in the design of the projection operators that define a complete set of properties \cite{Ghirardi:2004fk}, in the terminology of Section \ref{properties}. Given two particles prepared in a quantum state $\ket{\Psi}$ of identical bosons or fermions, we therefore say that one of the constituents \emph{possesses a complete set of properties} if, and only if, there is a one-dimensional projection operator $P$ on the single-particle Hilbert space $\mathcal{H}$ such that
\eq \bra{\Psi} \mathscr{E}_P \ket{\Psi}=1 \label{sepghirardi}, \en
 for 
\eq \mathscr{E}_P=P \otimes ( \mathbbm{1}-P) + (\mathbbm{1}-P) \otimes P~, \label{sepghirardi2} \en where the order of the operators reflects the respective Hilbert spaces they act on.

The above projection operator can be interpreted as the projection on the subspace in which one particle possesses the properties given by $P$, while the other particle is projected onto a subspace orthogonal to $P$. Similarly to the case of non-identical particles,  a particle cannot be entangled to any other particle if it possesses a complete set of properties.\footnote{We avoid the use of the term ``separable'', instead of ``non-entangled'', in the context of identical particles, since the term ``separable'' is often used to refer to the mere mathematical structure of the state, rather than to its possession of well-defined single-particle properties.} This definition rigorously adapts our discussion of Section \ref{properties} to the example (\ref{ABstate}) in Section \ref{entidp}. Indeed, for the state vector (\ref{ABstate}), the projection operator $P=\ket{A}\bra{A} \otimes \ket{1}\bra{1}$ has the required properties (\ref{sepghirardi},\ref{sepghirardi2}). 
 
Thus we come to the following conclusion \cite{Ghirardi:2003uq}: \begin{quotation}Identical fermions of a composite quantum state are non-entangled if their state is given by the antisymmetrization of a factorized state. \end{quotation}

The case of bosons is more subtle to treat \cite{Ghirardi:2003uq}: \begin{quotation}Identical bosons are non-entangled if either the state is obtained by symmetrization of a product of two orthogonal states, or if the bosons are prepared in the same, identical state. \end{quotation}

A special case arises when a product of two non-orthogonal states is symmetrized, as illustrated by the following example: Consider the state
\eq \ket{\Psi} &=& N(\alpha) \left( \ket{\phi}\otimes \left( \cos \alpha \ket{\phi} + \sin \alpha \ket{\psi} \right) \right. \nonumber  \\ &&\left. + \left( \cos \alpha \ket{\phi} + \sin \alpha  \ket{\psi} \right) \otimes \ket{\phi} \right) \label{bosonentan} \en
where we assume $\braket{\phi}{\psi}=0$, and $N(\alpha)$ is an appropriate normalization constant. The parameter $\alpha$ interpolates between a symmetrized state of a product of two identical single-particle quantum states, and the symmetrized product of two orthogonal states. For $\alpha=0$, it corresponds to the state $\ket{\phi}\otimes \ket{\phi}$, {\it i.e.}~a non-entangled state, since we can attribute the property $\ket{\phi}\bra{\phi}$ to both particles. For $0>\alpha>\pi/2$, two non-orthogonal states are symmetrized, and no projection operator $P$ which satisfies (\ref{sepghirardi}) can be found. In other words, no statement about ``at least one particle possesses a certain set of properties'' is possible. Thus, a physical reality can be attributed to neither one of the particles, and the state has to be considered entangled. For $\alpha=\pi/2$, the state corresponds to the symmetrized product of two orthogonal states, one particle possesses the property $P=\ket{\psi}\bra{\psi}$, the other particle possesses $Q=\ket{\phi}\bra{\phi}$, and the state is hence not entangled. 

\subsubsection{Subsystem properties and entanglement measures} \label{propertiesidpa}
The physical criteria based on the possession of realistic properties, as imported above from the case of distinguishable particles, can be directly related to entanglement measures such as the Slater rank introduced in Section \ref{slaterdecomp} \cite{Ghirardi:2004kx,Ghirardi:2004fk,Ghirardi:2003uq}. Also the von Neumann entropy of the reduced density matrix of either one of the particles can be used for the characterization of entanglement, though with some caution, as already mentioned in Section \ref{slaterdecomp}. Here, the interpretation for identical particles is distinct from the one established for distinguishable ones. One can finally formulate \cite{Ghirardi:2004fk} for fermions:
\begin{itemize}
\item[(i)] Slater rank of $\ket{\Psi}=1 \Leftrightarrow S(\varrho^{(1)})=1 \Leftrightarrow \ket{\Psi}$ is \emph{non-entangled} (the state is obtained by antisymmetrization of a product of two states). 
\item[(ii)] Slater rank of $\ket{\Psi}>1 \Leftrightarrow S(\varrho^{(1)})>1 \Leftrightarrow \ket{\Psi}$ is \emph{entangled} (the state is obtained by antisymmetrization of a sum of products of states). 
\end{itemize}
In contrast to the case of distinguishable particles, the entropy $S(\varrho^{(1)})$ is bounded from below by unity instead of zero: \emph{This residual value reflects the mere uncertainty in the particle label.} The analogy between the first case (i) and product states of distinguishable particles, and between the second case (ii) and entangled states of distinguishable particles is apparent. 
Note that the antisymmetrization of a product of two non-orthogonal states gives rise to an unnormalized and non-entangled state. 

For bosons, due to the possibility that the state is obtained by symmetrization of a product of two non-orthogonal states, the criterion is more elaborate \cite{Ghirardi-statphys}
\begin{itemize}
\item[(iii)] Slater rank of $\ket{\Psi}=1 \Leftrightarrow S(\varrho^{(1)})=0 \Rightarrow \ket{\Psi}$ is \emph{non-entangled} (both particles are in the same quantum state).
\item[(iv)] Slater rank of $\ket{\Psi}=2, 0<S(\varrho^{(1)}) <1 \Rightarrow \ket{\Psi}$ is \emph{entangled} (the state is obtained by symmetrization of a product state of non-orthogonal single-particle states).
\item[(v)] Slater rank of $\ket{\Psi}=2 , S(\varrho^{(1)})=1 \Rightarrow \ket{\Psi}$ is \emph{non-entangled} (the state is obtained by symmetrization of a product state of two orthogonal single-particle states). 
\item[(vi)] Slater rank of $\ket{\Psi}>2 \Rightarrow \ket{\Psi}$ is \emph{entangled} (the state is composed by symmetrizing a sum of more than one product states). 
\end{itemize}
Again, the entropy of the reduced density matrix $S(\varrho^{(1)})$ reflects the uncertainty in the particle label. Due to the higher occupation numbers allowed for bosons, it is, however, not bounded from below as for fermions. Note that the entanglement of a state is directly reflected neither by the Slater rank, nor by  the entropy alone: A Slater rank of 2 can correspond to, both, an entangled and a non-entangled state. The two latter cases (v) and (vi) are, again, analogous to the case of non-entangled and entangled distinguishable particles, as for fermions. The first two situations (iii) and (iv), however, can only occur for bosons and do not possess any analogy with distinguishable particles. They are realized by the example (\ref{bosonentan}) discussed above.

\subsection{Measurement-induced entanglement} \label{measurementinduced}
In the preceding Section, we elaborated on a notion of entanglement for identical particles that follows the spirit of the original EPR approach \cite{PhysRev.47.777}: If it is possible to assign a physical reality (subsystem properties) to the constituent particles of a composite quantum system, they are considered as \emph{not entangled}. When dealing with identical particles, however, another source of quantum correlations emerges. In measurement setups that delete which-way information \cite{Englerta:1999uq,Walborn:2002oj}, entanglement can be created between identical particles without any interaction, in contrast to distinguishable particles (see Section \ref{entint}). In order to understand this, consider, once again, the quantum state 
\eq  \ket{\Psi_{AB}}= \frac{1}{\sqrt{2}} \left(  \ket{A, 1} \otimes \ket{B, 0} \pm \ket{B,0} \otimes \ket{A,1} \right) .  \nonumber  \hspace{2.4cm} (\ref{ABstate})   \en
As we have shown above, this state is not entangled according to the criterion of \emph{particles which possess a complete set of properties.} Indeed, we can unambiguously assign the property $P=\ket{A, 1}\bra{A, 1}$ to one of the particles. 

Let us now assume, however, that we use two detectors to measure the particles, and that these detectors do not spatially project onto $\ket{A}$ or $\ket{B}$, but on a linear combination of these states: We decide to measure a particle in the external state $\ket{L}:=(\ket{A}+\ket{B})/\sqrt{2}$, and another one in the external state $\ket{R}:=(\ket{A}-\ket{B})/\sqrt{2}$, orthogonal to $\ket{L}$. Such measurement is not deterministic, {\it i.e.}~not in each realization of the experiment does one find one particle in each detector. The internal states of the particles which are possibly registered by the detectors are not determined any more, {\it i.e.}~no assignment of a complete set of properties is possible for particles located in $\ket{L}$ or $\ket{R}$ -- while \emph{before} the measurement the state (\ref{ABstate}) did not exhibit entanglement, according to our criterion defined earlier. A posteriori, the internal states of the particles detected in $\ket{L}$ and $\ket{R}$ are \emph{perfectly anti-correlated}, and the Bell inequality (\ref{CHSH}) can be violated \cite{Tichy:2009kl}. 

\begin{figure}[h] \center \includegraphics[width=6.5cm,angle=0]{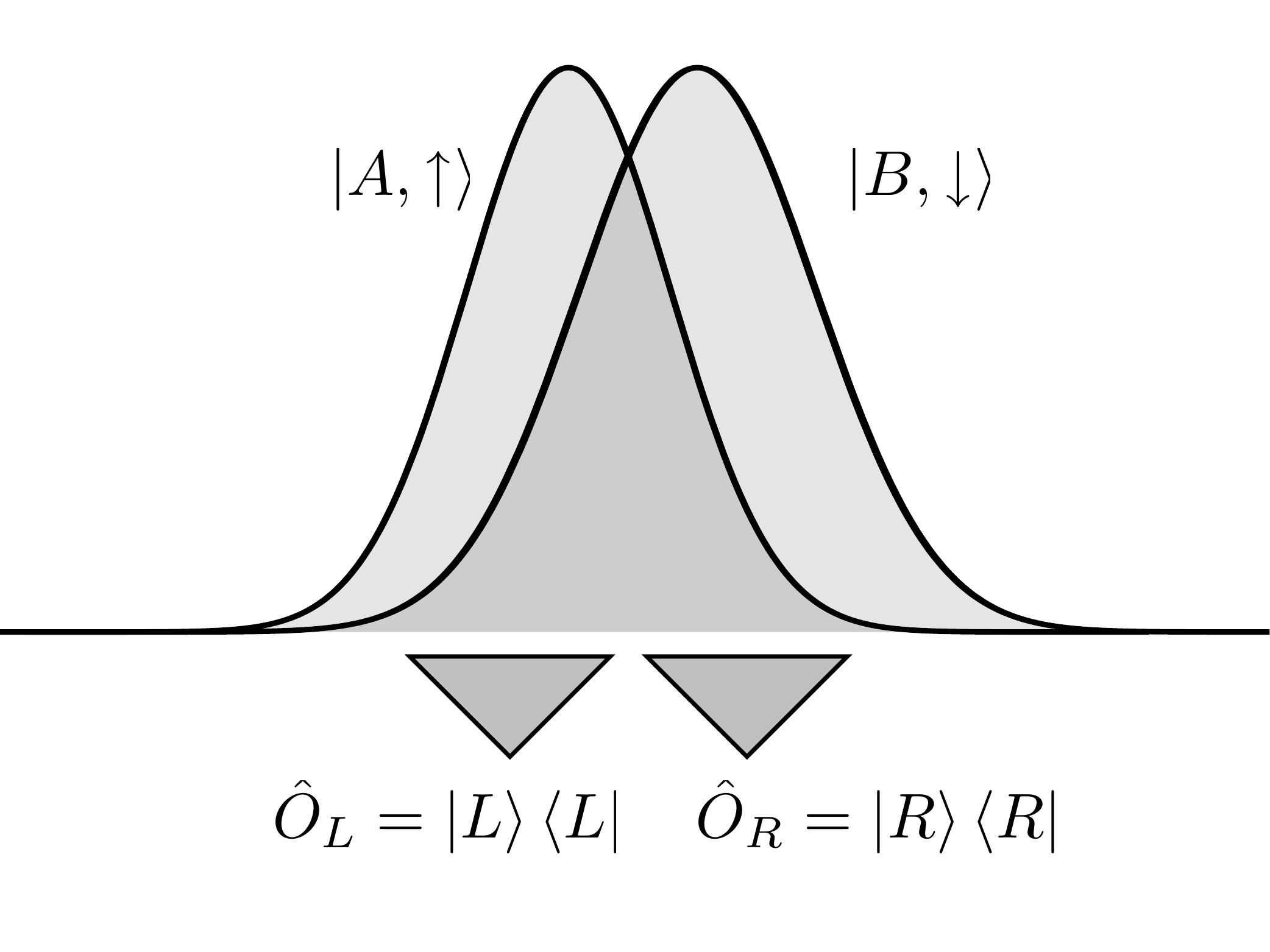} \caption{Two particles are initially prepared in two orthogonal quantum states $\ket{A}, \ket{B}$ with definite spin values $\ket{\uparrow}, \ket{\downarrow}$, which permits the attribution of unambiguous properties to each of them. They hence both possess a physical reality. However, for ambiguous detector settings with $\ket{B}\neq \ket{L}\neq \ket{A}$ and $\ket{B} \neq \ket{R} \neq \ket{A}$, the detection of one particle in each detector, $\ket{L}\bra{L}$ or $\ket{R}\bra{R}$, will come along with wide uncertainty regarding the state of the measured particles' spin, due to the deletion of which-way information in the course of the measurement.}  \label{ambiguoussetting}  \end{figure}

We call such choice of detectors $ O_L=\ket{L}\bra{L}$ and $ O_R=\ket{R}\bra{R}$ \emph{ambiguous}, since both particles initially prepared in the state given by (\ref{ABstate}) have a finite probability to trigger each detector, {\it i.e.}~the detectors have no one-to-one relationship to the external states of the particles. This situation is also illustrated in Figure \ref{ambiguoussetting}: The quantum state is initially non-entangled according to the above (EPR) criteria, but the individual spins of the particles measured by the two detectors are maximally uncertain, and strictly anti-correlated. Thereby, erasure of which-way information takes place: By the measurement of a particle in the state $\ket{L}$, the initial preparation of the particle -- whether in $\ket A$ or $\ket B$ -- is completely obliterated.

An ambiguous choice of the detector setting can thus induce quantum correlations between the measurement results at these detectors, even if the initial state (like the one in (\ref{ABstate})) is non-entangled, and no interaction between the particles has taken place. For an initially entangled state, the detected particles can result to be more, but also to be less entangled \cite{Tichy:2009kl}, depending on the details of the setup. Furthermore, the statistical behavior of identical particles upon detection strongly depends on whether and how they are entangled. For example, photons prepared in a maximally entangled $\ket{\Psi^-}$-state, see (\ref{Psimp}), behave like fermions when scattered on an unbiased beam splitter \cite{PhysRevLett.61.2921}: When one photon enters at each input port, they will always exit at different ports and never occupy the same output mode.

We denote this phenomenon -- non-vanishing entanglement upon detection of particles initially prepared in a non-entangled state (according to  \cite{Ghirardi:2004kx} and (\ref{sepghirardi})) --  \emph{measurement-induced} entanglement. This is the characteristic additional feature encountered when dealing with identical particles and their entanglement properties. 

We retain that, in contrast to distinguishable particles, which can be entangled by  deterministic procedures mediated by mutual interaction, measurement-induced entanglement is intrinsically probabilistic, {\it i.e.}~the success rate to find one particle in each detector is strictly smaller than unity. 

\subsubsection{Entanglement extraction} \label{entextr}
The abstract principle of deleting which-way information constitutes the basis for many applications which exploit the indistinguishability of particles to create entangled states. Such schemes are widely used, for example in today's quantum optics experiments with photons (see \cite{Prevedel:2009ec,Wieczorek:2009ff,Wieczorek:2009fe,Deb:2008xr,PhysRevLett.101.010503}, for an inexhaustive list of very recent state-of-the-art applications). For massive particles, however, no experiments have been performed yet which implement analogous ideas, but recent advances in current technology (see, {\it e.g.}~\cite{Karski:2009kx}) feed the hope that a realization analogous to the photon experiments will be performed in the near future. 

\begin{figure}[h] \center \includegraphics[width=6.5cm,angle=0]{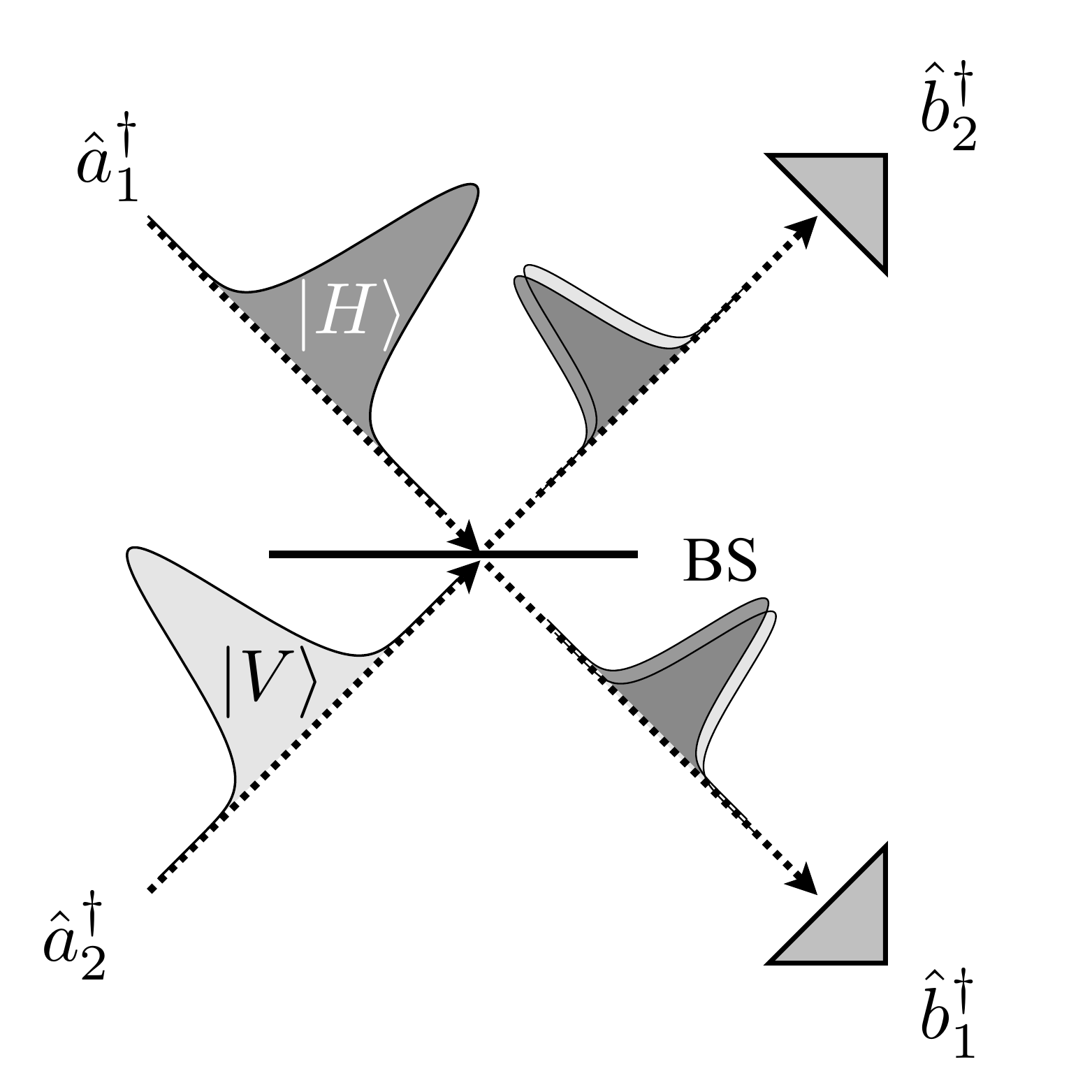} \caption{Entanglement creation with the Hong-Ou-Mandel (HOM) setup \cite{Hong:1987mz}. Two photons that are initially prepared in the modes $\hat a_{1(2)}^\dagger$ and which are horizontally (vertically) polarized fall onto the beam splitter (BS). Upon detection of one photon in either one of the two detectors, the polarization state of this photon is unknown, but fully anti-correlated to the polarization of the other one. Purely quantum correlations are detected when the photons have perfect overlap in space, time, and frequency. Slight deviations from this ideal case, as suggested in the figure where the overlap of the transmitted and reflected wave packets is not optimal, jeopardize the quantum nature of the correlations \cite{Tichy:2009kl}.}  \label{HOMfig}  \end{figure}

An ambiguous choice of detectors which implements the scheme of Section \ref{measurementinduced} is realized by a simple beam splitter in a Hong-Ou-Mandel configuration \cite{Hong:1987mz}: In the original experiment, two identical photons fall onto the opposite input ports of a balanced beam splitter and, due to two-particle interference, they always leave the setup together, at one output port. Instead of two identical photons, one can inject a horizontally and a vertically polarized photon \cite{PhysRevLett.61.2921}, as illustrated in Figure \ref{HOMfig}, {\it i.e.}~the initial state
\eq  \ket{\Psi_{\mathrm{ini}}}= \hat a^\dagger_{1,H}\hat a^\dagger_{2,V} \ket{0} \label{HOMINI} .\en
The final state after the scattering on the beam splitter reads
\eq \ket{\Psi_{\mathrm{fin}}}=\frac 1 2 \left(b_{1,H}^\dagger b_{1,V}^\dagger -b_{2,H}^\dagger b_{2,V}^\dagger + b_{1,H}^\dagger b_{2,V}^\dagger -b_{2,H}^\dagger b_{1,V}^\dagger  \right) \ket 0 \label{HOMFIN} .\en
Disregarding measurement results with two photons in the same port  -- described by the first two terms in $\ket{\Psi_{\mathrm{fin}}}$ --,  which is known as the \emph{post-selection} of those detection events with one particle in each output port (the two last terms in $\ket{\Psi_{\mathrm{fin}}}$), one effectively performs a projection onto the maximally entangled $\ket{\Psi^-}$ Bell state. 
Consequently, when analyzing the postselected subset of the total measurement record, perfect anti-correlations between the measured photons are encountered  \cite{Giovannetti:2006jh}. Such a combination of two-particle interferometry and which-way detection can be used to entangle any degree of freedom of indistinguishable particles, whether bosons or fermions \cite{Bose:2002le,Bose:2002vf}, or to distinguish fermionic from bosonic quantum states \cite{Bose:2003kx}. Recent experiments have shown that this scheme works even if the photons are created independently \cite{Beugnon:2006ec,Halder:2007th} or measured at very large distances from each other \cite{Marcikic:2004qr}. These effects lend themselves to manipulate entangled states by purely quantum-statistical means, {\it i.e.}~to change their entanglement properties without the need of any interaction between the constituents \cite{Omar:2002zr,al:2010lh,Paunkovic:2002jt}. 

\begin{figure}[h] \center \includegraphics[width=10.5cm,angle=0]{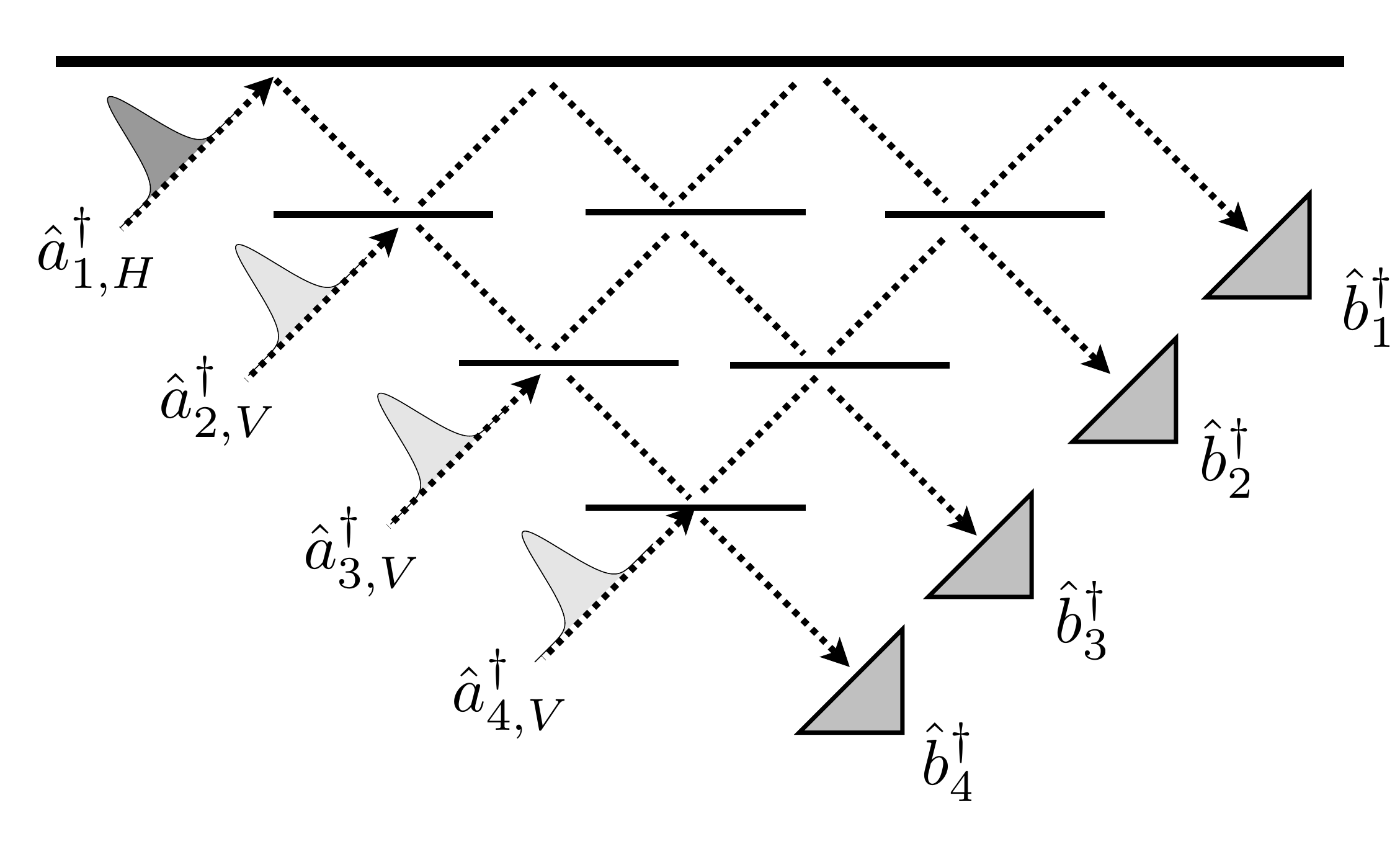} \caption{Multiport beam splitter  setup to create a four-particle W-state \cite{Lim:2005bf}. The beam splitters (short and thin horizontal lines) have reflectivities chosen such that the probability for any incoming photon to exit at any output port is always 1/4. When one particle is found at each output port, the four-photon polarization state is described by a four-particle W-state \cite{Lim:2005bf}. Note the strict analogy to the scheme in Figure \ref{HOMfig}.}  \label{multiport}  \end{figure}

The two-particle Hong-Ou-Mandel effect on which the above scheme relies can be generalized to an arbitrary number of input and output ports \cite{Lim:2005qt,Tichy:2010kx}. This also allows to generalize the above scheme to create entanglement within multipartite systems \cite{Zhang:2006yq,PhysRevA.71.013809}. Again, one prepares a non-entangled state of photons in which not all photons share the same internal state. The particles are scattered simultaneously on a multiport beam splitter, as illustrated in Figure \ref{multiport}. Detection is conditioned on one particle at each output port \cite{Lim:2005bf,Wang:2009ud}. The which-path information and thereby the information on the internal state of the photons that are found at each output port is deleted in the course of the scattering process, and the state obtained upon detection of  the predefined event (one particle in each output port) is a many-particle entangled state. Such situation is illustrated for four particles in Figure \ref{multiport}.
 It is thus possible to create a multitude of different multipartite entangled states \cite{Lim:2005bf,Sagi:2003qf}. Similar schemes, where the photons propagate in free space rather than through a beam splitter array, were shown to permit the preparation of multipartite entangled states through suitable settings and detection strategies \cite{PhysRevLett.99.193602,PhysRevLett.102.053601,Maser:2010fu,Maser:2009pi}. Indeed, with the local variation of the polarization states onto which the photons are projected, it is possible to tune through several families of entangled states \cite{PhysRevLett.101.010503}, including the large subclass of states that are symmetric upon exchange of any two subsystems \cite{PhysRevLett.99.193602,PhysRevLett.102.053601}. The induced correlations are not restricted to the polarization, but can also be created in motional degrees of freedom \cite{Guo:2008hc}. 

All the above scenarios rely on the fact that all particles are identical, and only distinguished through the mode and the internal state they are initially prepared in, {\it i.e.}~only through the property used for discrimination, and the degree of freedom in which they will be entangled in the final state. Whenever some discriminating information is available on output, {\it i.e.}~when the particles are partially distinguishable, {\it e.g.} due to different arrival times or different energies, the entanglement in the final state is jeopardized  \cite{Velsen:2005ve,branning1999eia,Velsen:2005nx,Jha:2008bs}. Instead of a fully quantum-mechanically correlated state, the state exhibits more and more classical correlations, and the more so the more distinguishable the particles are. Indeed, the transition from fully indistinguishable to fully distinguishable particles used in these schemes corresponds to a transition from purely quantum-mechanically entangled to purely classically correlated final states \cite{Tichy:2009kl}.

Besides the intrinsic indistinguishability which is the very basis of the entanglement strategies described above, it is also possible to exploit the Pauli principle, and thereby use the quantum-statistical behavior of fermions \cite{Cavalcanti:2007vn}, to create entangled states. Quantum correlations between the spins of two independent fermions in the conduction band of a semiconductor are enforced by the Pauli principle, and selective electron-hole recombination then transfers this entanglement to the polarization of the emitted photons. 

\subsubsection{Entanglement in the Fermi gas} \label{fermigasentanglement}
Different types of entanglement can be considered within a Fermi gas. On the one hand, electron-hole entanglement can be defined \cite{Beenakker:2006zt}. Here, we focus, however, on the entanglement between identical particles, specifically the electrons, in their spin degree of freedom. This entanglement can be understood as measurement-induced, similar to what we saw above. Different studies illustrate how measurement-induced entanglement is ubiquitous in a scenario in which the particles are measured in states different from those they were prepared in, in full analogy to the generic situation described in Section \ref{measurementinduced}: The Fermi gas is constituted of electrons prepared in momentum eigenstates, which are then detected in position eigenstates. 

The ground state of a Fermi gas of non-interacting particles at zero temperature reads 
\eq \ket{\phi_0}=\prod_{s=\uparrow,\downarrow} \prod_{|p|\le k_F} b_s^\dagger(p) \ket{0} , \en
where $\ket{0}$ represents the vacuum and $b_s^\dagger(p)$ creates an electron with spin $s$ and momentum $p$. This state corresponds to a single Slater determinant, and it is therefore not entangled according to our reasoning in Section \ref{subpropid}, but may exhibit measurement-induced entanglement. Indeed, when two particles are detected at positions $r$ and $r^\prime$, one observes entanglement between the spins of the detected electrons  \cite{Vedral:2003uq}. The spin correlations found between the fermions depend on their relative distance $x=|r-r^\prime|$: The two-body reduced density matrix of the particle pair reads
\eq \rho_{12} = \left( \begin{tabular}{cccc}
$1-f(x)^2 $& 0 & 0 & 0 \\
0 & 1 & $-f(x)^2$ & 0 \\
0 &  $-f(x)^2$ &1& 0 \\
0 & 0 & 0 &$1-f(x)^2 $ 
 \end{tabular} \right), \label{fermistate}
\en
in the basis \eq \{ \ket{\uparrow}_r \otimes \ket{\uparrow}_{r^\prime}, \ \ket{\uparrow}_{r}\otimes \ket{\downarrow}_{r^\prime}, \ \ket{\downarrow}_{r}\otimes \ket{\uparrow}_{r^\prime},\  \ket{\downarrow}_{r}\otimes \ket{\downarrow}_{r^\prime} \}, \en which describes the spin orientations of the two electrons found at the two locations, $r$ and $r^\prime$, with 
 \eq f(x) = \frac{3 j_1(k_F x)}{k_F x}, \en where $j_1(x)$ is the Bessel function of the first kind.
We can interpret (\ref{fermistate}) as a state of two effectively distinguishable particles \cite{Tichy:2009kl}, since it describes the spin state of the two electrons detected by two distinct detectors. Thereby, we can apply the notions of Section 2. 

The two-particle state (\ref{fermistate}) is a Werner state (see (\ref{WernerState})) with the particular property that it cannot violate any Bell inequality for a large range of  $f$, although it exhibits entanglement \cite{Werner:1989ve}. For inter-particle distances $x$ of the order or smaller than $\pi/(8 k_F)$, we find $f(x)^2\ge 1/2$, and the detected particles result to be entangled, by virtue of (\ref{purestatedecomp}). Physically, this corresponds to the situation in which the Pauli principle inhibits the detection of both particles with the same spin, and forces the electrons into an anti-correlated state.

The above consideration allows the extraction of a many-particle density matrix that describes the spin state of the detected particles, and permits to tackle entanglement between electrons in the Fermi gas under numerous distinct perspectives. Instead of detecting only two particles, it is immediate to study multipartite entanglement between many electrons at different locations \cite{Jie:2008oq,Lunkes:2005ao,Vertesi:2007qu,Habibian:2010ys}. Finite temperatures can be considered \cite{Oh:2004yq,Lunkes:2005et}, and also electron-electron interactions were included \cite{Hamieh:2009zr,Hamieh:2010ly}, although only as screened Coulomb interactions in an effective treatment that provides an approximation to the Fermi liquid. It is also possible to relax the assumption that the particles be projected onto position eigenstates \cite{Cavalcanti:2005vl}, and consider more realistic, coarse-grained measurement devices. All studies in this area share the conclusion that a rich variety of entangled states can be extracted, purely due to measurement-induced entanglement, and that such quantum correlations are not substantially affected when finite temperatures or screened Coulomb interactions are incorporated in the model. 
Hence, the measures for particle entanglement introduced in Section \ref{idparticlesee} are satisfactory in concept, but many dynamical situations occur in which the detection process itself indeed induces entanglement between particles, and a treatment that takes into account measurement-induced entanglement, as discussed in Section \ref{measurementinduced}, is more appropriate. This conclusion is not restricted to the case of a Fermi gas, but the scheme can be applied to any system of many identical particles.

\subsection{Mode entanglement} \label{modenen} 
While the above discussion of the entanglement of identical particles was centered around the correlations between degrees of freedom carried by \emph{particles}, one can also define the entanglement between modes. In this case, it is the very number of particles which becomes the degree of freedom in which the entities -- the modes -- can be entangled. For example, the state \eq \frac{1}{2} \left(  a^\dagger  a^\dagger \otimes \mathbbm{1}_{(B)}+ \mathbbm{1}_{(A)} \otimes \hat b^\dagger \hat b^\dagger \right)  \ket {0}_A \otimes \ket {0}_B \nonumber \\ = \frac{1}{\sqrt 2} \left( \ket{0}_A\otimes \ket{2}_B + \ket{2}_A\otimes\ket{0}_B \right) ,\en
where $\hat a^\dagger$ and $\hat b^\dagger$ represent particle creation operators of mode A and B,
is a state which exhibits entanglement between modes A and B (we make the tensor product structure explicit here): It describes a coherent superposition of two particles located in A and two particles located in B. The local particle number in $A$ and $B$ is unknown and exhibits correlations with the number of particles in the respective other mode. 

\subsubsection{The definition of modes and the problem of locality}
The Pauli principle results in a maximal mode occupation number one for fermions, so that  each mode can only reside in the state $\ket 0$ (no fermion) or $\ket 1$ (one fermion). That is to say, the Fock-space of $m$ fermionic modes is isomorphic to an $m$-qubit Hilbert-space, which is formalized by a map $\Lambda$ with the property \cite{Zanardi:2002uq}
\eq 
\Lambda: \prod_{l=1}^L \left(\hat c^\dagger_l \right)^{n_l} \ket{0}^{2nd quant.} \mapsto \otimes_{l=1}^L (\sigma_l^+)^{n_l} \ket{0,\dots,0}^{1st quant.} 
,\en
where $n_l \in \{0,1\}$, and we made the first and second-quantized formulation explicit: $c^\dagger_l$ is the fermionic creation operator for mode $l$, and $\sigma^+_l$ is the raising operator of the $l$th qubit, {\it i.e.}~$\sigma^+_l \ket{0}_l=\ket{1}_l$. On the level of operators, the Jordan-Wigner transformation \cite{Jordan:1928uq}  maps fermionic creation operators onto raising operators, such that 
\eq 
\hat c_l^\dagger \mapsto \sigma_l^+ \prod_{k=1}^{l-1}(-\sigma_k^z) , \ \ \sigma_l^\dagger \mapsto \hat c_l^\dagger \prod_{k=1}^{l-1} e^{i \pi \hat c_k^\dagger \hat c_k } .
\en
In other words, in order to take into account the fermionic anti-commutation relations for \emph{distinct} sites, $\{\hat c_k, \hat c^\dagger_l \}=\delta_{k,l}$, the mapping from the $l$th creation operator to the raising operator is \emph{non-local in the modes}, since it depends on the occupation numbers of all modes $k$ with $k<l$. Consequently, also the action on one mode depends on the occupation of other modes, and cannot be considered ``local'' anymore. The difficulties that arise due to this non-trivial mapping are aggravated by the impossibility of arbitrary rotations of qubits due to particle-number conservation (for massive and/or charged fermions, $\sum_{k} n_k= \mathrm{const.}$), or total parity conservation (for excitations, $\mathrm{mod}(\sum_{k} n_k,2)=\mathrm{const.}$). Fermionic mode entanglement can still be exploited, {\it e.g.}~for quantum computation purposes when suitable protocols are designed \cite{Bravyi2002210}, allowing applications in close analogy to the unrestricted form of particle entanglement.

Fermionic mode entanglement also possesses a physical interpretation, since it can be related to different types of superconductivity \cite{Zanardi:2002fk}. For example, in the BCS-model \cite{Bardeen:1957kx}, a finite value of the superconductivity order parameter is a necessary condition for entanglement, superconductivity and entanglement are thus related. 

Mode entanglement can be defined analogously for bosons \cite{Sanders:1992kx,Huang:1994uq}, with the difference that, due to the unconstrained occupation number, the resulting Hilbert space is much larger than in the case of fermions. 

In general, mode entanglement is strongly dependent on the way the modes are defined themselves, since any transformation of mode creation operators \eq \hat c^\dagger_i \rightarrow \tilde c^\dagger_i = \sum_j U_{ij} \hat c_j^\dagger \label{modetransfo} ,\en with unitary $U$ leaves the commutation relations of the creation operators invariant, while the reduced density matrix for a single mode will in general change considerably \cite{Zanardi:2002uq}. Indeed, a transformation as (\ref{modetransfo}) ought to be considered \emph{non-local}, since different modes are brought into coherent superposition of each other. 

While central to the definition of the entanglement of distinguishable particles, where the term ``local'' denotes operations which act on one subsystem alone, the very concept of locality is indeed difficult to define, let alone to apply consistently, when dealing with the entanglement of identical particles or with the entanglement of modes: The approaches to particle entanglement proposed in \cite{schliemann-cirac} -- particle entanglement according to complete sets of properties, as in Section \ref{entidp} --  and in \cite{Zanardi:2002uq} -- mode entanglement -- are compared in \cite{Gittings:2002gf}. The authors come to the conclusion that the definition based on the Slater rank presented in \cite{schliemann-cirac} does not meet the requirements for an entanglement measure, and is thus inappropriate for the quantification of entanglement of identical particles, since it is invariant under non-local operations, while it changes under operations restricted to one mode. This argument is based on the observation that a single-particle two-site unitary transformation does not affect the Slater rank of any non-entangled state and, thus, does not change its entanglement according to \cite{schliemann-cirac}, while such operation can be considered as nonlocal in the sites/modes. This line of thought shows that particle entanglement is a concept which does not go well along with the very concept of locality, since particles can be strongly delocalized and yet unentangled, in the sense that they individually possess a complete set of properties (in the EPR sense of Section \ref{properties}), as we show in the following: Consider, {\it e.g.}, the passage of two orthogonally polarized photons through a beam splitter, as already discussed in Section \ref{entextr} with Eqs.~(\ref{HOMINI}),(\ref{HOMFIN}), and illustrated in Figure \ref{HOMfig}. The quantum state and its evolution read \eq \ket{\Psi_{\mathrm{ini}}}=\hat a^\dagger_{1,H}\hat a^\dagger_{2,V} \ket 0 \mapsto \frac 1 2 (b_{1,H}^\dagger - b_{2,H}^\dagger)(b_{1,V}^\dagger + b_{2,V}^\dagger) \ket 0  \nonumber \\ =: \tilde a^\dagger_{1,H} \tilde a^\dagger_{2,V} \ket 0  \label{moderedefi}, \en
where $\hat a_{i,X}^\dagger$ ($\hat b_{i,X}^\dagger$) denote creation operators for particles in mode A$_i$ (B$_i$) with polarization $X$. While the initial state is clearly non-entangled, since the modes are spatially separated, the final state is entangled in the modes defined by the creation operators $b^\dagger_{k,Y}$. Since the photons never interact, the functional form of the initial state is preserved in the final state, which therefore can still be rewritten as a non-entangled state in terms of transformed modes $\tilde a_{k,Y}^\dagger$, as done in the second line of (\ref{moderedefi}). While the particles in the modes populated by $\tilde a^\dagger_j$ can still be attributed a full set of properties, local measurements in the modes $b_k$ exhibit quantum correlations. The choice of modes is clearly induced by the experimental setting: measurements take place on the modes defined by $\tilde b_j^\dagger$, and since entanglement is fundamentally defined by the correlations of measurement results (see Section \ref{Bellin}) the application of an entanglement measure ought to be performed on the state as described in terms of the $\tilde b_j^\dagger$, in the first line of (\ref{moderedefi}), while recasting the state into the second line's form is nothing but a formal manipulation, without a physical counterpart. 

When entanglement is \emph{not} considered within a measurement setup, the question  arises of which is the natural mode substructure to be considered in order to quantify the entanglement induced \emph{solely} by interaction, in a system of identical particles. A natural choice is given by the eigenstates of the single-particle Hamiltonian, in the absense of the interaction part \cite{Shi:2004la}. The resulting mode-entanglement, when the interaction is turned on again, then  originates uniquely from the interaction terms of the system Hamiltonian, and is not a by-product of the above redefinition of modes. However, given the indispensable, defining anchoring of entanglement to correlations of measurement results, such considerations have a rather formal flavor. 

\subsubsection{Particle entanglement}
The physical relevance of mode entanglement has been an object of intense debate \cite{Enk:2005gd,Enk:2006re,Drezet:2006gf,Drezet:2007by,Heaney:2009jl}, since a particle number superselection rule (SSR, see Section \ref{secSSR}) forbids, in principle, the detection of coherent superpositions of states of massive particles with different local particle numbers. While we show in Section \ref{modeentdet} that schemes exist which allow to circumvent this problem in some cases for both massless and massive particles, given a suitable experimental infrastructure, let us assume for the moment that such equipment is not present, and that particle-number correlations of massive particles cannot be regarded as entanglement, as proposed in \cite{Wiseman:2003mz}. 
The system under consideration allows its modes to have a variable number of particles, and the particles to carry an internal degree of freedom. The mode entanglement of a given state stems from both, the entanglement in the particle-number degree of freedom, {\it i.e.}~intrinsic mode entanglement, and entanglement in the internal degree of freedom of the particles. In order to assure that only the latter be capitalized as bona fide entanglement, only the entanglement from the state's projection onto states with locally well-defined particle numbers is computed \cite{Wiseman:2003mz}, thus neglecting the coherent superposition of different particle numbers. The result then reduces to the entanglement one would assign to distinguishable spins \cite{Shi:2003ys}. Simple models of two bosons or two spinless fermions distributed over four modes, the Hubbard dimer, and non-interacting electrons on a lattice are treated following this convention in \cite{Dowling:2006kx}, with a comparison to the application of a mode-entanglement measure. For the electrons on a lattice, the resulting entanglement in the spin degree of freedom agrees well with studies of entanglement in the Fermi gas as mentioned in Section \ref{fermigasentanglement} above. In general, the degree of particle entanglement is found to differ from the degree of mode entanglement. For example, the entanglement of particles for two non-interacting bosons in four sites vanishes, while the entanglement of modes of the same system adopts a finite value. Such discrepancy can be expected, since the very carriers of entanglement are different.

\subsubsection{Detection of mode-entanglement} \label{modeentdet}
The discussion of the previous Section followed the assumption \cite{Wiseman:2003mz} that coherent superpositions of different particle numbers of massive particles cannot be detected. This assumption itself is, however, under debate \cite{Enk:2005gd}, as we will discuss in the following. The key issue is best illustrated by the case of a single, delocalized particle. This simple setting has attracted much attention during the last decades, from very different perspectives \cite{Hardy:1994fu,Revzen:1996zt,Aharonov:2000mi,Dunningham:2007pi,Cooper:2008ff,Peres:1995mi,Pawlstrokowski:2006pi,Lo-Franco:2005ff,Lee:2003lh,Bjork:2001fu,Moussa:1998il}. The debate on whether such state bears entanglement or not has been intense, and only in the case of photons the issue can be regarded as thoroughly settled by experimental verification. Note, however, the fundamental difference between massive particle for which conservation laws imply a SSR, {\it e.g.} of charge, and massless particles such as photons, which can be locally created or destroyed \cite{Ashhab:2007udd}, under the condition that energy, momentum and angular momentum are conserved. 

The controversy arises from the fact that a state of a delocalized photon, which one obtains, {\it e.g.}, by  simple scattering  on a beam splitter, can be written, in the occupation number basis, as \cite{Enk:2005gd} 
\eq \ket{\Psi}_{A,B}=\frac{1}{\sqrt{2}}\left( \ket{0}_A \otimes \ket{1}_B + \ket{1}_A \otimes  \ket{0}_B \right) \label{singlephotonnonlocal} \en
where $A,B$ denote the modes, and $\ket{k}_{A/B}$ the occupation number, or Fock state. The state seems to exhibit entanglement between mode $A$ and mode $B$, but it is fully equivalent to the application of the following linear combination of creation operators on the vacuum
\eq \ket{\Psi}_{A,B}=\frac{1}{\sqrt 2}\left( \hat a^\dagger_A \otimes \mathbbm{1} +  \mathbbm{1} \otimes \hat a^\dagger_B \right) \ket{\mathrm{0}}_A \otimes \ket{\mathrm{0}}_B ,\en
for which the entanglement is not as obvious. When expressed in first quantization, there is, due to the irreducible structure of the Hilbert space, no place for entanglement at all: 
\eq \ket{\Psi}_{A,B}=\frac{1}{\sqrt 2} \left( \ket{A}+\ket{B} \right) .\en The choice of modes as the carriers of entanglement suggests, however, that indeed (\ref{singlephotonnonlocal}) captures the essential physics of the problem.

The apparently intuitive argument brought up against the feasibility to test the non-locality of such a state was that the detection of the particle at one mode -- a necessary requirement in the experiment --  would prohibit the measurement of any property associated with this very same particle at the other mode. However, by its construction does the verification of non-locality rely rather on the \emph{wave}-like rather than the \emph{particle}-like features, which indeed can be probed, as we will see now. 

Proposals to verify a single particle's mode entanglement mainly follow two lines, and we review one exemplary setup for each. \begin{figure}[h] \center \includegraphics[width=8.5cm,angle=0]{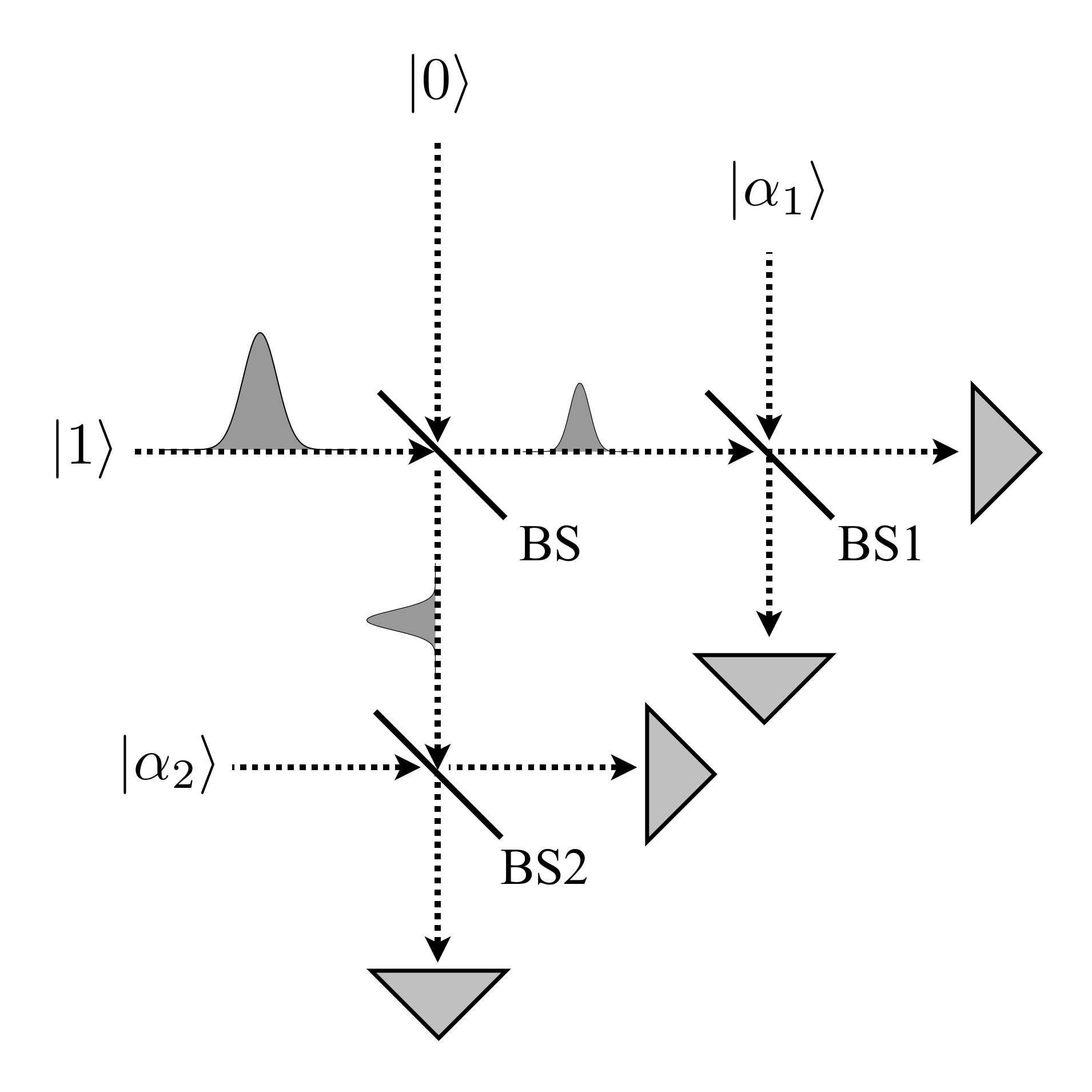} \caption{Mode entanglement detection with coherent states \cite{Hardy:1994fu,PhysRevLett.66.252}. The non-local single photon state is superposed with the coherent states $\ket{\alpha_1}$ and $\ket{\alpha_2}$, respectively, at the two beam splitters BS1 and BS2. The four detectors are depicted as triangles. }  \label{coherentStateModeentdet}  \end{figure}

\paragraph{Phase reference via external ancilla states}
 The first method relies on coherent states as phase reference \cite{PhysRevLett.66.252}: one of the modes in (\ref{singlephotonnonlocal}) is superposed, at a beam splitter, with a coherent state $\ket{\alpha}$ in the other mode. A photon detector that is located at one output port of the beam splitter cannot distinguish photons which  originate from either input mode. Therefore, the amplitudes for a single photon detection event, which stem from the single photon mode and from that fed with the coherent state, are added. Variation of the phase of the coherent state allows to project, upon detection of a photon at one output port of the beam splitter, onto different coherent superpositions of one and no photon in the input mode. Thus, the relative phase of the one-particle component $\ket{1}$ with respect to the no-particle component $\ket{0}$ can be measured, {\it i.e.}~one is not restricted to measurements in the basis $\{ \ket{0}, \ket{1} \}$.
This procedure can be performed on both modes and thereby permits a full Bell experiment, as illustrated in Figure \ref{coherentStateModeentdet}. 
The setup works analogously when single photon states are used instead of the coherent states  \cite{Lee:2000dk,Sciarrino:2002kl}, provided a phase reference is available. 

While the very implementation of that setup was not considered an issue and also acknowledged by critics, it was argued that the exhibited non-locality is not a property of the single photon, but intrinsically relies on the coupling to the coherent states which may bring in additional non-locality \cite{Vaidman:1995ye,Greenberger:1995qo,Hardy:1995tw}. This problem can be circumvented by modifying the setup such that the coherent states that are used are totally uncorrelated \cite{Cooper:2008ff}: If the reference states are created by the observers themselves in an independent way, no additional quantum correlations can be induced. The phase relation between the coherent states is crucial for the functioning of the scheme, but can also be achieved by purely classical communication. 

Experimentally, mode entanglement was finally verified, closely following the original proposal \cite{PhysRevLett.66.252}, in \cite{Hessmo:2004bs,Babichev:2004fv} with photons delocalized in space, and in \cite{DAngelo:2006dz} for photons delocalized in time.

Note, however, that this verification does not rely on the fact that one uses massless particles: Instead of a coherent state of light, a reservoir such as a BEC can also be used as a reference state, when massive particles are under consideration. Thereby, the fundamental SSR for the particle number which states that no coherent superpositions of a different number of particles can be directly measured, can be circumvented \cite{Ashhab:2007if,Heaney:2009jl,Heaney:2009kh,Goold:2009ys}. 

\paragraph{Transfer of mode entanglement onto ancilla particles}
The second method for the verification of mode-entanglement does not require coherent states as a reference, but introduces ancilla particles, which acquire the information whether the mode was populated or not, in a coherent way.  A subsequent readout of the state of the ancilla particles can then violate locality. In contrast to the above setup using coherent reference states, the issue of creation or annihilation of massive particles  (ruled out by a fundamental SSR) is more intricate here, as we will see below. The basic idea is strongly influenced by a proposal for entangling gates in quantum computation \cite{Cirac:1994qf}, and can be described as follows: 
\begin{figure}[h] \center \includegraphics[width=6.5cm,angle=0]{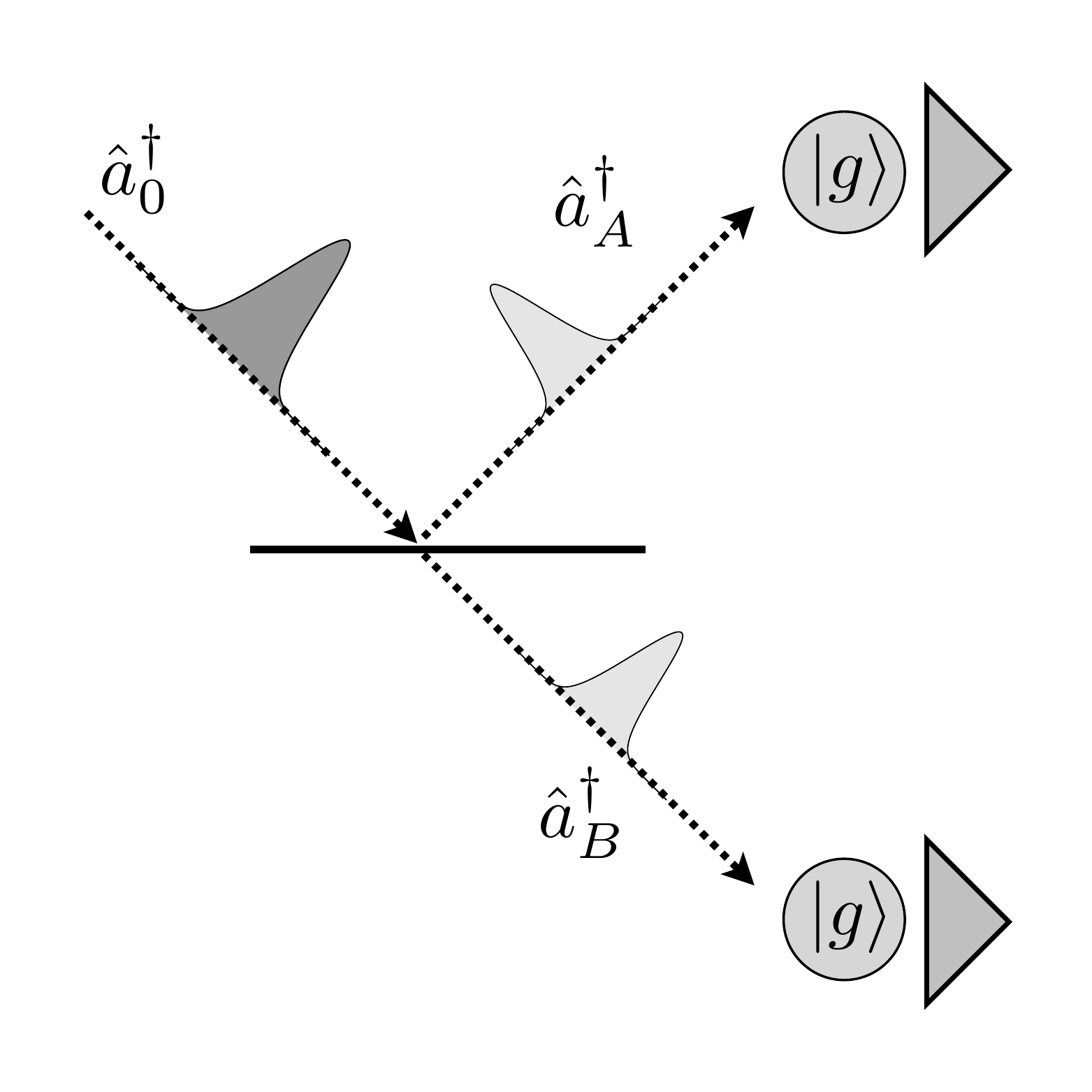} \caption{Detection of mode-entanglement with ancilla particles. A photon is scattered by the beam splitter (solid horizontal line), and excites coherently either the upper ancilla particle, or the lower one, which are both initially in their ground state $\ket{g}$. A subsequent read-out of the states of the ancilla particle in different bases reveals the quantum correlations between them.}  \label{Modeentanglementillu}  \end{figure}
Consider the state  \cite{Enk:2005gd,Cunha:2007fk,Ashhab:2007udd} 
\eq \frac{1}{\sqrt 2}\left( \ket{0,1}_{A,B}+\ket{1,0 }_{A,B} \right)\otimes \ket{g,g} ,\en where $\ket{i,j}_{A,B}$ represent the state of the modes in which $i$ particles are prepared in mode $A$, and $j$ particles are prepared in $B$. The state $\ket{g,g}$ describes two ancilla particles in their ground state, locally coupled to the modes $A$ and $B$, respectively. If an interaction between the particles prepared in the modes and the ancilla particles takes place, such that the latter become excited upon the presence of a particle in the respective mode, the full system's quantum state will evolve into
\eq \frac{1}{\sqrt 2}\ket{0,0}_{A,B} \otimes \left( \ket{g,e}   + \ket{e,g} \right) , \en
due to the coherent absorption of the delocalized particle by the ancilla particle coupled to  A or to B. The setup is schematically depicted in Figure \ref{Modeentanglementillu}.
The entanglement which was present as mode-entanglement is transferred to entanglement between the two ancilla particles which can be verified by the usual means, under the assumption that no SSR inhibits the measurement of coherent superpositions of $\ket{e}$ and $\ket{g}$ \cite{Benatti:2010fv}. Hence, by simple interaction with ancillae, the inaccessible mode entanglement can be transferred to an unconstrained degree of freedom. A realistic scenario based on this idea was proposed for cavity-QED experiments in \cite{Gerry:1996qa}, and successfully implemented, {\it e.g.}, in \cite{Maitre:1997bh} (see \cite{Raimond:2001bs} for a review of related cavity-QED experiments). Other methods for the realization of the proposal include the storage and retrieval of quantum information of one single delocalized photon in an atomic ensemble \cite{Choi:2008fk}.

As insinuated above, it is important to note that the absorption of the particle is indeed crucial for the entanglement between the ancilla particles. If the particle is not absorbed, the final state reads
\eq \frac{1}{\sqrt 2}\left( \ket{0,1}_{A,B} \otimes  \ket{g,e}  + \ket{1,0}_{A,B} \otimes  \ket{e,g} \right) , \en
and the density matrix that describes the state of the ancilla particles is a fully mixed state which contains only classical correlations, due to the remaining entanglement with the delocalized particle. One may argue that this problem can be circumvented if that particle is again measured in a coherent superposition of the modes $\frac{1}{\sqrt 2}\left( \ket{A} \pm \ket{B} \right)$, \emph{after} it has interacted with the ancilla particles. In this case, the ancilla state again remains in an entangled state $\frac{1}{\sqrt{2}} \left(\ket{e,g} \pm \ket{g,e} \right)$. The last projective measurement, however, can only be performed after the modes have been superposed at a beam splitter, and constitutes a \emph{global} measurement on the system.  Hence, the requirements for a rigorous proof of non-locality by the violation of a Bell inequality are not given \cite{Ashhab:2007if}.

It is to be retained that a rigorous experimental proof for the non-locality of mode-entanglement of photons has been achieved, as discussed above  \cite{Hessmo:2004bs,Babichev:2004fv}. Moreover, single-photon entanglement has also been purified \cite{PhysRevLett.104.180504}, and theoretical schemes for the detection of entanglement in one-particle many-mode entangled states are available \cite{al:2009oq}. 

For massive particles, the issue is more intricate, since the SSR which inhibits the local rotation of bases is of fundamental nature (see \cite{Cunha:2007fk} for an illustration of the analogies between massless and massive particles). As a way out, a pure reservoir state with broad particle distribution can be shown to be a good reference frame to circumvent the particle number SSR \cite{Ashhab:2009rq,White:2009vn,Heaney:2009jl}, in general. Since the reservoir state in the scheme in \cite{Heaney:2009jl} is shared by the two parties to ensure phase coherence, no strict non-locality conditions are established. Furthermore, the quantum correlations that are measured may have their origin in the entanglement between spatial regions of the reference state, and not necessarily in the coherent delocalization of a single massive particle. Also in this latter case, however, a Bell inequality violation signals the  mode-entanglement of massive particles, either stemming from the single delocalized particle, or from the reservoir. Other proposals circumvent the particle-number SSR by using two identical copies of the state under consideration, such that one copy provides a phase reference for the other one \cite{Heaney:2010cr}, and no coupling to a particle reservoir is necessary. The advancement of experimental techniques in the field of ultracold atoms \cite{Bakr:2009oq,Sherson:2010kl,Karski:2009tg,Treutlein:2006ij}, where mode entanglement is naturally present \cite{Heaney:2007nx}, feed the hope that the issue of mode entanglement of massive particles will be resolved  in the near future.

\subsubsection{Creation of mode entanglement}
Since a simple transformation of modes can create a mode-entangled state, it is not surprising that a panoply of scenarios have been designed to produce highly mode entangled multipartite states \cite{PhysRevA.55.2564,Pryde:2003fu,Fiurasek:2002ye,Lee:2002rt,Fiurasek:2003lh,PhysRevA.55.2564}. All these schemes build on linear optics and on single-photon detectors, hence consist of transformations of the modes and of projective measurements. In principle, the dimensionality of the subsystem states and the number of parties of a multipartite quantum state which can be obtained with a limited number of particles is unbounded \cite{Fiurasek:2003lh}. Indeed, entanglement was verified, {\it e.g.}, for a four-partite state created with only one single photon \cite{Papp:2009kl}.

\subsection{Entangled degrees of freedom} \label{photontypesent}
Given the physical carriers of entanglement, it remains to choose the degrees of freedom which define the subsystem Hilbert spaces. The spectra associated with these degrees of freedom can be finite and discrete, infinite and discrete, continuous, or composed of, both, a discrete and a continuous part. In general, any degree of freedom which can be measured in different bases, {\it i.e.}~which is not subjected to an uncircumventable SSR (see Section \ref{secSSR}), is a candidate to exhibit measurable entanglement. The actual choice is, however, often determined by the actually achievable experimental control over the respective degrees of freedom. 

Photons are particularly versatile carriers of entanglement, since they were shown to exhibit entanglement in various, distinct degrees of freedom. The most prominent example is certainly their polarization degree of freedom. Already before the prevalence of parametric down-conversion \cite{Kwiat:1995mw} which represents today's most versatile source for polarization-entangled photons, the implementation of quantum-informational protocols \cite{Bouwmeester:1997dz} was mainly tested with polarization-entangled photons stemming from the two-photon decay of excited atomic states \cite{Freedman:1972fk,Aspect:1981zr}. Other degrees of freedom  which support higher dimensional Hilbert spaces can equally well exhibit entanglement: The orbital angular momentum \cite{Mair:2001fv,Calvo:2007oq}, the position and momentum \cite{Howell:2004fu}, the time \cite{Marcikic:2004qr}, or the frequency axis \cite{Olislager:2010fc}. 

Entanglement needs not involve the same degree of freedom in each subsystem, even though these may be represented by identical physical entities (such as photons, ions, molecules). If the degree of freedom that is entangled differs between the subsystems, one talks of \emph{hybrid entanglement}. For example, the polarization of a photon can be entangled to the momentum of another one  \cite{PhysRevA.80.042322,PhysRevA.79.042101}, the arrival time to the polarization \cite{fujiwara}, or the orbital angular momentum to the polarization \cite{Nagali:2010ss}.   

\emph{Hyperentanglement} (or \emph{multiparameter entanglement}) denotes quantum correlations that involve several, or even all different degrees of freedom of a \emph{single} physical particle: Any particle in a hyperentangled state is correlated not only in one degree of freedom such as polarization, but also in several ones. Hyperentanglement was successfully demonstrated for photons, including two photons that are hyperentangled in three degrees of freedom, namely in their polarization, in their longitudinal momentum, and in their mode \cite{Vallone:2009qc}. Analogous experiments with photons entangled in frequency, wave vector and polarization were reported in  \cite{PhysRevA.66.023822,Barreiro:2005ve}. The use of several degrees of freedom of each particle naturally allows to realize high-dimensional states with few particles, hence more information can be encoded in few carriers. One impressive example is given by the 1024-dimensional, ten-qubit hyperentangled state of photons in \cite{Gao:2010tg}, where both the polarization and the mode degree of freedom define the subsystem structure.

\section{Atoms and molecules} \label{atommol}
In terms of the different facets entanglement can exhibit, as well as of the related conceptual issues which need to be dealt with, atoms and molecules represent fascinating physical objects which bear all of the difficulties that were discussed in Sections 2 and 3. In  most hitherto existing experiments, however, these systems were isolated and shielded from their environment, and most degrees of freedom were neglected and decoupled from the ones of interest, such as to realize the quantum-information abstraction of two qubits, with great success. However, atoms and molecules naturally offer several other ways to encode entanglement, ranging from atoms that are entangled with each other in their center of mass degree of freedom (see Section \ref{twoatomsent}), over the photons emitted in the course of a de-excitation or of a recombination process (see Sections \ref{atomphotonent}, \ref{photonphoton}), to the very constituents of the atoms, {\it i.e.}~to electrons entangled with nuclei (Section \ref{electronsnuclei}), or electrons entangled with each other (Section \ref{elecelec}). Thereby, they provide unique testing grounds for the direct application of the conceptual tools we discussed above. 

The studies that we will review hereafter employ the same physical object, but they illuminate very different aspects of entanglement, and they are motivated by different incitements. For unbound systems such as the products of a decay process, means were established to quantify correlations and experimentally accessible observables were proposed. For bound systems, the direct verification of entanglement is out of reach: Prior to any measurement, the fragmentation of the system has to be induced by some mechanism which necessarily impacts heavily on the quantum correlations of the constituents, and thereby strongly changes their entanglement properties. Alternatively, consequences of entanglement between the constituents of bound systems for other physical phenomena such as the bosonic or fermionic compound behavior were identified, and thereby indirect means for the verification of entanglement were established. Finally, entanglement provides a benchmark for approximation techniques in bound many-body systems, {\it e.g.} electronic entanglement assesses the strength of the exchange-correlation energy in density functional theory \cite{Grobe:1994fk}.

A particular case is the motional entanglement between atomic or molecular fragments. The conservation of momentum and energy, and the isotropy of the system enforces highly correlated fragmentation dynamics in three-dimensional configuration space. The kinematic nature of these correlations in atomic and molecular fragmentation processes provides a rather intuitive understanding of their very existence and their strength. On the other hand, the advances of experimental technologies have brought coherent phenomena in atomic processes  on the experimentalists' agenda \cite{Becker:2009oq,Corkum:2007fk,Krausz:2009ve}. Today, a complete mapping of the momenta of all fragmentation products is possible, and very strong and particular correlations  have indeed been recorded (see Section \ref{elecexp}). The question whether the interaction between fragments or between fragments and the environment cause the deterioration of entanglement to merely classical correlations in the final (detected) state, or whether \emph{rigorously quantum} correlations persist is, however, open. It constitutes both a challenge for the theory which needs to account for decoherence, and for the experiment, in  which one currently cannot perform a rotation of basis in momentum/position space. While several means are available for quantifying correlations, the \emph{nature} of those, classical or quantum, has to be assessed with Bell inequalities or by novel proposals such as the persistence of Cohen-Fano interferences for fragmenting molecules, discussed in Section \ref{elecelec} \cite{Chelkowski:2010hc}. 

\subsection{Entanglement between atoms} \label{twoatomsent}
Dichotomic, or qubit-like, entanglement between the \emph{internal} (electronic) states of trapped ions is well established and documented (see {\it e.g.} \cite{Hagley:1997bd}, or \cite{Haffner:2008wo} for a review on the entanglement of trapped ions), while entanglement in the motional degrees of freedom has only started to receive substantial attention \cite{Opatrny:2001ye,Opatrny:2003fk,Savage:2007qo,Guo:2008hc,Sancho:2009zt}. 

\begin{figure}[h] \center \includegraphics[width=10cm,angle=0]{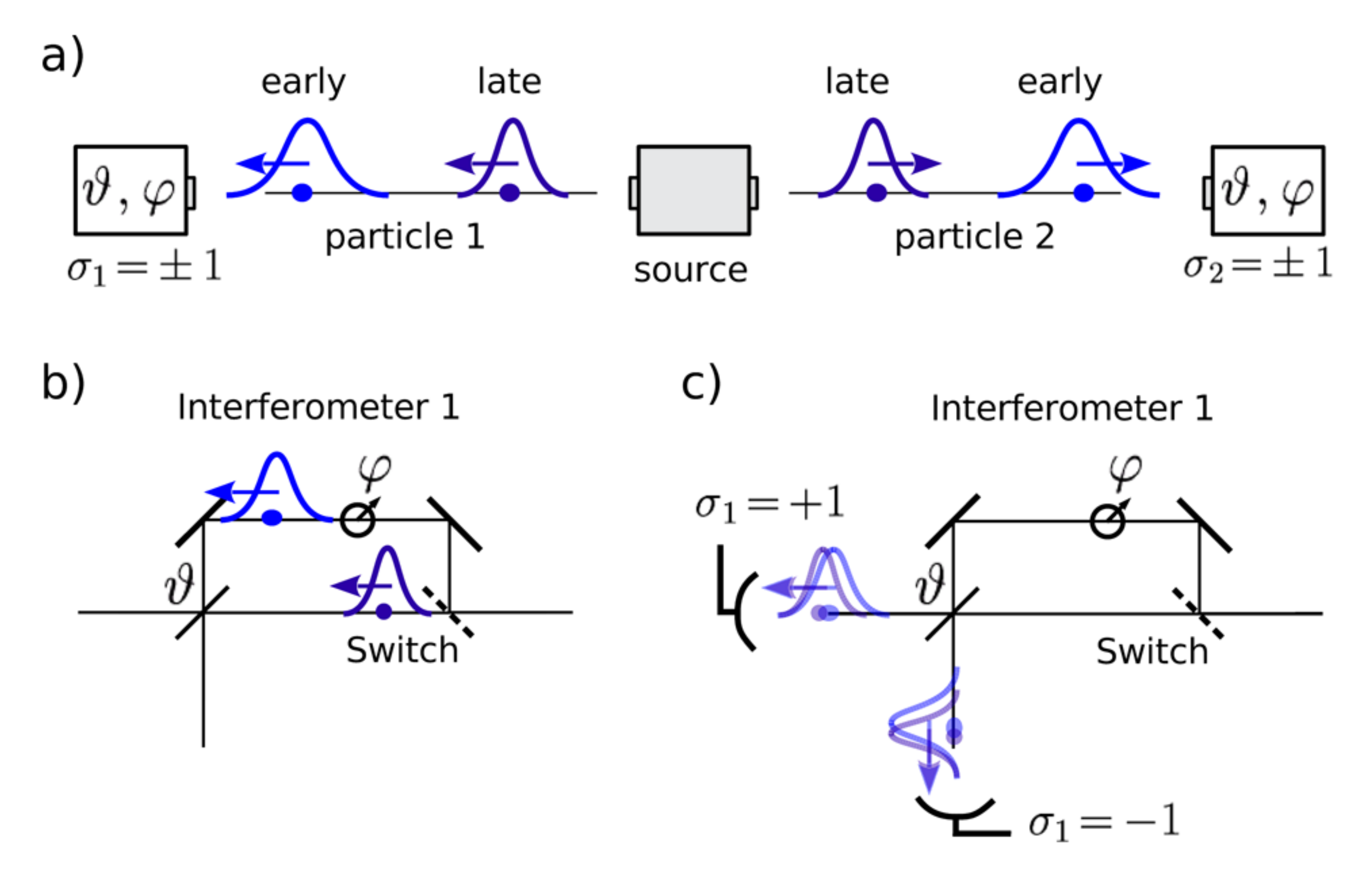} \caption{Courtesy of C. Gneiting and K. Hornberger \cite{Gneiting:2008bh}. Dissociation-time entanglement scheme \cite{Gneiting:2010qf,Gneiting:2008bh}. (a) A series of pulses creates a coherent superposition of particle pairs with early and late dissociation times. (b,c) With the help of a switch which deflects the early wave component into the long arm of an unbalanced Mach-Zehnder interferometer and leaves the late wave component in the short arm, the wave packets are brought to interference. By changing the phase $\varphi$ and the splitting ratio $\vartheta$, projective measurements onto arbitrary basis states can be performed. The detection of a particle in either one of the detectors is recorded as $\sigma_1=\pm 1$ and corresponds to the projection onto one of the states of the basis specified by the choice of $
\vartheta$ and $\varphi$. Copyright 2008 by The American Physical Society. }  \label{Gneitingfigure}  \end{figure}

One proposal to create and verify entanglement in atomic momenta involves the Feshbach dissociation of a diatomic, ultracold molecule \cite{Gneiting:2008bh,Gneiting:2010qf}. Two subsequent magnetic field pulses coherently prepare a molecule in a dissociating state which is given as the superposition of an ``early'' and a ``late'' particle pair (one single pulse does not suffice to dissociate the molecule deterministically), as illustrated in Figure  \ref{Gneitingfigure}. The resulting quantum state of the atomic fragments after the two pulses can thus be written as 
\eq \ket{\Psi}=\frac{1}{\sqrt 2} \left(\ket{e,e}+\ket{l,l} \right), \en
where $\ket{e}$ and $\ket{l}$ denote the quantum states of an atom released by the early, or by the late pulse, respectively. This state has the form of the Bell state given by (\ref{Phimp}), and can be considered as \emph{time-bin} entangled, a type of entanglement that was already verified for photons \cite{Marcikic:2004qr}. This \emph{dissociation time entangled} state can be probed by correlation measurements in different bases, in a Mach-Zehnder interferometer \cite{springerlink:10.1134/S0030400X10020062,springerlink:10.1007/s00340-009-3457-4}, such as to test a Bell inequality. 
 The interferometer can be substituted by proper adjustment of the magnetic field pulse shapes, such that the late pair has a larger momentum than the early one, and interference between the early and late wave packets occurs naturally. By increasing the delay between the dissociating pulses and thereby between the wave-packets, the technique bears the potential to investigate the length scales over which massive particles can be coherently delocalized. With more intricate pulse protocols, {\it e.g.} the dissociation of the molecule with three or more subsequent pulses into three or more pairs, also entanglement of massive particles involving larger dimensions should in principle be accessible. A bottleneck for the experimental realization of such protocols is the reproducibility of the magnetic pulse sequence for the dissociation of the molecules: This is a crucial ingredient, to ensure a stable relative phase between the early and the late component. On the other hand, the two interferometers do not need to be have a fixed relative phase, which allows macroscopic separations. Additionally, no spatial nor temporal resolution is required for the particle detection. For the proposed molecular BEC produced from a balanced spin mixture of fermionic $^6$Li, a temporal separation of $\tau=1$s between the early and the late component and a dissociation velocity of $v_{\mathrm{rel}}=1$cm/s are feasible \cite{Gneiting:2010qf}.

\subsection{Atom-photon entanglement} \label{atomphotonent}
While the entanglement between particles of different kinds represents the conceptually simplest case (see Section \ref{properties}) and naturally occurs in all interacting systems, it was experimentally verified only recently. A paradigmatic example is provided by two electronic energy levels of an atom (a ``two-level atom'') and a photon which populates a single mode of the quantized electromagnetic field. Such bipartite qubit-entanglement has been realized in cavity-QED \cite{Raimond:2001bs} and in cold atom experiments \cite{PhysRevLett.93.090410,Nature428,PhysRevLett.96.030404,PhysRevLett.101.260403,Matsukevich10222004,PhysRevLett.95.040405}. Since experiment and theory of this conceptually (not experimentally, though!) rather elementary scenario is well established and documented \cite{Raimond:2001bs}, we focus here on advances in high-dimensional entanglement in the external degrees of freedom, {\it e.g.} in the momenta of atoms and photons.

In contrast to molecules that first need to be dissociated by some external field, in order to provide two separate subunits, already the simple spontaneous decay of an electronically excited atom under emission of a single photon constitutes an elementary scattering process in which entanglement between the light field and the atom can be studied. 

\subsubsection{Decoherence of atoms due to photon emission}
Such correlations in the momentum of a spontaneously emitted photon with the momentum of the remaining atom, which originate from momentum conservation in the spontaneous-emission process, were measured in \cite{Kurtsiefer:1997kh}. In general, when photons are spontaneously emitted by atoms, the latter effectively loose their coherence, which leads to an observable loss of fringe visibility \cite{Pfau:1994dq,Chapman:1995cr} in interference experiments. This decoherence of the atomic states can be explained by the entanglement of the atoms with the emitted photons and the subsequent loss of these photons and of the information they carry. Formally, this loss  corresponds to a partial trace over the photonic degree of freedom, leaving the atom in a mixed state without the potential to exhibit interference. This situation should be contrasted to the scenario discussed in Section \ref{measurementinduced}, in which the  emitted  photons are detected in a way that their which-way information is lost, and the atoms remain mutually entangled. 

\subsubsection{Occurrence and strength of atom-photon entanglement} Under the assumption that the atom has infinite mass and that it is described by a $\delta$-localized wave-function \cite{springerlink:10.1007/BF01336768}, no motional entanglement between the photon and the atom can arise, simply because the atom is fixed in space.  A theoretical treatment of the spontaneous decay of an excited atom with emission of one photon shows that  entanglement is naturally present in the atom-photon system \cite{Chan:2002gb,Chan:2003ly} when the aforementioned idealization is given up. Like in the case of a bound hydrogen atom discussed in Section \ref{substruct}, the two-particle wave-function can be written in a factorized, {\it i.e.}~unentangled, form when a suitable transformation to collective coordinates is performed \cite{Fedorov:2005bh}:
\eq 
\ket{\Psi_{\mathrm{sys}}} &= & \ket{\Phi_{\mathrm{CM}}} \otimes \ket{\phi_{\mathrm{rel}}} \in \mathcal{H}_{\mathrm{CM}} \otimes  \mathcal{H}_{\mathrm{rel}}  \nonumber \\
&=& \sum_{j} \sqrt{\lambda_j } \ket{\chi_{\mathrm{at}}^{(j)}}\otimes \ket{\chi_{\mathrm{ph}}^{(j)}} \in \mathcal{H}_{\mathrm{at}} \otimes  \mathcal{H}_{\mathrm{ph}} \label{choiceofsub} ,
\en
where $\ket{\Phi_{\mathrm{CM}}} $ and $\ket{\phi_{\mathrm{rel}}}$ denote wave-functions that describe the center-of-mass-like and relative coordinates of the atom-photon system, respectively (the collective coordinates). On the other hand, $\ket{\chi_{\mathrm{at}}^{(j)}}$ and $\ket{\chi_{\mathrm{ph}}^{(j)}}$ refer to the atom and the photon wave-function, respectively. For a physical choice of subsystems, namely of the atom and the photon, however, entanglement is exhibited, as also immediate from the second line of (\ref{choiceofsub}). Once again, the choice of subsystems is the key issue, as already discussed in Section \ref{substruct}.  Such motional entanglement, as quantified by the Schmidt number $K$, (see Eq.~(\ref{Schmidtnumber})), is, in principle, unbounded since position and momentum are continuous degrees of freedom \cite{Keyl:2003uq}. In practice, however, only a finite number of discrete Schmidt modes are occupied, {\it i.e.}~the entanglement is typically rather low (in terms of Section \ref{quantif}, only few Schmidt coefficients $\lambda_j$ do not vanish). This can be understood from a kinematic argument which takes into account the mass discrepancy between the atom and the emitted photon, and the restriction of available phase-space due to momentum conservation. The Schmidt number of the resulting atom-photon entangled state turns out to be inversely proportional to the line width of the transition \cite{Chan:2002gb}. This can be understood intuitively: The narrower the transition, the better the energy of the compound photon-atom system is defined, and the stronger the photon is correlated to the recoiling atom. Specific three-level scattering schemes  \cite{Zhu:1995nx} could, in principle, enhance the resulting entanglement between atom and photon  \cite{Guo:2006qf,Chan:2003ly,Guo:2007fv}.

As an elementary scattering process, the atom-photon system constitutes hence a rather well understood scenario in which correlations are naturally present, and their strength can be understood from kinematic arguments and conservation laws. Due to this kinematic nature of their genesis, it is, again, the probing of the coherences of the many-particle state that constitutes the biggest experimental challenge. 

\subsubsection{Coincidence measurements and wave-packet narrowing}
Under the assumption that all observed correlations are of quantum nature, {\it i.e.}~that the quantum state under consideration is a pure state, correlations in position can be used to quantify the entanglement between the atom and the photon \cite{Fedorov:2005bh,al:2006fu}. This very assumption is, however, debatable, since an analysis based on correlations recorded in one specific basis does not allow to exclude a merely classically correlated state (see Section \ref{Bellin}).

Anomalous narrowing and broadening of the atomic and photon wave packets can be quantified by coincidence and single-particle measurement schemes \cite{Fedorov:2005bh}. In other words, when the position of the nucleus is known, the state of the photon is largely determined, whereas without the information on the nucleus' state, the width of the observed photonic wavepacket is considerably larger. 

This argument can be made quantitative by studying the coincident probability distribution $P(r_{\mathrm{ph}}, r_{\mathrm{at}},t)$ of the photon (ph) and the atom (at), which are obtained by fixing the position of one detector while varying the position of the other, and by counting only coincident signals from both detectors. From the joined atom-photon probability distribution $P(r_{\mathrm{ph}}, r_{\mathrm{at}},t)$, the marginal distributions can be obtained, 
\eq P(r_{\mathrm{ph}},t)=\int \mathrm{d} r_{\mathrm{at}} P(r_{\mathrm{ph}}, r_{\mathrm{at}},t) , \\
P(r_{\mathrm{at}},t) =\int \mathrm{d} r_{\mathrm{ph}} P(r_{\mathrm{ph}}, r_{\mathrm{at}},t) , \en
which describe the wavepackets of the photon and the atom, respectively, when no assumptions on the remaining particle is made, such that an integration over the coordinate not under consideration is effectuated. 

We denote the width, {\it i.e.}~the variance of the probability distributions, by
\eq 
\Delta r_{\mathrm{ph}}^{(c)/(s)}(t) , \ \ \  \Delta r_{\mathrm{at}}^{(c)/(s)}(t) ,
\en
for the photon and the atom, respectively, in the coincidence scheme (c) or for the marginal distribution (s). The ratio 
 \eq R(t)=\frac{\Delta r_{\mathrm{ph}}^{(s)}(t) }{\Delta r_{\mathrm{ph}}^{(c)}(t)} =\frac{\Delta r_{\mathrm{at}}^{(s)}(t) }{\Delta r_{\mathrm{at}}^{(c)}(t)}, \en 
then  quantifies the correlations of the wave-functions and, thereby, the \emph{spatial wave-packet narrowing}: A smaller width in coincident detection as compared to the  marginal distribution is a consequence of correlations; by construction, $R(t) \ge 1$. Wave-packet narrowing can also be understood as ``measurement-induced localization'' \cite{Freyberger:1999fk}.  For $t=0$, $R$ corresponds to the Schmidt number $K$ (see Eq.~(\ref{Schmidtnumber})): $K=R(t=0)$. The entanglement inscribed into the system, and thereby the value of $K$, necessarily remain constant, since the particles are non-interacting. However, the spreading of wave packets leads to temporally  increasing $R(t)$ \cite{al:2006fu}.   The presented scheme probes, however, no coherences, but only correlations, as immediate from the probability distributions given only in terms of position coordinates. Only if both the momentum and the position can be measured, a full realization of the EPR paradox becomes feasible \cite{Reid:2009bs}.

\subsubsection{Decoherence of the photon-atom system}
The interaction of the atom with other background photons subsequent to photon emission  leads to decoherence of the entangled atom-photon system \cite{Guo:2007fv}. The timescale $\Delta t_{\mathrm{dis}}$ of disentanglement can be estimated by modeling the environment as background photon bath. It turns out to be inversely proportional to the average number of resonant photons in the heat bath and can thus be increased by decreasing the temperature of the environment. On the other hand, the disentanglement time $\Delta t_{\mathrm{dis}}$ is also inversely proportional to the Schmidt number $K$. Hence, the timescale depends itself on the initial entanglement of the state, and the more entangled the state, the faster its entanglement is lost. The inclusion of other single body coherent effects like the dispersion of the atomic wave packet are shown to be of no relevance for the entanglement in the system \cite{Chan:2003ly}.

\subsection{Photon-photon entanglement} \label{photonphoton}
Two-photon emission in de-excitation transitions in atoms and ions has been the first source of polarization-entangled photons \cite{Freedman:1972fk,Aspect:1981zr,Aspect:1982ly,Haji-Hassan:1989cr,Perrie:1985fk}. Due to conservation of energy, momentum, and angular momentum, photon pairs that are emitted in such processes are, however, not only entangled in polarization, but also highly correlated in energy and angular direction ({\it i.e.}~in frequency and linear momentum). 
Correlations are not only present within identical degrees of freedom, but the system also exhibits hybrid entanglement (see Section \ref{photontypesent}), {\it i.e.}~entanglement between different degrees of freedom such as the polarization of one photon and the angular direction of the other. Furthermore, the strength of polarization-entanglement may depend on the emission angle, on the fraction of the available energy carried by one photon, and on the nuclear charge, as shown in the theoretical studies reviewed hereafter  \cite{Radtke:2008eu,2010arXiv1006.4799F,Surzhykov:2001ly,Surzhykov:2005ly,Borowska:2006nx}. 

\subsubsection{Experimental progress}
Experimental studies of the correlations of photons emitted in a two-photon decay have recently concentrated on highly charged ions, for which relativistic effects become important. For such systems, only the total decay rates and excited state life times were available until two decades ago \cite{Dunford:1989ve,Drake:1986qf}. The spectral distribution of the emitted photons was resolved later \cite{Schaffer:1999bh,Derevianko:1997dq,Jentschura:2008dq}, and today the spectrum of the two photons emitted, {\it e.g.}, in the recombination of $K$-shell vacancies of silver \cite{Mokler:2004bh} and gold \cite{Dunford:2003tg} can be measured, and are found in agreement with theoretical predictions \cite{Tong:1990ve}. Indeed, the experimental accuracy is sufficient to verify relativistic effects, {\it e.g.} in the two-photon spectrum of He-like tin \cite{Trotsenko:2010nx}. 

This important progress in the experimental capabilities \cite{al:2005ys,Mokler:2004cr} permits today to detect photons in a wide range of energies, and to extract differential cross-sections as a function of the energy redistribution among the photons, and of the opening angle, even for photons in the X-ray regime. Also the polarization of hard X-rays, for example in radiative electron capture into the K-shell of highly-charged uranium ions \cite{al:2009vn,Tashenov:2006fk}, becomes accessible. In the next years, the observation of polarization-correlations of photons in the X-ray domain is therefore likely to become possible. Thus, experiments analogous to those which measure correlations of low-energetic photons in transition processes of light atomic species, as already performed three decades ago \cite{Freedman:1972fk,Aspect:1981zr,Aspect:1982ly}, will soon be feasible for heavy nuclei, and for photons in the X-ray regime. These developments have triggered an extensive research agenda with the goal to fully characterize the emitted photon pair \cite{al:2007kx} in general atomic two-photon decay processes. 

\subsubsection{Theoretical studies}
\begin{figure}[h] \center \includegraphics[width=14.5cm,angle=0]{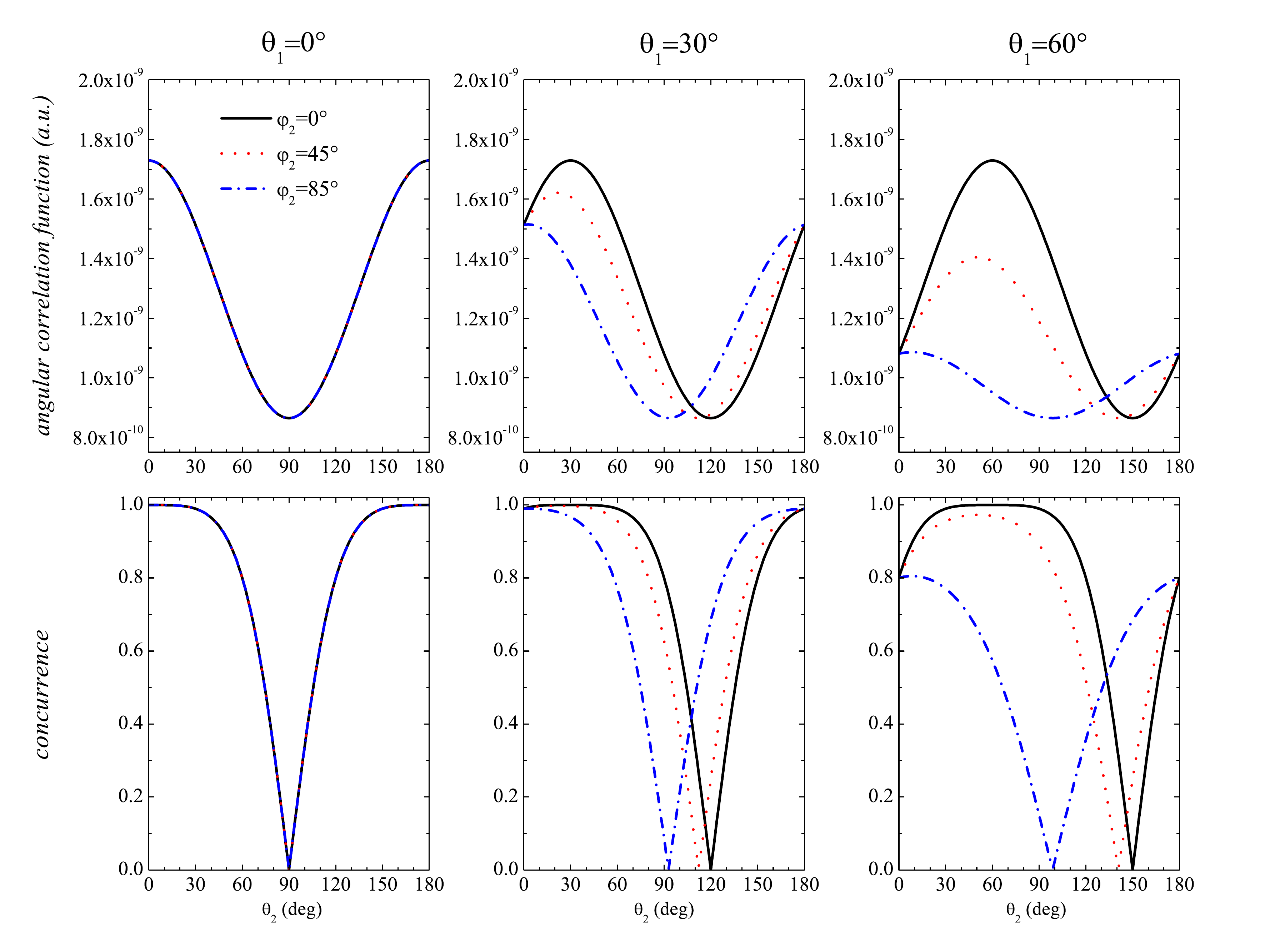} \caption{Courtesy of T. Radtke \cite{Radtke:2008eu}. Two-photon angular correlation (upper panel) and polarization entanglement quantified by the concurrence (lower panel), in the $2s_{1/2}\rightarrow 1s_{1/2}$-decay of initially polarized atomic hydrogen, {\it i.e.}~for which only one magnetic sublevel is populated. All values are given as a function of the angle $\Theta_2$ between the polarization axis of the atom (the quantization axis) and the first photon. Data are shown for different values of $\Theta_1$ -- the angle between the second photon and the polarization axis of the atom, and $\varphi_2$ -- the emission angle of  the second photon with respect to the axis perpendicular to the plane spanned by the atom polarization and the emission direction of the first photon. Angular correlations and the entanglement depend strongly on the particular decay geometry defined by the three angles $\Theta_1, \Theta_2, \varphi_2$. The probability to find photons at a given emission angle is proportional to the angular correlation function, which is shown here in arbitrary units. Copyright 2008 by The American Physical Society.}  \label{surzy}  \end{figure}

The first issue that was addressed by theoretical studies is the dependence of the photon polarization properties on the emission properties, {\it i.e.}, instead of back-to-back emission, an arbitrary geometry was considered. Already the polarization of a single emitted photon depends on its emission angle with respect to the polarization axis of the atom \cite{Surzhykov:2001ly}. If two photons are emitted, the entanglement between their polarization degrees of freedom strongly depends on the relative angle of emission  \cite{Radtke:2008eu,2010arXiv1006.4799F}, as also shown in Figure \ref{surzy}. It is also apparent that the photons also exhibit angular correlations, {\it i.e.}~they are not emitted isotropically. 

In the non-relativistic dipole approximation, the Schr\"odinger-equation is solved under the assumption of the interaction of the electrons with a spatially homogeneous electric field (the  wavelength of the photon is much larger than the dimension of the atom) \cite{Goppert-Mayer:1931fk}. In this model, the 
polarization-state of the two photons can simply be written as \cite{2010arXiv1006.4799F,1011.5816}
\eq \ket{\Psi_{2\gamma}} = \frac{1}{1+\cos^2 \theta} \left( \ket{H,H}+\cos \theta \ket{V,V} \right) \label{nonrel} \en
where $\ket{H}$ and $\ket{V}$ denote horizontal and vertical polarization with respect to the emission plane of the photons, $\theta$ is their opening angle, and the factor $1/(1+\cos^2 \theta)$ ensures normalization. For back-to-back emission, {\it i.e.}~$\theta=\pi$, the state is maximally entangled, while perpendicular emission, {\it i.e.}~$\theta=\pi/2$, leads to a non-entangled two-photon state. The entanglement properties of a photon pair emitted in such extreme configuration ($\theta=\pi$ or $\theta=\pi/2$) can also be derived by taking into account uniquely the conservation of angular momentum, they hence do not provide information about microscopic details of the decay process. 

Whereas, in practice, the total two-photon decay rates are dominated by the dipole-transition and higher multipoles do not contribute significantly, the latter can have a considerable influence on the \emph{correlations} that are carried by the emitted photons. The inclusion of relativistic and non-dipole effects -- in an approach based on the relativistic Dirac-equation in which the electric field is not assumed to be spatially constant \cite{Au:1976zr} -- leads to a prediction that significantly differs from (\ref{nonrel}) \cite{Surzhykov:2005ly,Borowska:2006nx,Radtke:2005oq}. In particular, the symmetry of the emission with respect to $\theta=\pi/2$, evident from (\ref{nonrel}), is broken by non-dipole contributions. This effect is further enhanced for large nuclear charges $Z$. Hence,  multipole and relativistic effects do not only manifest in strictly dipole-forbidden decay channels \cite{1011.5816}. 

In an experiment, the initial state preparation of the atoms or ions is typically not under perfect control, and one needs to account for an incoherent mixture of the magnetic sublevels. This mixedness can jeopardize the entanglement in the final photonic state. The 2$s_{1/2}\rightarrow 1s_{1/2}$ transition, however, is unaffected by the incoherent population of magnetic sublevels: Due to angular momentum conversation, the final two-photon state remains pure. In contrast, the incoherent preparation of the initial state leads to a mixed final state for the $3d_{5/2}\rightarrow 1s_{1/2}$ transition. The resulting, mixed two-photon sate is of Werner type (see Section \ref{Bellin}), under certain geometries \cite{Radtke:2008gf}. 

Beyond correlations in the polarization degree of freedom, entanglement also prevails in  other degrees of freedom. For example, the photon-photon emission angle is correlated with the polarization of one photon - independently of the polarization of the other one \cite{Surzhykov:2009ve}, {\it i.e.}~the system exhibits hybrid entanglement, similarly as in the studies discussed in Section \ref{photontypesent}. As concerns the verification of the quantum nature of such correlations, a change of basis is achievable in the polarization degree of freedom, though in the angle a change of basis was realized so far only for photons in the optical range \cite{DAngelo:2004fk,Howell:2004uq}. 

The above studies have shown that photon-photon entanglement may display a much more intricate behavior than one would expect by considering the conservation of angular momentum alone. Relativistic effects significantly alter the symmetry properties of the system, and thereby lead to observable consequences in the angular correlations and the polarization entanglement. An inclusion of many-body effects \cite{Surzhykov:2010uq} into the theory promises the future design of probes of effects like parity violation \cite{Dunford:1996dq}. 

\subsection{Entanglement between electrons and nuclei} \label{electronsnuclei}
The choice of electrons and nuclei as carriers of entanglement allows to consider bound systems as well as fragmented ones. The fragmented systems exhibit close analogies to the arguments presented in Section \ref{atomphotonent}, where photon-atom entanglement is discussed, adding final-state interaction and of the finite mass of the electron to the problem. Again, kinematic intuition can explain the strength of momentum-correlations: The discrepancy of the masses of the electron and the nucleus leads to a rather low value of entanglement, whereas an analogous scenario of dissociating molecules with balanced fragment masses results in stronger entanglement \cite{Fedorov:2004kx}. In this investigation, besides the momentum of the incident photon, also the Coulomb interaction in the final state is neglected, hence the effect of the possible deterioration of entanglement due to interaction is not taken into account. The evolution of the state and of the entanglement between the two particles is thereby constrained by the free-particle two-body momentum and energy conservation.

\subsubsection{Bound systems: Composite behavior}
Electrons bound by a nucleus are naturally entangled with it \cite{Tommasini:1998qf} in their external degrees of freedom. As we will discuss hereafter, such entanglement between the constituents of any bound system can be related to the compound's bosonic or fermionic behavior and thereby provides an important indicator for the strength of effects due to the composite nature of particles. With the experimental realization of Bose Einstein Condensation (BEC) with bosonic atoms, the question naturally arises to which extent composite particles can be treated as elementary bosons \cite{Avancini}, and under which conditions the underlying constituents, {\it i.e.}~the electrons, will manifest themselves and jeopardize the bosonic behavior of the compound. 
Intuitively, one would relate the \emph{density} of such composite particles to their bosonic or fermionic behavior: Roughly speaking, if the wave-functions of the electrons of different atoms start to overlap considerably, one would expect their fermionic character to inhibit the composite system to behave as a boson. An upper limit for the occupation number of composite boson states was derived formalizing this idea \cite{Rombouts:2002uq}, and applied to the problem of trapped hydrogen atoms. Typical maximal occupation numbers of the ground state are shown to lie in the range of $10^{13}$, while current experiments reach $10^6-10^9$, well below this bound. Composite bosonic behavior is, on the other hand, not restricted to systems in which the constituents are close in space and bound by interaction: Also non-interacting and spatially separated biphotons can exhibit composite-particle properties, as demonstrated by experiments measuring the de Broglie wave-length of an ensemble of entangled photons, defined as $\lambda/n$ where $\lambda$ is the average wave-length of the photons, and $n$ the average number of photons in the ensemble \cite{Fonseca:1999uq,Jacobson:1995kx}. Hence, compositeness is not limited to spatially bounded particles; the density of particles and the overlap of the wave-functions of the constituents cannot constitute a universal criterion for composite behavior. 

The entanglement between the constituent particles provides such a key quantity that defines to what extent the bosonic/fermionic commutation relations for creation operators of composite particles are valid \cite{Law:2005dq,Chudzicki:2010bh}. Given a composite system $C$ of two particles of type $A$ and $B$, which are either both fermions or bosons, the compound quantum state can be written in Schmidt decomposition as
\eq \ket{\Psi_C}= \sum_{n=0}^\infty \sqrt{\lambda_n} \ket{\phi_{A,n}} \ket{\phi_{B,n}} .\en
The creation operator for a composite particle then reads 
\eq \hat c^\dagger = \sum_{n=0}^\infty \sqrt{\lambda_n} \hat a^\dagger_n \hat b^\dagger_n  ,\en
where $\hat a^\dagger_n$ creates a particle in the state $\ket{\phi_{A,n}}$, and analogously for $\hat b^\dagger_n$. 
The commutation relation of creation and annihilation operators $\hat c$ and $\hat c^\dagger$ becomes \cite{Law:2005dq}
\eq [\hat c, \hat c^\dagger ] = 1 + s \Delta ,\en
where $s=1 (-1)$ when $A$ and $B$ are bosons (fermions). Due to the term \eq \Delta= \sum_{n=0}^\infty \lambda_n(\hat a_n^\dagger \hat a_n + \hat b_n^\dagger \hat b_n)\ , \label{DeltaD}\en the creation/annihilation operators $\hat c^\dagger$ and $\hat c$ are not strictly speaking bosonic operators. By analyzing quantum states of many composite $C$-particles, it turns out that the Schmidt number $K$, (\ref{Schmidtnumber}), is a direct indicator for the composite behavior: The quotient $N/K$, where $N$ is the number of composite C-particles in one quantum state, needs to be very small in comparison to unity in order for the composite particles to behave as bosons, {\it i.e.}~the larger the entanglement, the better the composite particle behaves as a boson. This can also be retraced intuitively from the structure of $\Delta$ in (\ref{DeltaD}): This expression is indeed a combination of the Schmidt coefficients $\lambda_n$ with the number operators $\hat a^\dagger_n \hat a_n$ and  $\hat b^\dagger_n \hat b_n$. The larger the Schmidt number $K$, the smaller are the individual Schmidt coefficients and, consequently, also the expectation value of $\Delta$.

For the hydrogen atom, it can be shown that the particle density above which bosonic behavior breaks down, as obtained by such entanglement-based analysis \cite{Chudzicki:2010bh} roughly corresponds to the value previously derived in \cite{Rombouts:2002uq}. The formalism can also be applied to other composite bosons such as Cooper pairs \cite{PhysRevA.75.043613}, and to the behavior of non-elementary fermions \cite{Sancho:2006nx}. 

The above treatment of the composite behavior has shown that certain problems can have a universal solution under a quantum-information perspective that is not restricted to a particular situation. Furthermore, the composite behavior can, in principle, be probed experimentally, and thereby the entanglement between constituents becomes accessible even without breakup of the whole system.

\subsection{Electron-electron entanglement} \label{elecelec}
Electrons as potential carriers of entanglement provide a rich Hilbert space structure with bound and unbound spectra of compounds. They can carry  entanglement in their spin and spatial degrees of freedom, and entanglement can result from interactions, or it can be induced by measurements  \cite{Tichy:2009kl,Vedral:2003uq}. Due to the prevalence of the long-range, inter-electronic Coulomb-interaction over spin-spin effects, however, most theoretical studies concentrate on the entanglement in spatial (external) degrees of freedom. Since the electron-electron-interaction varies in relative strength with respect to the electron-nucleus-interaction, depending on the charge of the nucleus, both theory and experiment can, in principle, access very different regimes. Indistinguishability plays an important role, since the electronic wave-functions typically overlap in space and the (anti)symmetrization of the spatial part of the wave-function thereby becomes relevant. 

Both the rigorous theoretical description and the experimental verification of entanglement remains, however, a difficult challenge in this setting: In the experiment, the spins of ejected  electrons cannot be measured since the methods known from solid-state physics \cite{Burkard:2007ly} cannot be directly implemented for free electrons. For entanglement verification in the external degrees of freedom, a measurement in a second basis in addition to  momentum cannot be implemented with state-of-the-art technology, since the realization of a basis rotation analogous to the one realized for photons in \cite{DAngelo:2004fk,Howell:2004uq} needs the implementation of electron lenses with strengths much superior than currently available. One is therefore restricted to measurements of mere momentum-correlations.

In theory, the many-body wave-function for many-electron atoms is necessary for entanglement studies. While first steps were performed which allow to compute the degree of  entanglement via the electron density in special models such as the fermionic Hubbard model \cite{Franc-ca:2008zr,franca}, such treatment is, up to now, not available for atoms. The computation of the realistic many-particle wave-function requires, however, large if not prohibitive numerical effort. Due to the non-integrability of any non-hydrogenlike atom, theoretical studies of multi-electron systems have, so far, mainly focused on exactly solvable model atoms. While such models differ strongly from real multielectron atoms - as concerns the interelectronic interaction and the definition of the confining potential, they allow insight in some qualitative features. They also constitute a testing ground for approximate techniques \cite{Amovilli:2003zr,Coe:2008kx,Maiolo:2006kx,Dehesa:2010dq} which may help to study entanglement in more realistic atomic models in the future.

\subsubsection{Harmonic confinement and interaction: The Moshinsky atom}
One exactly solvable model-system for two bound electrons is the Moshinsky atom \cite{Moshinsky:1968nx,Moshinsky:1968ly} which was initially introduced to characterize the reliability of the Hartree-Fock approximation in interacting systems. In the model, both the confinement  and the interaction are harmonic. The Hamiltonian of the system reads (with masses and actions measured in units of $m$ and $\hbar$, respectively)
\eq
\hat H=\frac{1}{2} \left(\Omega^2 \hat x_1^2 + \hat p_1^2 \right)  + \frac{1}{2} \left(\Omega^2 \hat x_2^2 + \hat p_2^2 \right) + \frac{\omega^2}{2} (\hat x_1-\hat x_2)^2 ,
 \label{MoshinskyHamiltonian}\en
where we denote the spring constant of the confining (inter-electronic) interaction with $\Omega$ ($\omega$).  In center-of-mass and relative coordinates,
\eq 
\hat X=\frac{1}{\sqrt 2} \left( \hat x_1+\hat x_2 \right), \ \ \hat x=\frac{1}{\sqrt 2}\left(\hat x_1- \hat x_2\right) ,
\en
the Hamiltonian factorizes in two independent harmonic oscillators in $\hat X$ and $\hat x$, with frequencies $\omega$ and $\sqrt{2\Omega^2+\omega^2}$, respectively, which permits a closed analytic solution of the problem: The wave-function factorizes in these coordinates, and can be written as 
\eq \ket{n_{\mathrm{rel}}, n_{\mathrm{CM}}} = \ket{n_{\mathrm{rel}}} \otimes \ket{n_{\mathrm{CM}}} ,\en
where the quantum numbers $n_{\mathrm{rel}}$ and $n_{\mathrm{CM}}$ denote the excitation in the relative  and in the center-of-mass coordinate, respectively. The relative strength $\tau=\omega/\Omega$ of the interelectronic interaction determines the reliability of the Hartree-Fock calculation. For $\tau=1$, the ground-state energy in the Hartree-Fock approximation amounts to approximately 95\% of the exact value \cite{Moshinsky:1968nx}. This is consistent with an overlap of 94\% between the exact and the Hartree-Fock wave-function \cite{Moshinsky:1968ly}. 

Some basic intuition for the particle entanglement in this system is gained under the assumption that the electrons are distinguishable and do not carry spin \cite{Amovilli:2004fk}. Then only the relative strength of the interelectronic interaction $\tau$ affects the entanglement of the electrons. The entropy of the one-electron reduced density matrix, (\ref{eoff}), can be extracted for the ground state by purely analytical means and grows monotonically with $\tau$. 

A slightly more complex model is realized by taking into account the spin degree of freedom and the indistinguishability of electrons \cite{Yauez:2010fk}. Due to vanishing spin-spin and spin-orbit interaction, the total wave-function  factorizes  in a spatial, $\phi(x_1,x_2)$, and a spin-component, $\chi(\sigma_1,\sigma_2)$. The symmetry of the spin-part of the wave-function directly governs the symmetry of the spatial part, since the two terms must be of opposite parity in order to obtain an antisymmetric compound  wave-function. Due to this dependence, one would expect that the parity of the spin part thereby governs the entanglement of the system, but it turns out that it is the relative orientation of the spins (parallel or antiparallel) which is a relevant parameter for entanglement, as evaluated with the help of the properties of the single-particle reduced density matrix.  The electron-electron entanglement vanishes in the ground state for $\tau \rightarrow 0$, and it is maximal for $\tau \rightarrow \infty$. For the excited states, however, the entanglement shows a discontinuous behavior at $\tau=0$: In the limit of very small interactions, {\it i.e.}~$\tau \rightarrow 0$, the entanglement remains finite in the case of antiparallel spins, only for $\tau=0$ the system is non-entangled again. In this case, the system is non-entangled due to the degeneracy of energy-eigenstates. In contrast, for parallel spins, the system shows always a continuous behavior in the limit of small relative interaction strengths and vanishes for $\tau\rightarrow 0$. As demonstrated by this study \cite{Yauez:2010fk}, the indistinguishability of particles adds qualitatively new features as compared to the model of distinguishable particles. A physical explanation for the exhibited, rather intricate behavior has, however, not yet been achieved.

\subsubsection{Coulombic interactions}
The harmonic interactions considered in the last Section permitted an analytical treatment of the model-atoms and a  straightforward evaluation of entanglement measures. The next step towards a more realistic scenario consists in including the Coulomb interaction into the model. Already for two electrons bound by one nucleus, however, an analytical treatment is impossible, and the numerical treatment is difficult due to the triple collision singularity, {\it i.e.}~the attractive fixed point of the classical phase-space flow for which both electrons fall into the nucleus, symmetrically, along the collinear axis.  Only strongly simplified models have been in the focus of recent studies. Since the nonregularizable triple collision singularity constitutes one of the dynamical key features of the three-body Coulomb problem \cite{0953-4075-40-8-F01,Tanner:2000vn,Choi:2004ys,Byun:2007zr}, its neglect eliminates one of the sources of classically chaotic dynamics in the three body dynamics. Chaos, however, implies the abundance of avoided crossings on the spectral level, and the latter have been shown to be associated with entanglement in interacting many-particle systems \cite{Venzl:2009kl}.

One-dimensional helium is studied in \cite{Carlier:2007uq}, where the two electrons are restricted to a one-dimensional space, such that they are always on opposite sides of the nucleus (the $eZe$-configuration). Effectively, one deals with a one-dimensional atom with charge $Z=2$ and two spin-free electrons. The Hamiltonian thus reads
\eq 
H= \frac{\hat p_x^2}{2}-\frac 2 {\hat x} + \frac{\hat p_y^2}{2} - \frac 2 {\hat y} + \frac{1}{\hat x+\hat y} ,
\en
where $x,y>0$ denote the distance of the two electrons to the nucleus, and  the Coulomb interaction between the electrons is described by $1/(\hat x+\hat y)$. The two-particle wave-functions, on which the analysis of the entanglement in the system is based, are obtained by using the analytically exact solutions of the single-electron problem, \eq 
\left( \frac{\hat p_x^2}{2} - \frac{2}{x}  \right) \phi_n(x) = E_n \phi_n(x) .
\en
The Hamiltonian $H$ is then evaluated in the truncated product basis, 
\eq 
\Psi_{i,j}(x,y)=\phi_i(x) \phi_j(y) , \en
and it is diagonalized numerically. The electrons always remain spatially separated, no overlap of their wave-functions in space can occur, and thereby the indistinguishability does not play any role. The singularity between the electrons and their vanishing overlap does not prevent entanglement to be generated, and, again, the entanglement grows with the state's energy. 

Other simplified model atoms include the Crandall atom (harmonic confinement and ${1}/{r^2}$-interaction), and the Hooke atom (harmonic confinement and Coulomb interaction, a good description for two electrons in a quantum dot) \cite{al:2010ys}. The indistinguishability and the spin degree of freedom of the particles were included, but no spin-orbit interaction. The features already encountered for the Moshinsky atom in \cite{Yauez:2010fk} were reproduced qualitatively by both models: The entanglement grows with relative interaction strength and with the excitation. For antiparallel spins, the limit of vanishing interaction strengths does not necessarily yield a non-entangled state. The entanglement measures in \cite{al:2010ys} are identical to those used in \cite{Yauez:2010fk}, and it remains open whether this discontinuity effect has to be considered an artifact of the entanglement measures that are used, or whether a physical explanation will be provided in future. 

A numerical study of entanglement \cite{al:2010ys} can also be performed on helium-like atoms using eigenfunctions of the Kinoshita type \cite{PhysRev.105.1490}, which are a modification of the Hylleraas expansion \cite{Hylleraas:1928uq}. In the study \cite{al:2010ys}, the wave-function is approximated with functions of the following form \cite{Koga:1996ij}, 
\eq 
\ket{\Psi_N}=e^{- \chi s} \sum_{j=1}^N c_j s^{l_j/2} \left( \frac{t}{u} \right)^{m_j} \left(\frac u s\right)^{n_j/2} ,
\en 
where $s=|\vec r_1|+|\vec r_2|$, $t=|\vec r_1|-|\vec r_2|$ and $u=|\vec r_1-\vec r_2|$ naturally satisfy the triangle inequality, $s\ge u \ge |t|$. The exponent $\chi$, the mixing coefficients $\{c_j \}$ and the non-negative integers $\{l_j, m_j, n_j \}$ are parameters that are determined numerically \cite{Koga:1996ij}. The entanglement of the ground state is again lower than the entanglement of the triplet or singlet excited states. The ground-state entanglement decreases with increasing nuclear charge $Z$, which can be understood, again, as caused by the relative decrease of the electron-electron-interaction strength.

\subsubsection{Quantum critical points}
A connection between the behavior of entanglement at quantum critical points and at the ionization threshold in a few-electron atom is drawn in \cite{Osenda:2007kx} for the entanglement in the ground state of a spherical helium model in which the Coulombic repulsion between electrons is replaced by its spherical average. In this, again, effectively one-dimensional model, the von Neumann entropy (see (\ref{eoff})) of the reduced density matrix of one electron in the Helium ground state exhibits singular behavior at the critical point, {\it i.e.}~for the value of $Z$ for which the system becomes unbound. For the excited triplet state \cite{Osenda:2008fk}, the scaling properties of the von Neumann entropy is shown to be qualitatively different: Due to the opposite symmetry of the spatial part of the wave-function (antisymmetric ground state and symmetric triplet state), the entanglement is a continuous function of the nuclear charge. This can be understood from the fact that the ionized system possesses the same symmetry as the triplet state, while the symmetry of the ground state and the unbound system are distinct.

\subsubsection{Conclusions}
In general, while the quantum information tools presented in Section 2 are today available to study the entanglement of electrons in atoms, the complexity of many-electron atoms has not yet permitted a quantitative and realistic treatment. Still, a consistent picture that provides kinematic intuition emerges from the available studies: The entanglement grows, as expected, with the interaction between the electrons and with the energy, or principal quantum number, of the state. This observation is consistent with the fact that the Hartree-Fock approximation deteriorates when one considers highly excited states \cite{Moshinsky:1968nx}. These properties are contained in all models that were considered so far. Also, a growing spatial separation of two particles is not a direct indicator for a smaller degree of entanglement they exhibit: Indeed, in the above models, the typical distances grows with larger energies, and so does the entanglement. Given the qualitative agreement of all existing studies this tendency is expected to persist in future studies with more accurate electronic wave-functions.  However, with the onset of chaotic dynamics, qualitatively new features can be expected \cite{Venzl:2009kl,Garcia-Mata:2007kl,Benenti:2009oq,Gopar:2008tg,Wang:2004ij}. 

The state energy and entanglement were also shown to be not always monotonously dependent on each other, even in these simplified models \cite{Yauez:2010fk}. Entanglement thus offers an additional, independent analytical quantity in the study of atoms. The interplay of interaction and indistinguishability of particles has lead to interesting first results which we, however, do not yet have intuitive understanding for. It is unclear how the indistinguishability will manifest in systems for which more degrees of freedom, spin-orbit-coupling and other relativistic effects are incorporated. 

\subsubsection{Entanglement of ionized electrons} \label{ionizedelectrons}
In the process of single-photon double-ionization of atoms \cite{Chandra:2004ij}, a Werner state in the spins \cite{Werner:1989ve} emerges (see Eq.~(\ref{WernerState})). In the case of vanishing coupling between spins and angular momenta, the entanglement of the final state is fully determined by conservation laws and thereby totally independent of the details of the process. It can thus be inferred by the sole knowledge of the electrons' energies. In contrast, spin-orbit interaction induces dependences of the entanglement on the incident photon's polarization \cite{Chandra:2006kl}. The same argument applies also for the spin correlations of electrons ejected in the photoionization of linear molecules \cite{Chandra:2004dq}: They turn out to be widely independent of their spatial correlations, and, in the absence of spin-dependent interactions, the degree of entanglement can be predicted purely from conservation laws. It does, however, sensitively depend on the kinematic details of the process when spin-orbit interactions are included and thereby spatial symmetries are broken.

\subsection{Experiments with electrons: From two-center interference to two-particle correlations} \label{elecexp}
The above theoretical studies on entanglement between the spins of ejected electrons show again that a simple picture based on conservation laws has to be refined when a more realistic and intricate interaction is considered. The verification of spin-correlations between ejected electrons is, however, out of reach for current technology. In contrast, experimental progress has permitted detailed measurements of momentum-correlations between electrons, as we discuss hereafter. 

\subsubsection{Cohen-Fano interference}
The two-center interference of \emph{single} electrons has been a subject of intense theoretical and experimental studies. With the advance of reaction microscopes, the implementation of the original proposal of the Cohen-Fano interference in the photo-ionization of molecules has finally become possible. Oscillations were predicted in the cross-section which depend on the ejection angle of the photoelectron with respect to the molecular axis \cite{PhysRev.150.30}. These can be interpreted as the result of the interference of two inequivalent paths induced by the presence of the two nuclear centers and the previous delocalization of the electron at these two centers. 
Interference patterns, predicted by theoretical calculations of the photoionization of homonuclear molecules (see, {\it e.g.}, \cite{al:2009kl,Fernndez:2009qa} for recent discussions) could be verified in experiments that use photoionization \cite{Okunishi:2009fu} or other scattering mechanisms such as impact ionization \cite{al:2010mi,Hargreaves:2009dz,Galassi:2002ly,Sarkadi:2003qf}. 
The mechanisms which lead to the breakdown of the interference pattern are also rather well understood: For asymmetric molecular configurations, partial localization of the emitted electrons takes place \cite{al:2010fv}, as also probed experimentally for CO$_2$ in \cite{Sturm:2009ij}. Due to the breaking of the symmetry, a preferred direction emerges which also leads to an asymmetric electron emission. Such asymmetry can also be induced for homonuclear molecules simply by the polarization of the absorbed photon \cite{Martin:2007pi}. 

In such processes, also decoherence of the single-electron states plays an important role \cite{Zimmermann:2008dq}. In order to observe the interference pattern, the wavelengths of the directly ejected and the scattered wave need to coincide, and an interaction with one center which affects one of the pathways can jeopardize this coherence. Indeed, for large kinetic energies, the interaction at one center is strong enough to inhibit interference of the two pathways, and the interference pattern is lost. 

Despite being a single-particle interference effect, a connection of Cohen-Fano-interference to entanglement was recently proposed in \cite{Chelkowski:2010hc}. The scheme aims at probing the delocalization of the electronic wave-function over separated nuclei and thereby at probing the persistence of mode entanglement. The pump-probe-protocol first dissociates a $H_2^+$ molecule. After a certain delay time, in which the fragments fly apart, the probe pulse eventually also photoionizes the fragments and the ejected electron is recorded. The persistence of Cohen-Fano interferences for large delay times would show that even at large internuclear distance the electron can be well delocalized over two separated nuclei, whereas its absence would indicate that the state needs to be described by a classical mixture. In this sense, Cohen-Fano interferences witness mode-entanglement of a single delocalized electron: Such state of two separated nuclei and a single electron that is in a coherent superposition of being bound by either one of them realizes single-particle mode entanglement (see Section \ref{modenen}), this time, however, with a massive, charged particle. The experimental proposal hence opens a route to probe whether there are bounds to the distances over which massive particles can be delocalized, and its realization bears the potential to contribute to the debate on mode-entanglement mentioned in Section \ref{modenen}. 

\subsubsection{Two-electron measurements} \label{twoelectronsmol}
\begin{figure}[h] \center \includegraphics[width=10.5cm,angle=0]{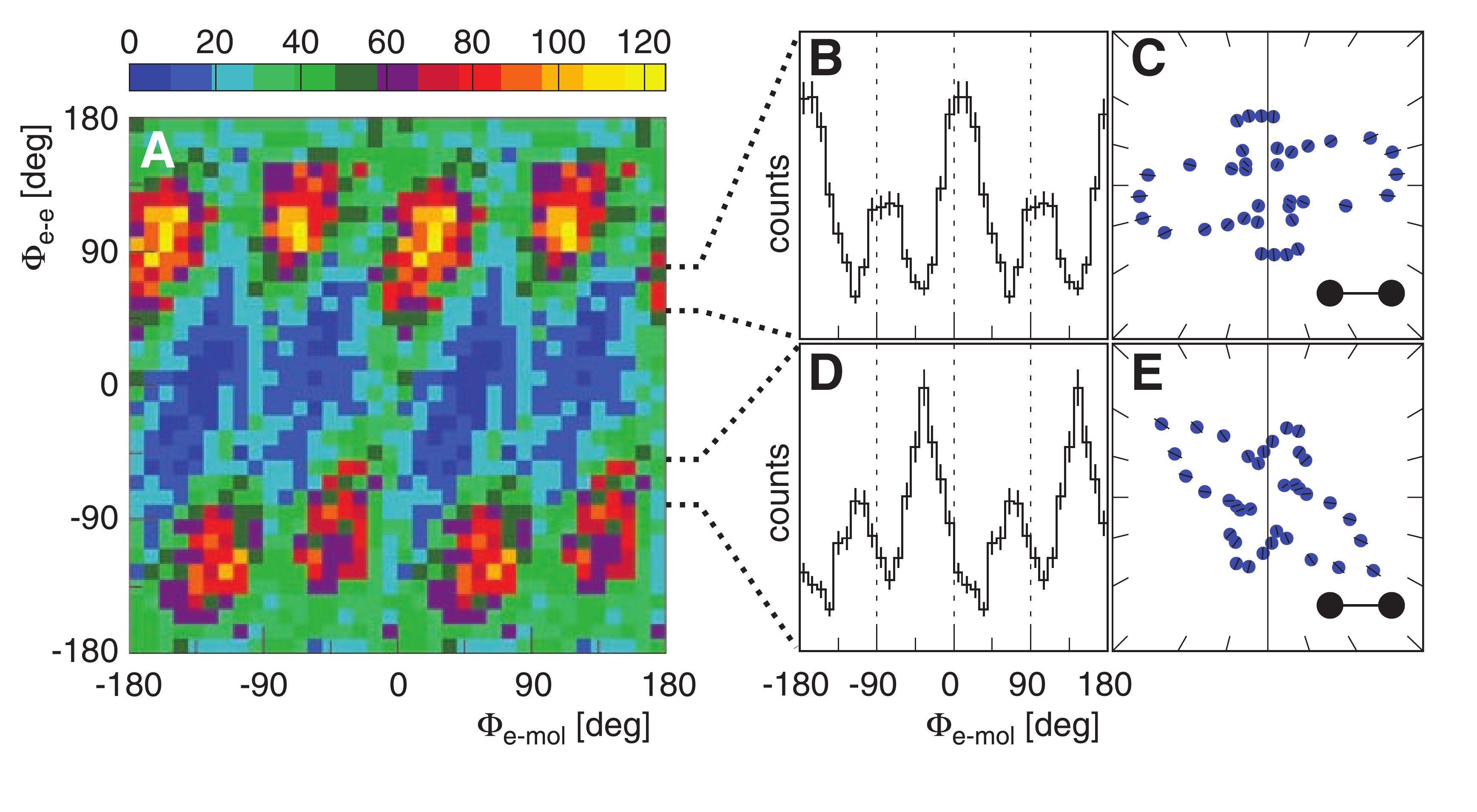}\caption{Courtesy of R. D\"orner \cite{al.:2007df}. Electron-electron correlations in the double-photo-ionization of molecular hydrogen \cite{al.:2007df}, for photon-energy $E_\gamma=160$ eV, and the energy of the second electron conditioned in the range 5 eV$<E_2<$25 eV, corresponding to $E_1\approx$ 85 eV to 105 eV. 
Panel A: Event number as a function of the angle between molecular axis and fast electron $\Phi_{\mathrm{e-mol}}$, and between both electrons $\Phi_{\mathrm{e-e}}$. Panel B: Event distribution in $\Phi_{\mathrm{e-mol}}$ conditioned on 50$^{\circ}< \Phi_{\mathrm{e-e}} < 80^{\circ} $. Panel D: Same for -80$^{\circ}< \Phi_{\mathrm{e-e}} < -50^{\circ} $. Panels C,E: Polar data corresponding to B and D, respectively. Note that when one selects the whole range of $\Phi_{\mathrm{e-e}}$, the resulting distribution in $\Phi_{\mathrm{e-mol}}$ results to be homogeneous and looses the structure exhibited in B-D.  From \cite{al.:2007df}. Reprinted with permission from AAAS.}  \label{doerner}  \end{figure}

The success in the verification of coherent \emph{single}-particle effects \cite{Zimmermann:2008dq} has recently also lead to a consideration of the impact of coherent \emph{many}-particle phenomena. The treatment of the photo-induced breakup of the H$_2$-molecule was proposed in \cite{W.-Vanroose:2005rt} and experimentally performed in \cite{al.:2007df} using the Cold Target Recoil Ion Momentum Spectroscopy (COLTRIMS) \cite{Drner:2000kl} technique. In this experiment, molecular hydrogen is doubly photoionized, and all reaction products are collected, such that a full reconstruction of all momenta takes place. This permits also to measure the orientation of the molecule in space, and thereby the emission angles of the electrons with respect to the molecular axis. The data show that when energy sharing between the electrons is very unbalanced, {\it i.e.}~one electron acquires a much larger kinetic energy than the other one, both electrons exhibit a Cohen-Fano-like interference pattern in their direction of emission. If, however, the electrons share the kinetic energy in a more balanced way, the single-electron interference pattern vanishes. This can be understood from the interaction between the electrons that is important in such range of energy sharing. If the data is filtered (postselected) to a given emission angle of one electron, an interference pattern re-emerges for the other one, {\it i.e.}~the electrons are strongly correlated. The data, reproduced here in Figure \ref{doerner}, thereby suggests that entanglement between the outgoing electrons is responsible for the correlations that are visible between their emission angles.
Such interference was observed for the same setting also in the sum of electron momenta \cite{Kreidi:2008bs}. It is visible in the momentum distribution of the individual electrons only for extremely asymmetric energy sharing, while it always persists for the sum of moments. 

The evidence of entanglement in these experiments is rather strong: Instead of probing a purely kinematic effect, the persistence of single-particle interference is assessed. Indeed, particles loose their ability to exhibit perfect interference fringes when they are entangled to other particles. By the selection of the electrons in certain ranges of energy sharing, one can effectively select different regimes of mutual interaction strength. Thereby, it is probed how the interaction between the electrons affects the interference pattern. On the other hand, only observables which commute with the momenta of the electrons are measured, and, strictly speaking, the correlations that are found can also be reproduced by local realistic theories, similarly as discussed below in (\ref{localrealisticion}). 

\subsubsection{Entanglement and symmetry breaking}
\begin{figure}[h] \center \includegraphics[width=5cm,angle=0]{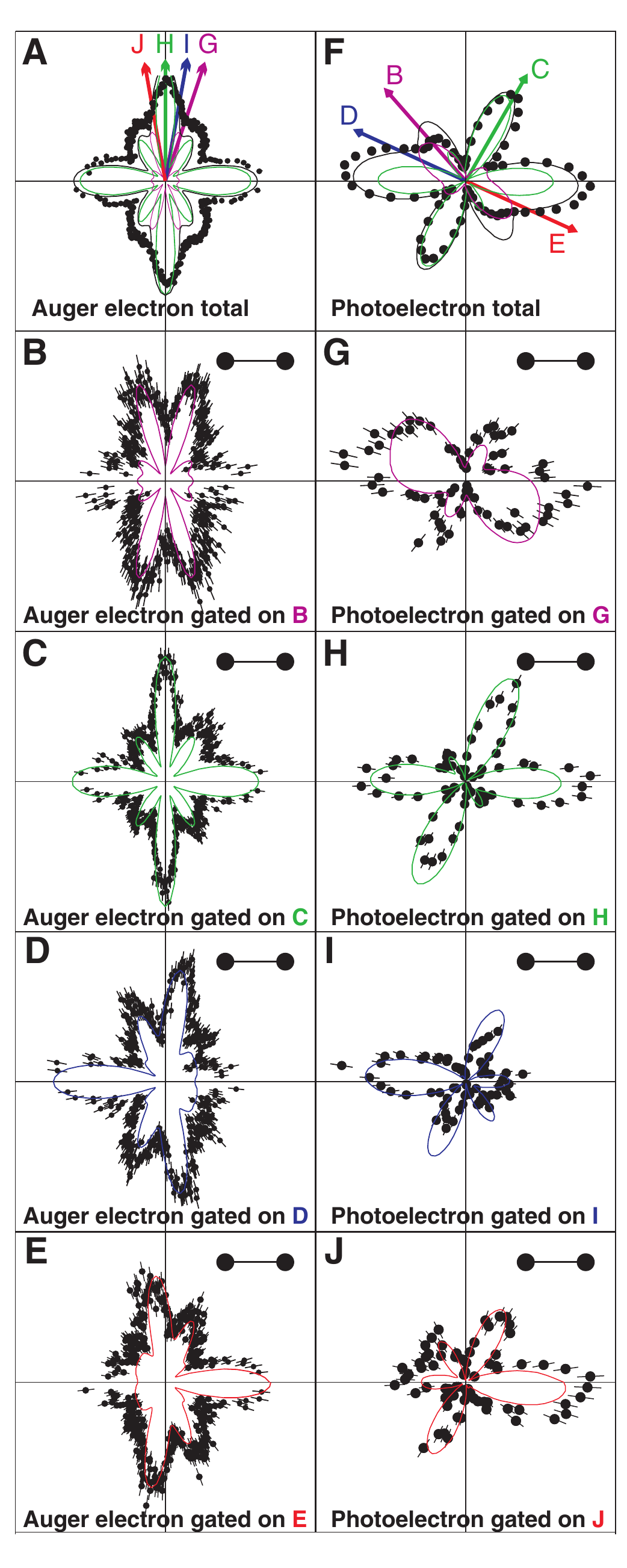}\caption{Courtesy of M. Sch\"offler \cite{Schoffler:2008cr}. Auger electron and photoelectron angular distributions in the molecular frame, for circularly polarized light with $E_\gamma=419$ eV. Dots denote experimental data, lines the theoretical prediction, Eq. 1 in \cite{Schoffler:2008cr}. The molecular axis is depicted by the barbell, the photon propagates into the figure plane. A and F show the non-conditioned data for Auger electron and photoelectron. By conditioning on selected angles of the Auger electron as in A, the distributions G-J result for the photoelectron. On the other hand, by selecting the photoelectron as shown in F, the Auger electron adopts the distributions B-E.   From \cite{Schoffler:2008cr}. Reprinted with permission from AAAS.
}  \label{corehole}  \end{figure}

An entanglement-based study proposes a resolution to the question whether a core vacancy created in a diatomic homonuclear molecule by ionization is localized at one center, or delocalized. By photo double ionization of N$_2$ and collection of both the photoelectron and the subsequently ejected Auger electron, it was shown \cite{Schoffler:2008cr} that the two electrons are, again, highly correlated: If conditioned on the photoelectron to be emitted in the direction of the molecular axis, the remaining Auger-electron is left distributed asymmetrically. In contrast, if the photoelectron is detected perpendicularly to the molecular axis, {\it i.e.}~without any preferred direction, also the Auger electron seems to be ejected from a delocalized state, as suggested by its angular distribution. This is clearly observed in the data shown in Figure \ref{corehole}. 
Hence, the electrons are anti-correlated, independently of the choice of the emission direction of the first one. The filter that fixes the emission direction to a certain angle can be interpreted as a choice of basis from either left and right localized states $\left\{ \ket{L}, \ket{R} \right\}$ or even and odd, delocalized states $\{ \ket{E}, \ket{O} \}$. The latter are defined as follows for Auger (A) and photo (P) electrons:
 \eq \ket{E}_{A/P}&=&\frac{1}{\sqrt 2} \left( \ket{L}_{A/P}+ \ket{R}_{A/P} \right) ,\\ 
\ket{O}_{A/P}&=&\frac{1}{\sqrt 2} \left( \ket{L}_{A/P}- \ket{R}_{A/P} \right) .\en
The electrons seem to be neither localized nor delocalized, but in an entangled state such that the condition of (de)localization imposed on one electron implies the (de)localization of the other. As a simple model, we can consistently describe the observed data with the following quantum state 
 \eq \ket{\Psi}&=& \frac{1}{\sqrt{2}} \left( \ket{R}_A\ket{L}_P -\ket{L}_A\ket{R}_P  \right) \nonumber \\ 
 &=&  \frac{1}{\sqrt{2}} \left( \ket{E}_A\ket{O}_P - \ket{O}_A\ket{E}_P  \right) .\en
One has to retain, however, that this ``change of basis'' performed by the condition on certain emission angles relies on strong model-assumptions: Effectively, only momenta are, again, measured. No information on any expectation value of observables that do not commute with the momentum is obtained, and the a-posteriori conditioning on a certain angle does not correspond to an active choice of different, non-commuting observables. The requirements for the violation of a Bell inequality (see Section \ref{Bellin}) are therefore \emph{not} met. Strictly speaking, the data currently cannot rule out a separable mixed state of the following form:
\eq \rho &\propto & p_1 \ket{L}_A\ket{R}_P \bra{L}_{A}\bra{R}_P + p_2 \ket{R}_A\ket{L}_P \bra{R}_{A}\bra{L}_P  \nonumber \\
&+& p_3 \ket{O}_A\ket{E}_P \bra{O}_{A}\bra{E}_P  + p_4 \ket{E}_A\ket{O}_P \bra{E}_{A}\bra{O}_P , \label{localrealisticion} \en
where the $p_i$ are the probability weights of the respective pure states. This state describes a purely probabilistic mixture of anti-correlated electron pairs, and corresponds to a local-realistic model which simulates the acquired data, and does not require entanglement. These equations reflect again very directly the problem of distinguishing a classical mixture, as given in (\ref{eq:psimm}), from a coherent superposition as in (\ref{sepmixed}), when only the correlations in one basis are measured.

\section{Conclusions and outlook}
In fact deeply rooted in atomic physics that initially stimulated the very development of quantum mechanics, entanglement has now largely overcome its restriction to the reductionist form of bipartite qubit correlations of some internal degree of freedom. It is a phenomenon sought under natural conditions which has become a tool for the deeper understanding for naturally occurring, typically complex systems, and thereby returns to its very conceptual origins. Simultaneously, it provides an analytic tool for many-body phenomena that are hard to understand in terms of single-particle observables. 

Given a fixed subsystem structure, a complete mathematical apparatus, insinuated in Section 2, is available today for the characterization of entanglement, despite the computational difficulties for the characterization of entanglement in mixed states with many parties and dimensions. \emph{Conceptual} difficulties like the problem of identical particles, discussed in Section 3, turned out not to possess a universal solution which can be applied like an all-purpose tool to all possible situations. Knowledge on the physical setup and the observables under consideration and, especially, on the restrictions which potentially apply for measurements are necessary to find the suitable treatment in a given scenario.

First studies on entanglement in bound systems of electrons have emerged, in which, however, the combined effects of the particles' indistinguishability, the long-range character of the Coulomb-interaction and the spin-orbit coupling have not yet been fully incorporated. 
Where the electrons' indistinguishability was taken into account, the very intrinsic feature of identical particles, however, {\it i.e.}~measurement-induced entanglement (see Section \ref{measurementinduced}), has not yet found applications in interacting systems apart from a Fermi gas model with screened Coulomb interaction which does not affect the spins, {\it i.e.}~the degree of freedom in which entanglement is considered \cite{Hamieh:2009zr,Hamieh:2010ly}. While, conceptually, the possession of a complete set of properties defines a physical reality and characterizes a separable subunit, measurement-induced entanglement beyond the engineered examples mentioned in Section \ref{entextr} can be expected in many situation, {\it e.g.}, between electrons ejected from the same orbital quantum state. In systems in which both the particles' indistinguishability and their mutual interaction play a prominent role, effects that can not be explained in terms of one of these aspects alone and hence require the understanding of their subtle interplay can be expected \cite{Bose:2005cl}. 

As opposed to bound systems in which entanglement is present due to a permanent binding interaction, the unbound decaying systems we have reviewed typically share the same physical reason for the existence of quantum correlations: Conservation laws for energy and momentum leave the fragments entangled in these very degrees of freedom, at any range of energy, from Feshbach-resonance induced decay of ultracold molecules \cite{Gneiting:2010qf} in the very low energy range, to electron-positron pairs created at very high photon energies \cite{Krekora:2005rq}. A quite intuitive feature is that the degree of entanglement between products which are very unbalanced in mass tends to be smaller than between constituents of similar properties, simply due to kinematics and  conservation of momentum. This rule of thumb can, however, be circumvented with suitable schemes \cite{Guo:2006qf,Chan:2003ly}, which shows that entanglement adds a qualitatively new feature to the description of dynamical systems which \emph{cannot} be completely reduced to kinematical quantities. The understanding of entanglement is hence by far not completed by considering the kinematics of processes: The very distinction between classical and quantum correlations and the mechanisms which lead to the breakdown of coherences remain the most important issues that need to be addressed. The quantification of wave-packet narrowing emerged as tool for the verification of correlations between the fragments. It holds the disadvantage that it only provides a clear signature for entanglement under the assumption of pure quantum states, {\it i.e.}~it cannot distinguish classical and quantum correlations. Correlations were indeed recently verified \cite{Schoffler:2008cr,al.:2007df}, however, in the momentum probability distribution rather than position.

Beyond the application in atomic and molecular physics, the conceptual issues we have discussed also arise in the presently very active field of quantum effects in biological systems. The question whether coherent effects play a relevant role for biological phenomena has arisen and is under debate  \cite{EisertBio,Briegel:2008ff,Arndt:2009uq}. Experimental evidence for the role of coherence in light-harvesting complexes, responsible for the functioning of photosynthesis, was recently obtained \cite{Ishizaki:2010kx,Collini:2010vn,Sarovar:2010ys} and has triggered intensive research activities in theory \cite{Fassioli:2010gb,Caruso:2009ys,scholak3560,Scholak:2011fk,Scholak:2010fk,Mulken:2010uq,Muhlbacher:2009kx,Hoyer:2010vn,Thorwart:2009ys}. Simultaneously, theoretical results support the idea of surviving dynamical entanglement at room temperature in an environment prohibitive for  entanglement in static systems \cite{Cai:2010qf,Cai:2010fu,Galve:2010kx}, which feeds the hope to encounter coherent effects in other noisy, wet and warm systems. The importance of multipartite entanglement \cite{scholak3560} as well as of molecular vibrations \cite{asadian} for efficient energy transport in networks was also shown recently, and contributes to the picture that quantum effects in biology could play an important functional role.

All this is evidence that the fields of quantum information and of atomic and molecular physics have only started to interact. The reason is twofold: On the one hand, the experimental capabilities that permit to resolve and verify coherent phenomena have only emerged recently \cite{Becker:2009oq}, and, on the other hand, conceptual issues inhibited the direct application of the highly abstract and idealized notions of quantum information science to this field. First successful applications of the concepts borrowed from quantum information yielded interesting results and new insight in the dynamics of atoms and molecules and provided answers to fundamental questions, {\it e.g.} regarding the delocalization and entanglement of massive particles in nature. The study of many-particle quantum coherence in atomic and molecular physics promises further interesting results in the next years, and will eventually lead to vital feedback to the field of quantum information itself. 

\ack The authors enjoyed fruitful and stimulating discussions with Celsus Bouri, Vivian Fran\c ca, Dominik H\"orndlein, Pierre Lugan, Fernando de Melo, Benno Salwey, Torsten Scholak, Markus Tiersch and Hannah Venzl, and are indebted to Vivian Fran\c ca and Pierre Lugan for careful proofreading. Financial support by the German National Academic Foundation (MCT), by DFG (FM), and by DAAD withiin the PROCOPE program (MCT and AB) is gratefully acknowledged. 

\clearpage
\section*{Bibliography}

\providecommand{\newblock}{}

%\bibliographystyle{iopart-num}
%\bibliography{bibliography}
\end{document}